\documentstyle[12pt]{article}
\begin{document}
\textheight 22 truecm
\textwidth 15 truecm
\topmargin -1 cm
\begin{titlepage}
\title{Complete On-Shell Renormalization Scheme for the
Minimal Supersymmetric Higgs Sector
\thanks{Supported in part by the Polish Committee for Scientific
        Research under the grant 2~0165~91~01}}
\author{
Piotr H. Chankowski
\thanks{On leave of absence from Institute of Theoretical Physics, Warsaw
University, Ho\.{z}a~69, 00-681 Warsaw, Poland.}\\
Istituto Nazionale di Fisica Nucleare, Sezione di Padova\\
Dipartimento di Fisica "Galileo Galilei", University of Padova
\and
Stefan Pokorski$^{~\dagger}$\\
Max-Planck-Institute f\"ur Physik, Werner-Heisenberg-Institute\\
\and
Janusz Rosiek\\
Institute of Theoretical Physics, Warsaw University\\
}

\date{December 1992}
\maketitle
\begin{abstract}
Systematic on-shell renormalization programme is carried out for the
Higgs and gauge boson sectors of the Minimal Supersymmetric Standard
Model. Complete 1-loop results for the 2- and 3-point Green's
functions are explicitly given. The Higgs boson masses and the cross
sections for the neutral scalar production in the $e^+e^-$ col\-li\-ders
are calculated.
\end{abstract}

\end{titlepage}

\renewcommand{\thesection}{Section~\arabic{section}.}
\renewcommand{\theequation}{\arabic{section}.\arabic{equation}}

\setcounter{equation}{0}
\section{Introduction}
\vskip 0.2cm

One-loop radiative corrections to the Higgs sector in the
Minimal Supersymmetric Standard Model (MSSM) have been extensively
studied by a number of authors \cite{B,HMASS,EPA,RGE,MY1,MY3}.
The motivation for this effort is
clear. The Higgs sector in the MSSM is strongly constrained (at the tree
level, in terms of only two free parameters one predicts a large number
of masses and couplings of the physical scalars) and it may play
important role in understanding the mechanism of the spontaneous
 ~$SU(2)\times U(1)$ ~symmetry breaking as well as in discovering (or
ruling out) supersymmetry. Indeed, if the soft supersymmetry breaking
scale is relatively high, say, ${\cal O}$(1 TeV), then the lightest
supersymmetric Higgs boson may well be the first trace of supersymmetry.
It is, therefore, important to study in detail the potential
experimental signatures of the MSSM Higgs boson and to understand
if one can experimentally distinguish the lightest supersymmetric Higgs
boson from the Minimal Standard Model one, without discovering the
other scalar states.

It is now well established and understood that, in presence of a heavy top
quark, 1-loop corrections in the MSSM Higgs sector are large. They
significantly alter the strategies for the experimental search, even for
the top quark mass close to the present experimental limit (e.g. the
channel ~$h^0\rightarrow A^0A^0$ ~may be open even for ~$m_t=90$ GeV).
In addition, 1-loop corrections introduce the dependence of the
predictions on a number of unknown parameters of the MSSM. There have
been used three main approaches to calculation of 1-loop corrections:
1) the effective potential approach (EPA) \cite{EPA}, 2) the renormalization
group approach (RGE) \cite{RGE} (with the heavy sparticles decoupled),
3) diagrammatic calculation of all the relevant Green's
functions \cite{MY1,MY3,BRI}.

The purpose of this paper is to give full details on the introduced
earlier \cite{MY1,MY3} systematic renormalization programme for the Higgs and
gauge boson sectors of the MSSM. Our approach is a straightforward
extension of the on-shell renormalization of the Standard Model \cite{HOL}.
The results of the complete 1-loop corrections to the scalar mass
spectrum calculated in this framework were given in ref. \cite{MY1} and those
for some of the Higgs boson production cross sections at the ~$e^+e^-$
{}~colliders, with all the 1-loop corrections to the Higgs boson couplings
included, in ref. \cite{MY3}. Here we also extend and systematize those
results, always with emphasize on phenomenology of the MSSM Higgs
production at the ~$e^+e^-$ ~colliders. Systematic discussion of the Higgs
boson decays is presented elsewhere \cite{MY4}.

A systematic renormalization programme for the Higgs sector in the MSSM
is useful in several ways. Firstly, our calculation provides a "reference
frame" for the other two methods which use certain approximations (e.g.
they neglect the contribution of the gauge sector, of the ~$p^2$
{}~dependent effects in the scalar self energies, of 1-loop corrections
to the vertices etc.). Thus, the accuracy of different methods in various
regions of the parameter space is now well checked. Secondly, our
formalism can be consistently extended to calculate loop corrections to
other sectors of the MSSM. And finally, comparison of our renormalization
framework with the other two methods as well as with the diagrammatic
calculation based on different renormalization conditions \cite{BRI} helps to
clarify certain points on the choice of independent parameters for the
Higgs sector in the MSSM and on its gauge and renormalization scheme
dependence.

It is well known that the MSSM Higgs sector is conveniently
parameterized in terms of ~$\tan\beta\equiv v_2/v_1$ ~(where ~$v_i$ ~are
the $VEVs$ of the two Higgs doublets present in the model) and the
pseudoscalar mass ~$M_{A^0}$. ~These two parameters are directly related
to the free (and unconstrained) parameters in the scalar Higgs potential.
In addition, ~$\tan\beta$ ~is an important parameter for model building,
with theoretical prejudice for ~$1\leq\tan\beta\leq m_t/m_b$ ~\cite{RSB}.
However, one should be aware of the fact, that ~$\tan\beta$ ~is not an
observable and beyond the tree level it is gauge and renormalization
scheme dependent. It is, of course, a standard feature of perturbation
theory that physical results (e.g. predictions for cross sections
expressed in
terms of measurable quantities such as masses and charges) are
renormalization scheme dependent up to the order higher than considered
(i.e. at 1-loop level this dependence is a 2-loop effect). However,
 ~$\tan\beta$ ~(as any other renormalizable parameter in the lagrangian),
when expressed in terms of the physical masses ~$M_{A^0}$ ~and
 ~$M_{h^0}$ ~($h^0$ ~is the lightest scalar), depends on the
renormalization scheme at the same order at which it is calculated (i.e.
at 1-loop level this dependence is also 1-loop effect). This implies, in
turn, that e.g. 1-loop corrected values of ~$M_{h^0}$, ~when expressed in
terms of ~$\tan\beta$ and ~$M_{A^0}$, ~are (for the same values of
 ~$\tan\beta$ ~and ~$M_{A^0}$) scheme dependent to 1-loop effects and the
same applies to cross sections for the ~$h^0$ production and its decay
widths.

In particular, the separation ~$\tan\beta\leq 1$ and ~$\tan\beta\geq 1$
is renormalization scheme dependent already to 1-loop effects and one
cannot use the theoretical limit ~$\tan\beta\geq 1$ ~unless in the same
(or at least effectively similar) scheme as the one used to get it. This
discussion becomes clearer when we compare the calculations in the
$R_{\xi}$ gauges and in the Landau gauge. As it will be seen in detail in
the next section, ~$R_{\xi}$ ~gauges require additional counterterms
$\delta v_i$ to renormalize tadpole diagrams, whereas in the Landau
gauge they can be renormalized with the mass and coupling constant
counterterms only. The additional freedom in renormalization conditions
for ~$\delta v_i$ ~ can make ~$\tan\beta$ ~arbitrarily different
(already at 1-loop level) from its definition used in the EPA
and RGE approaches (which use Landau gauge
and are consistently employed to get the limit ~$\tan\beta\geq 1$).

Our scheme has been deviced to be very close to the Landau gauge
calculations, although using the 't Hooft-Feynman gauge for easy work
with Feynman integrals. We can, therefore, use theoretically favoured
constraints for ~$\tan\beta$. ~One should remember, however, that they
are based on additional assumptions (like demanding radiative gauge
symmetry breaking in grand unification models)
which are much stronger than supersymmetry itself.
For phenomenological study it is therefore sensible to relax them.
On the other hand some limits on ~$\tan\beta$ are crucial for a meaningful
discussion of loop corrections. In particular, for
$\tan\beta\rightarrow 0$, because of the growing top quark Yukawa coupling,
the 1-loop corrected ~$M_{h^0}\rightarrow\infty$ ~(eventually
perturbation theory breaks down). In this paper we study the region
$1/2\leq\tan\beta\leq m_t/m_b$ of the renormalized ~$\tan\beta$ ~(in our
renormalization scheme). The lower limit is taken from constraints
on the tree level Yukawa couplings \cite{GRGU} - in our renormalization scheme
the Yukawa vertices receive only generic electroweak corrections (large
tadpole corrections are absorbed into Higgs boson renormalization
constants) and therefore the tree level lower limit for $\tan\beta$
can be used as a sensible constraint for our renormalized $\tan\beta$.
The upper limit is taken from the results on the mechanism of radiative
$SU(2)\times U(1)$ symmetry breaking studied in the Landau gauge in the
effective potential and renormalization group approaches; as mentioned
above our renormalized $\tan\beta$ is effectively very close to the one
used in the two other approaches.

Given the scheme dependence of the ~$(\tan\beta, M_{A^0})$ parameterization
it seems appropriate to complement it with the ~$(M_{h^0}, M_{A^0})$
parameterization of all the other physical quantities in the Higgs
sector. Here there is however the well known ambiguity: the tree level
mass spectrum of scalars is symmetric under the interchange ~$\tan\beta
\leftrightarrow\cot\beta$ ~whereas their couplings and, in consequence, cross
sections and decay widths are not. For fixed ~$m_t$ ~and ~$m_b$, ~this
symmetry is broken at the 1-loop level by Yukawa couplings which take
different values in the two regions but the two-fold ambiguity of
$(M_{h^0}, M_{A^0})$ ~parameterization remains to be present in certain
mass ranges. Still, we find the ~$(M_{h^0}, M_{A^0})$ ~parameterization
very convenient for phenomenological purposes and will use it heavily
in this paper.

The rest of the paper contains the details of our formalism and the
last chapter is devoted to an extensive phenomenological study for
the ~$e^+e^-$ ~colliders.

\setcounter{equation}{0}
\def\rvec#1{\vbox{\ialign{##\crcr
${\hspace{1pt}\scriptscriptstyle\rightarrow\hspace{-1pt}}
$\crcr\noalign{\nointerlineskip}
      $\hfil\displaystyle{#1}\hfil$\crcr}}}
\def\lvec#1{\vbox{\ialign{##\crcr
     ${\scriptscriptstyle\leftarrow}$\crcr\noalign{\nointerlineskip}
      $\hfil\displaystyle{#1}\hfil$\crcr}}}
\def\lrvec#1{\vbox{\ialign{##\crcr
${\hspace{1pt}\scriptscriptstyle\leftrightarrow\hspace{-1pt}}
$\crcr\noalign{\nointerlineskip}
      $\hfil\displaystyle{#1}\hfil$\crcr}}}

\section{Lagrangian and counterterms}
\vskip 0.2cm

\indent In this Section we define our renormalization scheme for the Higgs
sector of the MSSM.
In general we follow the notation and conventions of ref. \cite{ROS}, where the
full lagrangian and the complete set of Feynman rules for the minimal
supersymmetric standard model is given. Let us recall several most important
for us formulae. We denote the two Higgs doublets as $H_1 = (H^1_1, H^2_1)$,
$H_2 = (H^1_2, H^2_2)$, ~$<H_1> = (v_1/\sqrt 2, 0)$,
$<H_2> = (0, v_2/\sqrt 2)$. Scalar kinetic terms read:
\begin{eqnarray}
  L_{kin}=
  \overline H_1\left(\lvec{\partial_{\mu}}
  +{i\over 2}g_1 B_{\mu}-ig_2T^a W^a_{\mu}\right)
  \left(\rvec{\partial_{\mu}}
  -{i\over 2}g_1 B_{\mu}+ig_2T^a W^a_{\mu}\right)H_1\nonumber\\
  +\overline H_2\left(\lvec{\partial_{\mu}}
  -{i\over 2}g_1 B_{\mu}-ig_2T^a W^a_{\mu}\right)
  \left(\rvec{\partial_{\mu}}
  +{i\over 2}g_1 B_{\mu}+ig_2T^a W^a_{\mu}\right)H_2
\end{eqnarray}

\noindent ($T^a$ ~are the SU(2) generators normalized so that
${\rm Tr}(T^aT^b)={1\over 2}\delta^{ab}$).
Scalar potential has the well known form:

\begin{eqnarray}
  V_0&=&m^2_1 \overline H_1 H_1 +m^2_2
  \overline H_2 H_2 -m^2_{12}\left(\epsilon_{ab} H^a_1 H^b_2 +
  c.c.\right)\nonumber\\
  &+&{1\over 8}(g^2_1 +g^2_2) (\overline H_1 H_1 -\overline H_2 H_2)^2
  +{1\over 2}g^2_2 \mid \overline H_1 H_2 \mid^2
\end{eqnarray}

\noindent where $m^2_{12}$ is defined to be negative and
$\epsilon_{12}=-\epsilon_{21}=-1$.
The renormalization constants are defined as follows:
\begin{eqnarray}
  m^2_i \rightarrow Z^{-1}_{Hi} (m^2_i +\delta m^2_i),~~~
  m^2_{12} \rightarrow Z^{-1/2}_{H1}Z^{-1/2}_{H2}(m^2_{12}
  +\delta m^2_{12}),\nonumber
\end{eqnarray}
\begin{eqnarray}
  v_i\rightarrow Z^{1/2}_{Hi}(v_i-\delta v_i),~~~
  g_1\rightarrow Z_1 Z^{-3/2}_B g_1,~~~ g_2\rightarrow Z_2 Z^{-3/2}_W g_2,
\end{eqnarray}
\begin{eqnarray}
  H_i\rightarrow Z^{1/2}_{Hi} H_i,~~~ B_{\mu}\rightarrow Z^{1/2}_B B_{\mu},~~~
  W_{\mu}\rightarrow Z^{1/2}_W W_{\mu}\nonumber
\end{eqnarray}

Having renormalized all fields and parameters as indicated above, we introduce
gauge fixing term, expressed already in terms of the renormalized quantities
(in
this respect our approach differs slightly from
that employed in Hollik's paper \cite{HOL}):
\begin{equation}
  L_{gfix}=-{1\over 2\xi^B_1}~f^2
  -\sum_a{1\over 2\xi^{Wa}_1}~{\cal F}^a {\cal F}^a
\end{equation}
\noindent where
\begin{eqnarray}
  f=\partial^{\mu}B_{\mu}-{i\over {2\sqrt 2}}\sqrt {\xi^B_1\xi^B_2}g_1
  \left(\overline H^2_2v_2-H^2_2v_2
  +H^1_1v_1-\overline H^1_1v_1\right)\nonumber
\end{eqnarray}
\begin{eqnarray}
  {\cal F}^a = \partial ^{\mu} W^a_{\mu}&-&i\sqrt{\xi^{Wa}_1\xi^{Wa}_2}g_2
  \left( \overline H_2T^a<H_2>-<\overline H_2> T^a H_2\right.\\
  &+&\left.\overline H_1 T^a <H_1>-<\overline H_1> T^a H_1\right)
  \nonumber
\end{eqnarray}

\noindent We choose to work in the  gauge "infinitesimaly" different from the
't Hooft - Feynman gauge. Therefore we set
\begin{eqnarray}
  \xi^{B,W}_{1,2} \rightarrow 1+\delta \xi^{B,W}_{1,2}
\end{eqnarray}
and next use ~$\delta\xi^{B,W}_i$ ~as finite counterterms.
As usual, we define physical gauge boson fields ~$W^{\pm}$, ~$Z^0$
and ~$A^{\gamma}$ ~by the formulae
\begin{eqnarray}
  \left(\matrix{B_{\mu}\cr W^3_{\mu}}\right)
  =\left(\matrix{c_W&-s_W
              \cr s_W&c_W}\right)
  \left(\matrix{A^{\gamma}_{\mu}\cr Z^0_{\mu}}\right),~~~
  W^{\pm}_{\mu}={1\over \sqrt 2}\left(W^1_{\mu}\mp i~W^2_{\mu}\right)
\end{eqnarray}
where ~$c_W\equiv\cos\theta_W\equiv g_2/\sqrt {g^2_1+g^2_2}$,
$s_W\equiv\sin\theta_W$.
{}~We will also use the following combinations of the gauge fixing parameters:
\begin{eqnarray}
  \xi^Z_i\equiv\xi^{W^3}_i\cos^2\theta_W +\xi^B_i\sin^2\theta_W,~~~
  \xi^{\gamma}_i\equiv\xi^{W^3}_i\sin^2\theta_W+\xi^B_i\cos^2\theta_W,\nonumber
\end{eqnarray}
\begin{eqnarray}
  \xi^{\gamma Z}_i\equiv\left(\xi^{W^3}_i-\xi^B_i\right)
  \sin\theta_W\cos\theta_W, ~~~ \xi^W_i\equiv\xi^{W^{1,2}}_i
\end{eqnarray}

\noindent It is convenient to introduce the decomposition of the
renormalized Higgs fields:
\begin{equation}
  H^1_1 ={1\over \sqrt 2}(v_1+\phi_1+i~\varphi_1),
{}~~~ H^2_2 ={1\over \sqrt 2}(v_2+\phi_2+i~\varphi_2)
\end{equation}
and, next, the tree level mass eigenstates for the renormalized fields:
scalars ~$H^0_i\equiv\left(H^0,~h^0\right)$, ~pseudoscalars
{}~$H^0_{i+2}\equiv\left(A^0,~G^0\right)$ ~and charged Higgs bosons
{}~$H^{\pm}_i\equiv\left(H^{\pm},~G^{\pm}\right)$, where i=1,2 ($G^0\equiv
H^0_4$
{}~and  ~$G^{\pm}\equiv H^{\pm}_2$  ~are Goldstone bosons).
In what follows we will use compact matrix notation introduced in
\cite{ROS}. The $H^0_i$'s and $H^0_{i+2}$'s are related
to the original fields by the rotations \cite{HHG,ROS}:
\begin{eqnarray}
  \phi_i=Z^{ij}_R~H^0_j,~~~
  \varphi_i=Z^{ij}_H~H^0_{j+2}
\end{eqnarray}
where
\begin{eqnarray}
  Z_R=\left(\matrix{\cos\alpha&-\sin\alpha\cr\sin\alpha&\cos\alpha}\right)
  ,~~~
  Z_H=\left(\matrix{\sin\beta&-\cos\beta\cr\cos\beta&\sin\beta}\right)
\end{eqnarray}
For the charged Higgs bosons we have
\begin{eqnarray}
  \left(\matrix{H^2_1\cr\overline H^1_2}\right)
  =Z_{\pm}\left(\matrix{H^{-}\cr G^{-}}\right),~~~
  Z_{\pm}=\left(\matrix{\sin\theta & -\cos\theta\cr\cos\theta &
\sin\theta}\right)
\end{eqnarray}

\noindent For arbitrary values of ~$v_1$, $v_2$ ~(not necessarily
minimizing the tree level
potential; they may e.g. minimize the 1-loop potential) the angles ~$\alpha$,
{}~$\beta$, ~$\theta$ ~are given as follows. Let us first rewrite the
scalar potential
in terms of the fields ~$\phi_i$, ~$\varphi_i$ ~and $H^1_2$, ~$H^2_1$.
In this basis the symmetric mass matrices of the scalar fields have the form
\begin{eqnarray}
  L_{\phi_1\phi_2}=-{1\over 2}~\phi_i~\left({\cal M}^2_R\right)_{ij}~\phi_j
\end{eqnarray}
and similarly ~$\left({\cal M}^2_H\right)_{ij}$
{}~for ~$L_{\varphi_1\varphi_2}$ ~and ~$\left({\cal M}^2_{\pm}\right)_{ij}$
for $L_{H^2_1H^1_2}$.
We have:
\begin{eqnarray}
  \left({\cal M}^2_R\right)_{11}
  &=&m^2_1+{1\over 8}(g^2_1+g^2_2)(3v^2_1-v^2_2)\nonumber \\
  \left({\cal M}^2_R\right)_{22}
  &=&m^2_2+{1\over 8}(g^2_1+g^2_2)(3v^2_2-v^2_1)\\
  \left({\cal M}^2_R\right)_{12}
  &=&m^2_{12}-{1\over 4}(g^2_1+g^2_2)v_1 v_2\nonumber
\end{eqnarray}
\begin{eqnarray}
  \left({\cal M}^2_H\right)_{11}
  &=&m^2_1+{1\over 8}(g^2_1+g^2_2)(v^2_1-v^2_2)\nonumber\\
  \left({\cal M}^2_H\right)_{22}
  &=&m^2_2+{1\over 8}(g^2_1+g^2_2)(v^2_2-v^2_1)\\
  \left({\cal M}^2_H\right)_{12}&=&-~m^2_{12}\nonumber
\end{eqnarray}
\begin{eqnarray}
  \left({\cal M}^2_{\pm}\right)_{11}
  &=&\left({\cal M}^2_H\right)_{11}+{1\over 4}g^2_2 v^2_2\nonumber\\
  \left({\cal M}^2_{\pm}\right)_{22}
  &=&\left({\cal M}^2_H\right)_{22}+{1\over 4}g^2_2 v^2_1\\
  \left({\cal M}^2_{\pm}\right)_{12}
  &=&\left({\cal M}^2_H\right)_{12}+{1\over 4}g^2_2 v_1v_2\nonumber
\end{eqnarray}

\noindent The angle $\alpha$ is then given by
\begin{equation}
  \tan 2\alpha ={2\left({\cal M}^2_R\right)_{12}
  \over \left({\cal M}^2_R\right)_{11}-\left({\cal M}^2_R\right)_{22}}
\end{equation}
and the tree level scalar masses read:
\begin{eqnarray}
  m^2_H=\left({\cal M}^2_R\right)_{11}
  \cos^2\alpha+\left({\cal M}^2_R\right)_{22}
  \sin^2\alpha+\left({\cal M}^2_R\right)_{12}\sin2\alpha\\
  m^2_h=\left({\cal M}^2_R\right)_{11}
  \sin^2\alpha+\left({\cal M}^2_R\right)_{22}
  \cos^2\alpha-\left({\cal M}^2_R\right)_{12}\sin2\alpha
\end{eqnarray}
The analogous formulae hold for ~$\tan2\beta$ ~and ~$\tan2\theta$.
{}~If ~$v_1$, $v_2$ ~minimize the tree level potential we have ~$\theta=\beta$,
{}~ $Z_{\pm}~=~Z_{H}$ ~ with
\begin{eqnarray}
\tan\beta=v_2/v_1
\end{eqnarray}
{}~(this is the case with our renormalization conditions). Then
{}~$m^2_{G^0}=m^2_{G^{\pm}}=0$ ~but one has to add also the gauge fixing terms.
Note also that in this case the rotations remain unaltered by the inclusion
of the gauge fixing contributions to the mass matrices. They only shift the
Goldstone boson poles to ~$m^2_{G^0}=M^2_Z$, ~$m^2_{G^{\pm}}=M^2_W$
{}~(in the 't Hooft - Feynman gauge).
\vskip 0.5cm

The counterterms in the Higgs sector read:
\vskip 0.3cm

\noindent a) Counterterms linear in ~$H^0$ ~and ~$h^0$:
\begin{eqnarray}
  \delta L_{lin}=-~\delta_i~Z^{ij}_R~H^0_j
\end{eqnarray}
where
\begin{eqnarray}
  \delta_1&=&v_1~\delta m^2_1 -m^2_1~\delta v_1+v_2~\delta m^2_{12}
  -m^2_{12}~\delta v_2\nonumber\\
  &+&{1\over8}\left(\delta g^2_1+\delta
g^2_2\right)\left(v^3_1-v_1~v^2_2\right)
  +{1\over8}\left(g^2_2+g^2_1\right)
  \left[\left(v^2_2-3v^2_1\right)\delta v_1\right.\nonumber\\
  &+&\left.2~v_1~v_2~\delta v_2
  +\left(2v^3_1-v_1~v^2_2\right)\delta Z_{H_1}
  -v_1~v^2_2~\delta Z_{H_2}\right]
\end{eqnarray}
and $\delta_2$ is given by a similar equation with indices
$1$ and $2$ interchanged.

\vskip 0.3cm
\noindent b) Counterterms bilinear in the scalar fields (without derivatives):
\begin{eqnarray}
  \delta L_{SS} = -~{1\over 2}~Z^{lk}_R
  \left(\delta{\cal M}^2_R\right)_{lj}Z^{ji}_R~H^0_i H^0_k
\end{eqnarray}
with
\begin{eqnarray}
  \left(\delta{\cal M}^2_R\right)_{11}&=&\delta m^2_1
 +{1\over 8}\left(\delta g^2_1+\delta g^2_2\right)\left(3v^2_1-v^2_2\right)\\
  &+&{1\over 8}\left(g^2_1+g^2_2\right) \left[2 v_2\delta v_2-6 v_1\delta v_1
  -v^2_2\left(\delta Z_{H_1}+\delta Z_{H_2}\right) +6 v^2_1\delta
Z_{H_1}\right]\nonumber
\end{eqnarray}
\begin{eqnarray}
  \left(\delta{\cal M}^2_R\right)_{12}&=&\delta m^2_{12}
  -{1\over 4}\left(\delta g^2_1+\delta g^2_2\right) v_1 v_2\\
  &+&{1\over 4}\left(g^2_1+g^2_2\right)
  \left[v_1\delta v_2+v_2\delta v_1
  -v_1 v_2\left(\delta Z_{H_1}+\delta Z_{H_2}\right)\right]\nonumber
\end{eqnarray}
and ~$\left(\delta{\cal M}^2_R\right)_{22}$ ~given by the equation
with indices $1$ and $2$ interchanged
as compared to ~$\left(\delta{\cal M}^2_R\right)_{11}$. ~Similarly:
\begin{eqnarray}
  \delta L_{PP} = -~{1\over 2}~Z^{lk}_H \left(\delta{\cal M}^2_H\right)_{lj}
  Z^{ji}_H~H^0_{i+2}H^0_{k+2}~-~{1\over 2}~\delta\xi^Z_2~M^2_Z~H^0_4H^0_4
\end{eqnarray}
where
\begin{eqnarray}
  \left(\delta{\cal M}^2_H\right)_{11}&=&\delta m^2_1
  +{1\over 8}\left(\delta g^2_1+\delta g^2_2\right)\left(v^2_1-v^2_2\right)\\
  &+&{1\over 8}\left(g^2_1+g^2_2\right)\left[2 v_2\delta v_2-2 v_1\delta v_1
  -v^2_2\left(\delta Z_{H_1}+\delta Z_{H_2} \right)+2 v^2_1\delta
Z_{H_1}\right]\nonumber
\end{eqnarray}
\begin{eqnarray}
  \left(\delta{\cal M}^2_H\right)_{12}=-~\delta m^2_{12}
\end{eqnarray}
and ~$\left(\delta{\cal M}^2_H\right)_{22}$ ~is given by the equation with
indices $1$ and $2$ interchanged as compared to
{}~$\left(\delta{\cal M}^2_H\right)_{11}$. ~Finally:
\begin{eqnarray}
  \delta L_{\pm}=-~Z^{lk}_{\pm} \left(\delta{\cal M}^2_{\pm}\right)_{lj}
   Z^{ji}_{\pm}~H^{-}_i H^+_k~-~\delta\xi^W_2 M^2_W~H^+_2H^-_2
\end{eqnarray}
where
\begin{eqnarray}
  \left(\delta{\cal M}^2_{\pm}\right)_{11}&=&\delta m^2_1
  +{1\over 8}~\delta g^2_1\left(v^2_1-v^2_2\right)
  +{1\over 8}~\delta g^2_2\left(v^2_1+v^2_2\right)\nonumber\\
  &-&{1\over 4}~g^2_2 \left[2~v_2~\delta v_2
  -v^2_2 \left(\delta Z_{H_1}+\delta Z_{H_2}\right)\right]\\
  &+&{1\over 8}\left(g^2_2+g^2_1\right)\left[2v_2\delta v_2 -2v_1\delta v_1
  +2v^2_1 \delta Z_{H_1}
  -v^2_2\left(\delta Z_{H_1}+\delta Z_{H_2}\right)\right]\nonumber
\end{eqnarray}
\begin{eqnarray}
  \left(\delta{\cal M}^2_{\pm}\right)_{12}&=&-~\delta m^2_{12}
  +{1\over 4}~\delta g^2_2~v_1~v_2
  -{1\over 4}~g^2_2 \left(v_1~\delta v_2+v_2~\delta v_1\right)
  \phantom{aaaaa}\nonumber\\
  &+&{1\over 4}~g^2_2~v_1~v_2\left(\delta Z_{H_1}+\delta Z_{H_2}\right)
\end{eqnarray}
and ~$\left(\delta{\cal M}^2_{\pm}\right)_{22}$ ~is given by the equation with
{}~$v_1$, ~$\delta m^2_1$, ~$\delta Z_{H_1}$ ~and
{}~$v_2$, ~$\delta m^2_2$, ~$\delta Z_{H_2}$ ~interchanged as compared to
{}~$\left(\delta{\cal M}^2_{\pm}\right)_{11}$.
{}~In the equations above we have used the abbreviations:
\begin{eqnarray}
  \delta g^2_2 &\equiv&g^2_2\left(2~\delta Z_2-3~\delta Z_W\right)\\
  \delta g^2_1 &\equiv&g^2_1\left(2~\delta Z_1-3~\delta
Z_B\right)=-~g^2_1\delta Z_B
\end{eqnarray}
(the last equality holds
because ~$\delta Z_1 =\delta Z_B$ ~due to the ~$U(1)$ ~Ward identity).
\vskip 0.3cm

\noindent c) Counterterms to the scalar kinetic terms. Introducing
matrix notation ~$(\delta Z_H)_{ij}\equiv {\rm diag}(\delta Z_{H_1},
\delta Z_{H_2})$ ~we can write:
\begin{eqnarray}
  \delta L_{p^2}&=&{1\over 2}~Z^{lk}_R \left(\delta Z_H\right)_{lj} Z_R^{ji}
  ~\partial_{\mu} H^0_i \partial^{\mu} H^0_k \nonumber\\
  &+&{1\over 2}~Z^{lk}_H \left(\delta Z_H\right)_{lj} Z_H^{ji}~
  \partial_{\mu}H^0_{i+2} \partial^{\mu}H^0_{k+2}~\\
  &+& Z^{lk}_{\pm} \left(\delta Z_H\right)_{lj} Z_{\pm}^{ji}
  ~\partial_{\mu}H^-_i \partial^{\mu}H^+_k\nonumber
\end{eqnarray}
\vskip 0.3cm

\noindent d) Counterterms to the propagators which mix ~$A^0$, ~$G^0$
{}~with ~$Z^0$, $A^{\gamma}$,
and ~$G^{\pm}$ ~with ~$W^{\pm}$ ~have the form:

\begin{eqnarray}
  \delta L^0_{mix}&=&{1\over {2\sqrt {g^2_1+g^2_2}}}~Z^{ki}_H
  \left(\Delta^Z_k~Z^0_{\mu}+\Delta^{\gamma}_k~A^{\gamma}_{\mu}\right)
  \partial^{\mu}H^0_{i+2}\\
  &-&{1\over 2}M_Z\left(\delta\xi^Z_1
  -\delta\xi^Z_2\right)Z^0_{\mu}\partial^{\mu}H^0_4
  -{1\over 2}M_Z\left(\delta\xi^{\gamma Z}_1
  -\delta\xi^{\gamma  Z}_2\right)A^{\gamma}_{\mu}
  \partial^{\mu}H^0_4\nonumber
\end{eqnarray}
\begin{eqnarray}
  \delta L^{\pm}_{mix}={i\over 2}g_2\Delta^W_k
  Z^{ki}_{\pm}~W^{+}_{\mu}\partial^{\mu}H^{-}_i
  +{i\over 2}\left(\delta\xi^W_1-\delta\xi^W_2\right)
  M_W~W^+_{\mu}\partial^{\mu}H^-_2+{\rm c.c.}
\end{eqnarray}
where
\begin{eqnarray}
  \Delta^Z_k&=&(-1)^k\left[-g^2_1\left(v_k\left(\delta Z_1-\delta Z_B
  +\delta Z_{H_k}\right)-\delta v_k\right)\right.\nonumber\\
  &-&\left.g^2_2\left(v_k\left(\delta Z_2-\delta Z_W+\delta Z_{H_k}\right)
  -\delta v_k\right)\right]
\end{eqnarray}
\begin{eqnarray}
  \Delta^{\gamma}_k=(-1)^k~g_1~g_2\left(\delta Z_1-\delta Z_B
  -\delta Z_2+\delta Z_W\right)v_k
\end{eqnarray}
\begin{eqnarray}
  \Delta^W_k=(-1)^k\left[v_k\left(\delta Z_2-\delta Z_W
  +\delta Z_{H_k}\right)-\delta v_k\right]
\end{eqnarray}
Note that when ~$\tan\beta=v_2/v_1$ ~and
{}~$\delta v_1/v_1=\delta v_2/v_2$  ~(as will be the case
in our renormalization scheme) the counterterm for the
{}~$A^{\gamma}_{\mu}-A^0$ ~mixed propagator vanishes.
\vskip 0.3cm

\noindent e) Gauge bosons sector.
Counterterms to the gauge boson kinetic terms read
{}~($\partial^2_{\mu\nu}
\equiv g^{\mu\nu}\partial^2-\partial^{\mu}\partial^{\nu}$):
\begin{eqnarray}
  \delta L^{gb}_{kin}
  &=&{1\over 2}\left(c^2_W\delta Z_B+s^2_W\delta Z_W\right)
  A^{\gamma}_{\mu}\partial_{\mu\nu}^2A^{\gamma}_{\nu}
  -{1\over 2}~\delta\xi^{\gamma}_1~A^{\gamma}_{\mu}
  ~\partial^{\mu}\partial^{\nu}A^{\gamma}_{\nu}\nonumber\\
  &+&{1\over 2}\left(s^2_W\delta Z_B+c^2_W\delta Z_W\right)
  Z^0_{\mu}\partial_{\mu\nu}^2Z^0_{\nu}
  -{1\over 2}~\delta\xi^Z_1~Z^0_{\mu}~\partial^{\mu}\partial^{\nu}Z^0_{\nu}
  \nonumber\\
  &+&c_W~s_W\left(\delta Z_W-\delta Z_B\right)
  A^{\gamma}_{\mu}\partial_{\mu\nu}^2Z^0_{\nu}
  -\delta\xi^{\gamma Z}_1A^{\gamma}_{\mu}
  \partial^{\mu}\partial^{\nu}Z^0_{\nu}\\
  &+&\delta Z_W~W^{+}_{\mu}\partial^2_{\mu\nu}W^{-}_{\nu}
  -\delta\xi^W_1~W^+_{\mu}\partial^{\mu}\partial^{\nu}W^-_{\nu}\nonumber
\end{eqnarray}
Counterterms to the gauge boson masses read:
\begin{eqnarray}
  \delta L^{gb}_{mass}
  &=&{1\over 8}\left[\left(g^2_1+g^2_2\right)\delta X
  +2~g^2_2\left(v^2_1+v^2_2\right)\left(\delta Z_2-\delta Z_W\right)
  \right]Z^{0\mu}Z^0_{\mu}\nonumber\\
  &+&{1\over 4}~g_1~g_2\left(v^2_1+v^2_2\right)
  \left(\delta Z_2-\delta Z_W\right)Z^{0 \mu}A^{\gamma}_{\mu}\\
  &+&{1\over 4}~g^2_2\left[\delta X+2\left(v^2_1+v^2_2\right)
  \left(\delta Z_2-\delta Z_W\right)\right]W^{+\mu}W^{-}_{\mu}\nonumber
\end{eqnarray}
We have used here the equality ~$\delta Z_1=\delta Z_B$ ~to simplify the
formula and introduced the combination
\begin{eqnarray}
  \delta X=v^2_1~\delta Z_{H1}+v^2_2~\delta Z_{H2}
  -2~v_1~\delta v_1-2~v_2~\delta v_2
\end{eqnarray}
\vskip 0.3cm

\noindent f) Counterterms for the vertices $Z^0Z^0H^0_i$
and $A^{\gamma}Z^0H^0_i$ are:
\begin{eqnarray}
  \delta L_{VVS}={1\over 2}~\Delta_{ZZ}^k Z^{ki}_R
  ~Z^{0\mu}Z^0_{\mu}H^0_i+{1\over2}~g_1 g_2\left(\delta Z_2-\delta Z_W\right)
  C^i_R~Z^{0\mu}A^{\gamma}_{\mu}H^0_i
\end{eqnarray}
where
\begin{eqnarray}
  \Delta^k_{ZZ}=v_k~g^2_2\left(\delta Z_2-\delta Z_W\right)
  +{1\over2}\left(g^2_1+g^2_2\right)\left(v_k\delta Z_{H_k}
  -\delta v_k\right)
\end{eqnarray}
and for future use we have introduced the matrix ~$C^i_R$:
\begin{eqnarray}
  C^i_R\equiv v_1~Z^{1i}_R+v_2~Z^{2i}_R
\end{eqnarray}
\vskip 0.3cm

\noindent g) Counterterms for the vertices ~$Z^0H^0_iH^0_{j+2}$
and ~$A^{\gamma}H^0_iH^0_{j+2}$~ are given by:
\begin{eqnarray}
   \delta L^0_{VSP}&=&-~{1\over{2\sqrt{g^2_1+g^2_2}}}~
   Z^{lj}_H Z^{ki}_R \left(\overline\Delta\right)_{lk}~
   Z^0_{\mu} \left(H^0_{j+2}\lrvec{\partial^{\mu}} H^0_i\right)\nonumber\\
   &-&{g_1~g_2\over{2\sqrt{g^2_1+g^2_2}}}~A^{ij}_M
   \left(\delta Z_2-\delta Z_W\right)
   A^{\gamma}_{\mu}\left( H^0_{j+2} \lrvec{\partial^{\mu}} H^0_i\right)
\end{eqnarray}
where ~$\lrvec{\partial^{\mu}}=\rvec{\partial^{\mu}}
-\lvec{\partial^{\mu}}$,~$A_M^{ij}\equiv Z_R^{1i}Z_H^{1j}-Z_R^{2i}Z_H^{2j}$
 ~and ~$\left(\overline\Delta\right)_{lk}
={\rm diag}\left(\overline\Delta_1,~-\overline\Delta_2\right)$ ~with
\begin{eqnarray}
  \overline\Delta_k=g^2_1~\delta Z_{H_k}
  +g^2_2\left(\delta Z_2-\delta Z_W+\delta Z_{H_k}\right)
\end{eqnarray}
Note that although the tree level couplings $A^{\gamma}Z^0H^0_i$ and
$A^{\gamma}H^0_iH^0_{j+2}$ vanish in the initial
lagrangian, 1-loop corrections to those vertices are divergent and
their counterterms are given by the above expressions.
The apparent contradiction with the "theorem" that, if the vertex
does not exist at the tree level, corrections to it must be finite, can
be attributed to the ~$Z^0 - \gamma$ mixing. Indeed, as we will show in
the next section (eqs. \ref{eq_dzw}, \ref{eq_dz2}), the combination
{}~$\delta Z_2-\delta Z_W$ is entirely determined by ~$Z^0 - \gamma$ mixed
(unrenormalized) propagator at zero momentum:
\begin{eqnarray}
   \delta Z_2-\delta Z_W=-{1\over c_W s_W M^2_Z}~\Pi_{\gamma Z}(0)
\end{eqnarray}
which gets contributions only from gauge/Higgs bosons sector
and thus has the universal character; no new renormalization
conditions can be imposed on the $A^{\gamma}Z^0H^0_i$ and
$A^{\gamma}H^0_iH^0_{j+2}$ vertices.

\setcounter{equation}{0}
\section{Renormalization conditions and physical Higgs boson masses}
\vskip 0.2cm

\indent Counterterms defined in the preceding section can
be used to renormalize divergences
of the loop corrections in the gauge and Higgs boson sectors of MSSM.
Their finite parts must be fixed by choosing appropriate renormalization
conditions. The renormalization scheme we are going to define should
be compatible with supersymmetry. Since we want to calculate measurable
quantities, we are interested in preserving supersymmetry of the ~$S-$
matrix and not necessarily of the Green functions themselves.
With this restriction, our counterterms are sufficient
for the renormalization programme provided a supersymmetry preserving
regularization is used. (We will therefore use dimensional reduction
\cite{DRED} which preserves SUSY at least to one loop.)
This is  because the wave function renormalization constants (like
{}~$\delta Z_{H_i}$), which violate supersymmetry (e.g. the
infinite parts of the wave function renormalization constants of bosons and
fermions from the same supermultiplet are different when working
in the Wess-Zumino gauge with the physical fields only), cancel in the
{}~$S-$matrix. Below we specify our "on-shell" renormalization scheme.
Quantities with a hat (which are finite) are obtained from divergent quantities
(without a hat) by adding counterterms defined in section 2.

\noindent 1) Gauge bosons sector.

\noindent We write the 1$-$PI two point functions
of the massive gauge boson propagators as
\begin{eqnarray}
  i~\left(g^{\mu\nu}-{k^{\mu}k^{\nu}\over k^2}\right)\Pi^T\left(k^2\right)
  +i~{k^{\mu}k^{\nu}\over k^2}~\Pi^L\left(k^2\right)
\end{eqnarray}
and for the photon as
\begin{eqnarray}
i\left(g^{\mu\nu}k^2-k^{\mu}k^{\nu}\right)\Pi^{\prime}_{\gamma}\left(k^2\right)
\end{eqnarray}
and require  that
\begin{eqnarray}
\hat\Pi^{\prime}_{\gamma}\left(k^2=0\right)&=&0\nonumber\\
{\cal R}e~\hat\Pi^T_W\left(k^2=M^2_W\right)&=&0\nonumber\\
{\cal R}e~\hat\Pi^T_Z\left(k^2=M^2_Z\right)&=&0\\
{\cal R}e~\hat\Pi^T_{\gamma Z}\left(k^2=0\right)&=&0\nonumber
\end{eqnarray}
that is that the tree level $W^{\pm}$ and $Z^0$ masses are in
fact physical ones and that the real photon couples
without admixture of $Z^0$. Also the photon propagator has the
residuum equal to unity \cite{HOL}.
Explicitly these conditions give:
\begin{eqnarray}
\label{eq_dzw}
  \delta Z_W={\cal R}e\left(\Pi^{\prime}_{\gamma}\left(0\right)
  -2\cot\theta_W{\Pi^T_{\gamma Z}\left(0\right)\over M^2_Z}
  +{\Pi^T_W\left(M^2_W\right)-c^2_W\Pi^T_Z\left(M^2_Z\right)
  \over s^2_WM^2_Z}\right)
\end{eqnarray}
\begin{eqnarray}
  \delta Z_B={\cal R}e\left(\Pi^{\prime}_{\gamma}\left(0\right)
  +2\tan\theta_W{\Pi^T_{\gamma Z}\left(0\right)\over M^2_Z}
  -{\Pi^T_W\left(M^2_W\right)-c^2_W\Pi^T_Z\left(M^2_Z\right)
  \over c^2_WM^2_Z}\right)
\end{eqnarray}
\begin{eqnarray}
\label{eq_dz2}
  \delta Z_2={\cal
R}e\left(\Pi^{\prime}_{\gamma}\left(0\right)-{\left(2c^2_W+1\right)
  \over s_Wc_W}{\Pi^T_{\gamma Z}\left(0\right)\over M^2_Z}
  +{\Pi^T_W\left(M^2_W\right)-c^2_W\Pi^T_Z\left(M^2_Z\right)
  \over s^2_WM^2_Z}\right)
\end{eqnarray}
\begin{eqnarray}
\label{eq_dx}
  \delta X&=&{4\over g^2_2}{\cal
  R}e\left(M^2_W\Pi^{\prime}_{\gamma}\left(0\right)
  +2s_Wc_W\Pi^T_{\gamma Z}\left(0\right)\right.\nonumber\\
  &+&\left.{c^2_W-s^2_W\over s^2_W}\Pi^T_W\left(M^2_W\right)
  -{c^4_W\over s^2_W}\Pi^T_Z\left(M^2_Z\right)\right)
\end{eqnarray}

\noindent It is also possible to change slightly the finite parts of these
renormalization constants so that the $Z^0$ is renormalized on-shell
{\it including} the $Z^0 - \gamma$ mixing \cite{HOL2}:
\begin{eqnarray}
  {\cal R}e\left(\hat\Pi^T_Z\left(M^2_Z\right)
  +{\left(\hat\Pi^T_{\gamma Z}\left(M^2_Z\right)\right)^2
  \over
  M^2_Z\left(1-\hat\Pi^{\prime}_{\gamma}\left(M^2_Z\right)\right)}\right)=0
\end{eqnarray}
This amounts to small changes in the finite parts of counterterms
calculated in (\ref{eq_dzw}-\ref{eq_dx}).

\noindent The finite counterterms $\delta\xi^{B,W}_1$ are used to
locate poles of the
longitudinal parts of the gauge boson propagators at $M^2_W$, $M^2_Z$
and 0, respectively:
\begin{eqnarray}
  \delta\xi^W_1=-\delta Z_W-{1\over M^2_W}{\cal
R}e~\left[\Pi^L_W\left(M^2_W\right)-\Pi^T_W\left(M^2_W\right)
   +\hat\Pi^T_W\left(M^2_W\right)\right]
\end{eqnarray}
\begin{eqnarray}
  \delta\xi^Z_1&=&-\left(\cos^2\theta_W\delta Z_W
  +\sin^2\theta_W\delta Z_B\right)\nonumber\\
  &-&{1\over M^2_Z}{\cal R}e
  ~\left[\Pi^L_Z\left(M^2_Z\right)-\Pi^T_Z\left(M^2_Z\right)
  +\hat\Pi^T_Z\left(M^2_Z\right)\right]
\end{eqnarray}
\begin{eqnarray}
  \delta\xi^{\gamma}_1=0
\end{eqnarray}
Note that ~$\delta\xi_1^{W,Z}$ ~should be finite which provides a
useful test of the calculation.

\noindent 2) Tadpoles.

\noindent In order to make contact with the effective potential approach
of ref. \cite{EPA} (which uses Landau gauge) we want
to work with the fields $H^0$ and $h^0$ which have vanishing vacuum
expectation values order by order in perturbation theory. We therefore require
that the sum of the tree and
loop tadpoles vanishes. Furthermore, in our on-shell
renormalization scheme it is convenient to preserve at 1$-$loop
(and higher) the tree
level relations between $VEVs$ ~$v_i$ ~and the renormalized
parameters, i.e. we require
that the tree level tadpole vanishes by itself. Equivalently, tadpole
counterterms cancel the 1-loop tadpoles. In
Lorentz gauges, where $\xi^{B,W}_2=0$ (of which Landau gauge
with, in addition, $\xi^{B,W}_1=0$ is only
one example), one usually takes ~$\delta v_i=0$. ~It would be therefore
convenient to use $\delta v_i$ to cancel only this
contribution to tadpoles which
appear when $\xi^{B,W}_2$'s are switched on. However such separation
turns out to be
impossible due to the spurious infrared divergences which such
a procedure introduces.
Nevertheless, it is possible to do that for infinite parts only.
We fix therefore
$\delta v_i$ (in the dimensional reduction we employ) by the formula:
\begin{eqnarray}
  \delta v_i={1\over 4\left(4\pi\right)^2}
  \left(3g^2_2+g^2_1\right)v_i\left({2\over d-4}+\gamma-\log 4\pi\right)
\end{eqnarray}
and cancel the remaining parts of the tadpoles with the counterterms
$\delta m^2_1$ and $\delta m^2_2$.
This means that we eliminate these counterterms
everywhere in favour of tadpoles of the original scalar fields
{}~$\phi_i$:
\begin{eqnarray}
  T_{\phi_i}=Z^{ij}_R~{\cal T}^j
\end{eqnarray}
Notice, that we have ~$\delta v_1/v_1=\delta v_2/v_2\equiv\delta v/v$, which is
essential
for the definition of the electric charge by the Thomson limit. Without
this equality the on-shell photon would mix with ~$A^0$ ~and the Ward
identity \cite{HOL} leading to the correct definition of ~$e$ ~would be
spoiled.

After some rearrangements and substitutions, $\delta m^2_1$ and $\delta
m^2_2$ can be written down as:
\begin{eqnarray}
  \delta m^2_1 &=& -{{\cal T}_1\over v_1}- \tan\beta~\delta m^2_{12}
   +\left(m^2_1+\tan\beta~m^2_{12}\right){\delta v\over v}\nonumber\\
   &-&{1\over8}\left[g^2_2\left(2\delta Z_2-3\delta Z_W\right)
           - g^2_1\delta Z_B\right]\left(v^2_1-v^2_2\right)\nonumber\\
   &-&{1\over8}\left(g^2_2+g^2_1\right)\left[\left(v^2_2-5v^2_1\right)
   {\delta v\over v} + \left(3v^2_1-v^2_2\right)\delta Z_{H_1}-\delta
   X \right]\nonumber\\
   \delta m^2_2 &=& -{{\cal T}_2\over v_2}- \cot\beta~\delta m^2_{12}
   +\left(m^2_2+\cot\beta~m^2_{12}\right){\delta v\over v}\\
   &-&{1\over8}\left[g^2_2\left(2\delta Z_2-3\delta Z_W\right)
           - g^2_1\delta Z_B\right]\left(v^2_2-v^2_1\right)\nonumber\\
   &-&{1\over8}\left(g^2_2+g^2_1\right)\left[\left(v^2_2+5v^2_1
   -2\tan^2\beta v^2_1\right){\delta v\over v}\right.\nonumber\\
   &+&\left. v^2_1\left(\cot^2\beta-3\right)\delta Z_{H_1}
   +\left(2-\cot^2\beta\right)\delta X\right]\nonumber
\end{eqnarray}

\noindent 3) Mixing of pseudoscalars $A^0$ and $G^0$ with gauge bosons.

Because the mass of $A^0$ is usually chosen to parameterize the Higgs sector of
the model, we require that the on-shell $A^0$ does not mix with $Z^0$. Writing
the corresponding mixed propagator as
\footnote{This is as a compactified version of the standard
notation: $\Sigma^1_{ZP}\equiv\Sigma_{Z^0A^0}$,
$\Sigma^2_{ZP}\equiv\Sigma_{Z^0G^0}$,
$\Sigma^{11}_{PP}\equiv\Sigma_{A^0A^0}$,
$\Sigma^{12}_{SS}\equiv\Sigma_{h^0H^0}$ etc.}
\begin{eqnarray}
  -~p^{\mu}~\Sigma_{ZP}^i\left(p^2\right)~~~~~~i=1,2
\end{eqnarray}
and demanding ${\cal R}e~\hat\Sigma_{ZP}^1\left(M_A^2\right)~=~0$ we get:
\begin{eqnarray}
  \delta Z_{H1}=(1+\cos^2\beta){\delta v_1\over v_1}
  +\sin^2\beta {\delta v_2\over v_2}
  +{g^2_2\over 4}{\delta X\over M^2_W}
  -{\tan\beta\over M_Z}{\cal R}e~\Sigma_{ZP}^1\left(M^2_A\right)
\end{eqnarray}
\begin{eqnarray}
  \delta Z_{H2}=\cos^2\beta {\delta v_2\over v_2}
  +(1+\sin^2\beta){\delta v_1\over v_1}
  +{g^2_2\over 4}{\delta X\over M^2_W}
  -{\cot\beta\over M_Z}{\cal R}e~\Sigma_{ZP}^1(M^2_A)
\end{eqnarray}
The finite "counterterms"
$\delta\xi^{\gamma Z}_2$ can be used to cancel
the $\gamma- G^0$ mixing at $p^2=0$:
\begin{eqnarray}
\label{eq_spg2}
  {\cal R}e~\Sigma_{\gamma P}^2(0)-s_Wc_WM_Z\left(\delta Z_2-\delta Z_W\right)
  -{1\over2}M_Z\left(\delta\xi^{\gamma Z}_1-\delta\xi^{\gamma Z}_2\right)=0
\end{eqnarray}

\noindent 4) Pseudoscalar propagators.

\noindent To determine $\delta m^2_{12}$ we require:
\begin{eqnarray}
  {\cal R}e~\hat\Sigma_{PP}^{11}\left(M^2_A\right)=0
\end{eqnarray}
which gives:
\begin{eqnarray}
  \delta m^2_{12}=\sin\beta\cos\beta
  \left[{\cal R}e~\Sigma^{11}_{PP}\left(M^2_{A^0}\right)
  -M^2_A\left(\sin^2\beta\delta Z_{H1}
  +\cos^2\beta\delta Z_{H2}\right)\right.
\end{eqnarray}
\begin{eqnarray}
  &+&\sin^2\beta\left({1\over 8}\left(g^2_1+g^2_2\right)
  \left(v_1\delta v_1-v_2\delta v_2\right)
  +\left(m^2_1\delta v_1+m^2_{12}\delta v_2
  -T_{\phi_1}\right)/v_1\right)\nonumber\\
  &+&\left.\cos^2\beta\left({1\over 8}\left(g^2_1+g^2_2\right)
  \left(v_2\delta v_2-v_1\delta v_1\right)
  +\left(m^2_2\delta v_2+m^2_{12}\delta v_1
  -T_{\phi_2}\right)/v_2\right)\right]\nonumber
\end{eqnarray}

\noindent In addition we can arrange for
\begin{eqnarray}
  {\cal R}e~\hat\Sigma_{PP}^{22}\left(M^2_Z\right)=0,
{}~~~~{\cal R}e~\hat\Sigma_{+-}^{22}\left(M^2_W\right)=0
\end{eqnarray}
This, together with (\ref{eq_spg2}), determines the three independent constants
{}~$\delta\xi^B_2$, ~$\delta\xi^{W3}_2$ ~and ~$\delta\xi^{W}_2$.

\noindent 5) Renormalization of the standard model is usually
performed in such a way \cite{HOL,HOL2}
that $e=g_1g_2/\sqrt {g^2_1+g^2_2}$ is equal to the electric charge as measured
in the Thompson limit, that is $e^2/4\pi =1/137.036$.
This is achieved as follows.
Denote by $Z_{left}$ and $Z_{right}$ the wave function
renormalization constants of the
left and right handed electrons, respectively.
$Z_{left}$ and $Z_{right}$ can be used to give
the residues of the poles of the left and right handed electron
propagators equal to unity.
Then, writing the corrected electron-electron-photon vertex as
\begin{eqnarray}
  ie\gamma^{\mu}+ie\hat\Lambda^{\mu}_e
  =ie\gamma^{\mu}+ie\hat\Lambda^{\mu}_{eV}+ie\gamma^5\hat\Lambda^{\mu}_{eA}
\end{eqnarray}
one has
\begin{eqnarray}
  \hat\Lambda^{\mu}_{eV}&=&\Lambda^{\mu}_{eV}
  +{1\over 2}\left(\delta Z_{left}+\delta Z_{right}\right)
  +{1\over 4}\left(\delta Z_2-\delta Z_W\right)\nonumber\\
  &=&\Lambda^{\mu}_{eV}
  +{1\over 2}\left(\delta Z_{left}+\delta Z_{right}\right)
  -{1\over 4s_Wc_W}{\Pi^T_{\gamma Z}\left(0\right)\over M^2_Z}
\end{eqnarray}
where we have used (\ref{eq_dzw}, \ref{eq_dz2}). This is the equation
which, in the Standard Model, ensures that \cite{HOL,HOL2}:
\begin{eqnarray}
  \hat\Lambda^{\mu}_{eV}\left(q^2=0,p^2_1=p^2_2=m^2_e\right)=0
\end{eqnarray}
and the same condition can be used in MSSM, as a result of the residual ~$U(1)$
{}~gauge symmetry. This completes our renormalization scheme.

Basic formulae used to calculate the unrenormalized two point Green
functions, which are
necessary to determine the counterterms are collected in the Appendix A. Using
expressions displayed there and following the prescriptions given above
one can compute all the renormalization constants in the gauge and Higgs
sectors. They can subsequently
be used to renormalize all relevant self-energies for arbitrary
momenta $p^2$. The cancellation of divergences between counterterms
and bare two point Green functions provides a very strict test of the
calculations as well as of the computer programmes.

Having renormalized self energies one usually calculates one loop
corrected masses from the "perturbative" formula:
\begin{eqnarray}
  M^2_{H^0_i}=m^2_{H^0_i}+{\cal
R}e~\hat\Sigma_{SS}^{ii}\left(p^2=m^2_{H^0_i}\right)
\end{eqnarray}
However, since in MSSM the corrections to Higgs boson masses are expected
to be large, we do not use this formula and instead determine the masses as
the exact poles of the scalar Higgs boson ~$2\times2$ ~matrix
propagator \cite{MY1}. We thus have ~$M^2_{H^0_i}=p^2_{0i}$ ~where
{}~$p^2_{0i}$ ~are the two solutions of the equation:
\begin{eqnarray}
\label{eq_hmass}
{\cal R}e\left[\left(p^2-m^2_{H^0_1}-\hat\Sigma_{SS}^{11}(p^2)\right)
  \left(p^2-m^2_{H^0_2}-\hat\Sigma_{SS}^{22}(p^2)\right)
  -{\left(\hat\Sigma_{SS}^{12}(p^2)\right)}^2\right]=0\nonumber\\
\end{eqnarray}
In fact, we have checked numerically that the 1-loop masses computed from
the "perturbative" formula differ significantly from the
exact poles of the propagator. The difference between the two
is particularly big for small ~$\tan\beta$ ~as well as for large
{}~$\tan\beta$ ~and ~$M_A\sim M_Z$.
Similar difference has been also noticed recently in ref. \cite{BRI}.
The results for scalar Higgs boson masses obtained within the approach
just described have been extensively discussed in ref. \cite{MY1}.

Equation (\ref{eq_hmass}) can be solved only numerically. However
it is possible
to give approximate explicit expressions for the physical Higgs boson
masses. Their accuracy, as compared to the exact results, is typically 1-2
GeV. They read:
\begin{eqnarray}
M^2_{H^0_{1,2}} \approx {b\pm a\over 2\left(1-2\varepsilon\right)}
\end{eqnarray}
where
\begin{eqnarray}
a&=&\sqrt{\left(1-2\varepsilon\right)\left(\delta_1-\delta_2\right)^2
+ \left(\varepsilon\left(\delta_1+\delta_2\right) +
2\delta_3\left(1-\varepsilon\right)\right)^2}\nonumber\\
b&=&\left(1-\varepsilon\right)\left(
\delta_1+\delta_2\right)+2\varepsilon\delta_3
\end{eqnarray}
and
\begin{eqnarray}
{\delta}_1 = {\cal R}e~\hat{\Sigma}_{SS}^{11}\left(0\right)+m^2_{H^0_1}
{}~&\phantom{=}&~~~~~~~
{\delta}_2 = {\cal
R}e~\hat{\Sigma}_{SS}^{22}\left(0\right)+m^2_{H^0_2}\nonumber
\\
{\delta}_3 &=& {\cal R}e~\hat{\Sigma}_{SS}^{12}\left(0\right)
\end{eqnarray}
\begin{eqnarray}
\varepsilon = {3\alpha \over 16 \pi}~{s^2_W-c^2_W\over s^2_W}~{m^2_t\over
M^2_W}
\end{eqnarray}
$\varepsilon$ determines the dominant correction to the $p^2$ dependence
of the scalar self-energy.

The following important point is worth emphasizing. Defining the
masses as the exact poles of the full 1-loop $2\times 2$ matrix
propagator means that a summation has been performed over all 1-P
reducible self-energy dia\-grams with 1-loop segments. This accounts for
the leading higher order contributions to the Higgs boson mass
corrections (e.g. in the second order this contribution is ${\cal
O}(Y_t^8)$ whereas the irreducible 2-loop corrections are at most ${\cal
O}(Y_t^6)$ where $Y_t$ is the top quark Yukawa coupling). Another remark
is that, in our renormalization scheme, the on-shell residues of the
Higgs scalar propagators are not equal to unity. Thus, finite
renormalization constants for the external particles will appear in the
$S$-matrix (see next Section). Again, they are calculated from the
$2\times 2$ matrix propagator, i.e. with the 1-P reducible diagrams
included.

\setcounter{equation}{0}
\section{Cross sections for the Higgs particles production in the
$e^+e^-$ collisions}
\vskip 0.2cm

The main sources of the supersymmetric Higgs particles in the $e^+e^-$
colliders are the Bj{\o}rken process $e^+e^-\rightarrow Z^0h^0(H^0)$
($e^+e^-\rightarrow Z^0H^0_i$ in the matrix notation)
and the associated production of the scalar and pseudoscalar
$e^+e^-\rightarrow A^0h^0(H^0)$ (i.e. $e^+e^-\rightarrow H^0_iH^0_{j+2}$)
\footnote{ We do not discuss here the possibility of production of the Higgs
particles by {\it bremsstrahlung} off heavy quarks (e.g. $e^+e^-\rightarrow
\bar bbh^0(H^0)$), which can be significant for large $\tan \beta$
\cite{KAL}.}.
The set of diagrams we take into account for the Bj{\o}rken process
$e^+e^-\rightarrow Z^0H^0_i$ is shown in Figure 1 (boxes represent
loops of all possible virtual particles). Exactly analogous set
has been taken for the second process $e^+e^-\rightarrow H^0_iH^0_{j+2}$.
At the present level of accuracy we are interested in including
radiative corrections only in the Higgs sector and we neglect genuinely
electroweak corrections such as corrections to the electron self-energy,
to the $Z^0e^+e^-$ vertex, and the box diagrams.
Further simplification can be achieved when calculating matrix elements:
contribution from the diagrams 1e, f, i and from the longitudinal parts of
the $Z^0$ and $\gamma-Z^0$ mixing self energies in diagrams 1c, d, g are
proportional to the electron mass (or vanish). Therefore, in the (very
good) approximation $m_e~=~0$, the numerically important corrections can be
summarized as follows.

\noindent 1). Corrections to the $Z^0$, $\gamma$ and $\gamma-Z^0$
self-energies on internal and external lines. They are usually small, of
the order of few percent. The only exception is the vicinity of the
$Z^0$ resonance pole.

\noindent 2). Corrections to the scalar and pseudoscalar propagators.
They are the main source of the differences between the tree
and 1-loop results. As discussed in Section 2, they are responsible for
large changes in the physical masses $M_{h^0}$ and $M_{H^0}$ which are
predicted for given values of $\tan\beta$ and $M_A$, and also for finite
renormalization constants ${\cal Z}^{ext}$ for external particles in the
$S$-matrix elements. The latter are very important as they constitute the link
between the angle $\alpha$ which enters the tree level Higgs boson vertices
in the EPA method \cite{EPA} and our tree level angle $\alpha$.

The ${\cal Z}^{ext}$ factors are given as follows. Let $i'$ denotes the
index of the
"supplementary" Higgs particle: $i' = 2(1)$ when $i = 1(2)$ (formally
$i' = 3 - i$). Then the finite external line factor reads:
\begin{eqnarray}
  \left({\cal Z}^{ext}_{Si}\right)^{-1}
                      ={\cal R}e\left(1-{\partial\over\partial q^2}
                        \left.\left[\hat\Sigma_{SS}^{ii}(q^2)+{\left(
                        \hat\Sigma_{SS}^{ii'}\left(q^2\right)\right)^2
                        \over q^2-m^2_{H^0_{i'}}-
                        \hat\Sigma_{SS}^{i'i'}\left(q^2\right)}\right]
                        \right|_{M^2_{H^0_i}}\right)
\end{eqnarray}
and in the case of no mixing on the outgoing Higgs boson line the
$S$-matrix element is given by the truncated Green's function multiplied
by the $\left({\cal Z}^{ext}_{Si}\right)^{1/2}$. In presence of mixing
$i' \leftrightarrow i$ on the external line (with the scalar $i$ being
the final state particle) the respective factor on the external line
reads:
\begin{eqnarray}
\left.{\hat\Sigma_{SS}^{ii'}(q^2)\left({\cal
Z}^{ext}_{Si}\right)^{-1/2}\left(q^2 - M^2_{H^0_i}\right)\over
\left(q^2-m^2_{H^0_1}-\hat\Sigma_{SS}^{11}(q^2)\right)
  \left(q^2-m^2_{H^0_2}-\hat\Sigma_{SS}^{22}(q^2)\right)
  -{\left(\hat\Sigma_{SS}^{12}(q^2)\right)}^2}\right|_{M^2_{H^0_i}}
\end{eqnarray}
Therefore, the truncated Green's function is in this case effectively
multiplied by
\begin{eqnarray}
\left({\cal Z}^{ext}_{Si}\right)^{1/2}
{\hat\Sigma_{SS}^{ii'}\left(M^2_{H^0_i}\right)
 \over M^2_{H^0_i}-m^2_{H^0_{i'}}-
 \hat\Sigma_{SS}^{i'i'}\left(M^2_{H^0_i}\right)}\equiv
\left({\cal Z}^{ext}_{Si}\right)^{1/2}{\cal Z}^{ext}_{S_{mix}i}
\end{eqnarray}
In these expressions masses denoted by small letters are the tree level mass
parameters and capital letters stand for the physical masses.
Quantities with hat are, as usual, renormalized quantities.

To compactify the notation it will be useful to treat
formally the mixing constant as a two-element array:
\begin{eqnarray}
 \tilde {\cal Z}^{ext}_{S_{mix}i}= \left(1,{\cal Z}^{ext}_{S_{mix}i}\right)
\end{eqnarray}
Exactly analogous equations hold for the pseudoscalar
constants ${\cal Z}^{ext}_{Pi}$ and ${\cal Z}^{ext}_{P_{mix}i}$, but in that
case their effects are numerically much less important.

\noindent 3). Vertex corrections. For some choices of parameters they
can be numerically important, although only ${\cal O}\left(g^4
m_t^2/M_W^2\right)$. At the tree level the relevant Higgs boson vertices
read (assignment of momenta is shown in Fig.1):
\begin{eqnarray}
  V^{(0)\mu ij}_{ZPS}={e\over {2s_Wc_W}}A_M^{ij} \left(p~-~q\right)^{\mu}
  \equiv V^{(0)ij}_{ZPS}\left(p~-~q\right)^{\mu}
\end{eqnarray}
\begin{eqnarray}
  V^{(0)\mu\nu i}_{ZZS}={ie^2\over {2s^2_Wc^2_W}}~C_R^i~g^{\mu\nu}
  \equiv iV^{(0)i}_{ZZS}~g^{\mu\nu}
\end{eqnarray}
After including 1-loop corrections and adding the counterterms
defined in Section 3 the renormalized vertices have the form:
\begin{eqnarray}
  V^{\mu ij}_{ZPS}=
  p^{\mu}\left(\tilde V^{(0)ij}_{ZPS}+F_P^{ij}\right)
  -q^{\mu}\left(\tilde V^{(0)ij}_{ZPS}+F_S^{ij}\right)
\end{eqnarray}
\begin{eqnarray}
  V^{\mu\nu i}_{ZZS}=i~\left[g^{\mu\nu}\left(\tilde V^{(0)i}_{ZZS}+F^i_1\right)
  +p^{\mu}p^{\nu} F^i_2+q^{\mu}q^{\nu} F^i_3+
  p^{\mu}q^{\nu}F^i_4+q^{\mu}p^{\nu}F^i_5\right]
\end{eqnarray}
where ~$F_a$ ~are vertex formfactors.

The vertices $\tilde V^{(0)}$ are obtained from the tree level vertices
$V^{(0)}$ by inclusion of the 1-loop corrections on the external lines
and read:
\begin{eqnarray}
  \tilde V^{(0)ij}_{ZPS} &=& \left({\cal Z}^{ext}_{Si}
  {\cal Z}^{ext}_{Pj}\right)^{1/2}\times\nonumber\\
  &\times&\left( V^{(0)ij}_{ZPS} + {\cal Z}^{ext}_{S_{mix}i} V^{(0)i'j}_{ZPS}
  + {\cal Z}^{ext}_{P_{mix}j} V^{(0)ij'}_{ZPS}
  + {\cal Z}^{ext}_{S_{mix}i} {\cal Z}^{ext}_{P_{mix}j} V^{(0)i'j'}_{ZPS}
\right)\nonumber\\
  &\equiv& \left({\cal Z}^{ext}_{Si} {\cal Z}^{ext}_{Pj}\right)^{1/2}
\sum_{\alpha,\beta=1}^2
  \tilde {\cal Z}^{ext}_{S_{mix}i\alpha} \tilde {\cal Z}^{ext}_{P_{mix}j\beta}
  V^{(0) i\left(\alpha\right)j\left(\beta\right)}_{ZPS}
\end{eqnarray}
where
\begin{eqnarray}
  i\left(\alpha\right) = \left\{
\begin{array}{ll}
  i         & ~~~~~~\alpha = 1 \\
  i'=3-i    & ~~~~~~\alpha = 2
 \end{array}
\right.
\end{eqnarray}
and similarly for $j\left(\beta\right)$,
\begin{eqnarray}
  \tilde V^{(0)i}_{ZZS} = \left({\cal Z}_Z^{ext} {\cal
Z}^{ext}_{Si}\right)^{1/2}
  \left(V^{(0)i}_{ZZS} + {\cal Z}^{ext}_{S_{mix}i}V^{(0)i'}_{ZZS}\right)
\end{eqnarray}
with
\begin{eqnarray}
  {\cal Z}^{ext}_Z= {\cal R}e\left(1+{\partial\over\partial p^2}\left.
  \hat\Pi^T_{Z}(p^2)\right|_{M^2_Z}\right)
\end{eqnarray}
To include the photon exchange in the $s$-channel also the knowledge of
the vertices $V_{\gamma PS}$ and $V_{\gamma ZS}$ (vanishing at the tree level)
is necessary:
\begin{eqnarray}
  V^{\mu ij}_{\gamma PS}=p^{\mu}G_P^{ij}-q^{\mu}G_S^{ij}
\end{eqnarray}
\begin{eqnarray}
  V^{\mu\nu i}_{\gamma ZS}=i\left[g^{\mu\nu}G^i_1
  +p^{\mu}p^{\nu} G^i_2+q^{\mu}q^{\nu} G^i_3+
  p^{\mu}q^{\nu} G^i_4+q^{\mu}p^{\nu} G^i_5\right]
\end{eqnarray}
The 1-loop contributions to the vertices $V_{ZPS}$, $V_{ZZS}$, $V_{\gamma
PS}$ and $V_{\gamma ZS}$ can be found in Appendix B.

The cross sections (in the CMS system) for both processes have the form:
\begin{eqnarray}
  {d\sigma\over d\Omega}=
  {\lambda\left(s,M^2_{H^0_{j+2}(Z)},M^2_{H_i^0}\right)\over 64\pi^2
s^2\left|s-M^2_Z-\hat\Pi^T_{ZZ}\left(s\right)\right|^2}
  \left({\cal A}_1+{\cal A}_2\cos^2\theta_{CMS}\right)\nonumber\\
\lambda\left(s,m^2_1,m^2_2\right)=
\sqrt{s^2+m_1^4+m_2^4-2sm_1^2-2sm_2^2-2m^2_1m^2_2}
\end{eqnarray}
where
\begin{eqnarray}
\label{cr_sum_mat}
  {\cal A}_1+{\cal A}_2\cos^2\theta_{CMS}
  ={1\over 4}\sum_{pol}\left({\cal M}{\cal M}^{\dagger}\right)
\end{eqnarray}
\begin{eqnarray}
  {\cal M}_{PS}^{ij}=e\overline v(k_2)\gamma_{\mu}
  \left[V^{\mu ij}_{ZPS}\left(c_V-c_A\gamma^5\right)+\tilde V^{\mu
  ij}_{\gamma PS}
  +V^{(0)\mu ij}_{ZPS}Z_{\gamma Z}\right]u(k_1)
\end{eqnarray}
\begin{eqnarray}
  {\cal M}_{ZS}^i=e\overline v(k_2)\gamma^{\nu}\left[V^{\mu\nu i}_{ZZS}
  \left(c_V-c_A\gamma^5\right)+\tilde V^{\mu\nu i}_{\gamma ZS}
  +V^{(0)\mu\nu i}_{ZZS}Z_{\gamma Z}\right]u(k_1)\epsilon_{\mu}(p)
\end{eqnarray}
\begin{eqnarray}
Z_{\gamma Z} = {\hat\Pi^T_{\gamma Z}\left(s\right)\over
s\left(1-\hat\Pi^{\prime}_{\gamma}\left(s\right)\right)}
\end{eqnarray}
In the above expressions $u(k_1)$ and $v(k_2)$ are spinors of the
incoming electron-positron pair, $\epsilon_{\mu}(p)$ is the polarization
vector of the outgoing $Z^0$, $c_V=\left(-1+4s^2_W\right)/4s_Wc_W$ and
$c_A=-1/4s_Wc_W$. $\tilde V^{\mu ij}_{\gamma PS}$ and
$\tilde V^{\mu\nu i}_{\gamma ZS}$
are related to ~$V^{\mu ij}_{\gamma PS}$ ~and ~$V^{\mu\nu i}_{\gamma ZS}$
as follows:
\begin{eqnarray}
  \tilde V^{\mu ij}_{\gamma PS}&=&V^{\mu ij}_{\gamma PS}~
  {s-M_Z^2-\hat\Pi^T_{Z}(s)\over
  s\left(1-\hat\Pi^{\prime}_{\gamma}\left(s\right)\right)}
  \equiv p^{\mu}\tilde G_P^{ij} - q^{\mu} \tilde G_S^{ij}
\end{eqnarray}
\begin{eqnarray}
  \tilde V^{\mu\nu i}_{\gamma ZS}&=&V^{\mu\nu i}_{\gamma ZS}~
  {s-M_Z^2-\hat\Pi^T_{Z}(s)\over
  s\left(1-\hat\Pi^{\prime}_{\gamma}\left(s\right)\right)}\nonumber\\
  &\equiv& i\left[g^{\mu\nu}\tilde G^i_1
  +p^{\mu}p^{\nu} \tilde G^i_2 + q^{\mu}q^{\nu} \tilde G^i_3+
  p^{\mu}q^{\nu} \tilde G^i_4+q^{\mu}p^{\nu} \tilde G^i_5\right]
\end{eqnarray}
Performing the sum in eq. (\ref{cr_sum_mat}) one gets for the associated scalar
and
pseudoscalar production:
\begin{eqnarray}
  {\cal A}_{1PS}^{ij}=-~{\cal A}_{2PS}^{ij}={1 \over 8}~
  {\lambda}^2\left(s,M^2_{H^0_{j+2}},M^2_{H_i^0}\right)
  \left(\left|a^{ij}_{PS}\right|^2+\left|b^{ij}_{PS}\right|^2\right)
\end{eqnarray}
where
\begin{eqnarray}
  a^{ij}_{PS} &=& c_V \left(2\tilde V^{(0)ij}_{ZPS}+F_P^{ij}+F_S^{ij}\right)+2
  V^{(0)ij}_{ZPS}~Z_{\gamma Z} + \tilde G_P^{ij} + \tilde G_S^{ij}\nonumber\\
  b^{ij}_{PS} &=& c_A \left(2\tilde V^{(0)ij}_{ZPS}+F_P^{ij}+F_S^{ij}\right)
\end{eqnarray}
Analogous expressions for the Bj{\o}rken process are more complicated:
\begin{eqnarray}
  {\cal A}_{2ZS}^i &=& -~{{\lambda}^2\left(s,M^2_Z,M^2_{H_i^0}\right)\over
  32M_Z^2} \times \nonumber\\
  &\times& \left[ \left| 2 a^i_{ZS} -
  \left( s+M^2_Z-M^2_{H^0_i} \right) b^i_{ZS} \right|^2 - 4 s M_Z^2
  \left| b^i_{ZS}\right|^2 + \right. \nonumber \\
  &+&\left. \left| 2 c^i_{ZS}-\left(s+M^2_Z-M^2_{H^0_i}\right)d^i_{ZS}\right|^2
  - 4 s M_Z^2 \left| d^i_{ZS} \right|^2\right]
\end{eqnarray}
\begin{eqnarray}
  {\cal A}_{1ZS}^i =
s\left(\left|a^i_{ZS}\right|^2+\left|c^i_{ZS}\right|^2\right) -~{\cal
A}_{2ZS}^i
\end{eqnarray}
where
\begin{eqnarray}
  a^i_{ZS} &=& c_V \left(\tilde V^{(0)i}_{ZZS}+F_1^i\right)+
  V^{(0)i}_{ZZS}~Z_{\gamma Z} + \tilde G_1^i \nonumber\\
  b^i_{ZS} &=& c_V \left(F_3^i - F_4^i\right) + \tilde G_3^i - \tilde
  G_4^i \nonumber\\
  c^i_{ZS} &=& c_A \left(\tilde V^{(0)i}_{ZZS}+F_1^i\right) \nonumber\\
  d^i_{ZS} &=& c_A \left(F_3^i - F_4^i\right)
\end{eqnarray}

As it has already been discussed several times, the expressions for the cross
sections displayed above are not exactly
1-loop formulae: they have been designed to include the leading
terms in the higher order corrections.

\setcounter{equation}{0}
\section{Results}
\vskip 0.2cm

In this section we recall and extend the results obtained in ref.
\cite{MY1,MY3}. We focus on the predictions for the mass spectrum
for the two neutral scalars $h^0$ and $H^0$ and on their productions
rates in the $e^+e^-$ colliders. Two related questions are discussed:
the detectability of the scalars $h^0$ and $H^0$ as a function of the
$CMS$ energy and the possibility of distinguishing between the
supersymmetric and the minimal standard model (SM) Higgs bosons on the
basis of the production rates.

In Fig. 2a the allowed mass regions in the $(M_A,M_h)$ and $(M_A,M_H)$
planes are shown for $m_t =$ 120 and 180 GeV, and for several sets
of values for the soft supersymmetry breaking parameters. For the chosen set
of the central values of the parameters we also show the contours of fixed
$\tan\beta$ (Fig. 2b). The soft supersymmetry breaking parameters are:
universal
soft masses for the three generations of squarks and sleptons, $M_{sq}$
and $M_{sl}$, respectively, gaugino mass $M_{gau}$ (with the $SU(2)$ and
$U(1)$ gaugino masses related by the grand unification formulae
$M_{gau}^{U(1)}={5 \over 3} \tan^2 \theta_W M_{gau}^{SU(2)}$), universal
dimensionful parameter $A$ in the trilinear scalar
couplings $AY^I_f$ (where $Y^I_f$ is the Yukawa coupling for
the $I$ generation of fermions; $f=u,d,l$) and the Higgs mixing parameter
$\mu$ in the superpotential.
The general features of the results shown in Fig. 2 are already well
known. Here we illustrate their dependence on the values of the above
set of parameters and on the top quark mass. An interesting observation is
that changing $\mu$ and $A$ affects the mass spectrum differently for
different values of the top quark mass and $\tan\beta$. Larger values of $\mu$
and $A$ can
increase or decrease the corrections to $M_h$ and $M_H$, depending on
the values of $m_t$ and $\tan\beta$, what can be understood as the
effect of squark mixing. The difference between our results
and the results obtained by the EPA or RGE is typically ${\cal O}$(5 GeV).
A more detailed comparison has been presented elsewhere \cite{MY1}.

Two more remarks on Fig. 2 are important. One can see the well known
2-fold ambiguity: in part of the parameter space the same pair of values
$(M_A,M_h)$ corresponds to two different values of $\tan\beta$. Thus, in
two different versions of the model with different values of
$\tan\beta$, scalars with different couplings (i.e. different cross sections,
decay widths) happen to have the same mass. This has the following
implication: as seen in Fig. 3, plotting the production cross section
for the $h^0$ as a function of $M_h$ we get two branches of the
predictions, depending on the value of $\tan\beta$.

Secondly, there is also another approximate degeneracy: for large $M_A$
($> 100$ GeV) and large $\tan\beta$ ($\geq 5$) $M_h$ is almost
constant. As seen in Fig. 3, in this region the production cross section
in the process $e^+e^- \rightarrow Z^0 h^0$ also becomes independent of
$M_A$ and $\tan\beta$, and to a very good accuracy is equal to the cross
section for the production of the SM Higgs boson with the same mass
$M_h$. Similar remarks apply to the region $\tan\beta < 1$: for fixed
value of $\tan\beta$, the mass $M_h$ and the cross sections for the
$e^+e^- \rightarrow Z^0 h^0$ almost do not depend on the value of $M_A$.
Also, for large $M_A$, the dependence on $M_A$ of the $M_h$ and of the
cross sections is very weak for \underline{any} fixed value of
$\tan\beta$ (and in addition, as said above, the dependence on
$\tan\beta$ disappears for large $\tan\beta$). Finally, for large
$\tan\beta$ ($\geq 5$) and small values of $M_A$ ($\leq 100$ GeV) the
mass of the heavier scalar $H^0$ is almost constant. All those features
are important for experimental analysis: on the one hand the variables $M_A$
and $M_h$ (or $M_H$) are more direct and physical to use (in particular,
one should remember about renormalization scheme dependence of
$\tan\beta$). On the other hand they may be inconvenient in certain
regions of the parameter space and the use of the unconstrained
parameters ($M_A,\tan\beta$) provides a useful "blow up" of those
regions. The complementarity of both parameterizations will be further
illustrated below.

In Fig. 3 we show typical behaviour of the cross sections for the
processes $e^+e^- \rightarrow Z^0 h^0$ and $e^+e^- \rightarrow A^0 h^0$ as
a function of the $M_h$, for fixed values of $M_A$. The values of
$\tan\beta$ change along the curves, as marked on the Figures. We see
the well known complementarity of the two cross sections, which follows from
the tree level couplings and is preserved by radiative corrections
(as is clear e.g. in the EPA): the pair production is dominant for small
values of $M_A$ and large values of $\tan\beta$ , i.e. in the region
where the cross section for the Bj{\o}rken process is very small.

In Fig. 4 we show the regions in the ($M_A,M_h$) and ($M_A,M_H$)
variables in which at least one of the cross sections for the two
production processes considered here ($e^+e^- \rightarrow Z^0 h^0$ and
$e^+e^- \rightarrow A^0 h^0$) is larger than $\sigma_0$, for different
values of $\sigma_0$, different values of the $CMS$ energy and the top
quark masses $m_t = $ 140 and 180 GeV. In Fig. 5 we show analogous
plots in the ($M_A,\tan\beta$) planes. It should be pointed out that for
$\sqrt{s} = 500$ GeV there is one additional very important production
process, namely the $WW$ fusion \cite{WWFUS}, which we do not consider here.
The experimental signatures for the Higgs boson production via this
mechanism are, however, different from the signatures in the two
processes discussed here, so one can consider them independently.

The effective regions in Fig. 5 are superpositions of the regions
where $\sigma_{Ah(H)}$ or $\sigma_{Zh(H)}$ are dominant. For instance,
in the ($M_A,\tan\beta$) plane and at high enough energies, the observed
structure with the dip for $M_A = 100$ GeV reflects the complementarity
of the two cross sections: the first one dominates to the left of the
dip (apart from the very small values of $\tan\beta$) and the second one
to the right of the dip. Notice that for realistic values of $\sigma_0$
the region of unobservability of the $h^0$ does not disappear with
increasing energy. Fortunately, in this region of the parameter space
the heavier neutral MSSM scalar $H^0$ becomes observable at high
enough energy. For $m_t = 140$ GeV, $max(\sigma_{Zh},\sigma_{Ah})$ or
$max(\sigma_{ZH},\sigma_{AH})$ are larger then 0.05 pb in the whole
range of ($M_A,\tan\beta$) for $\sqrt{s} = 200$ GeV and for $m_t = 180$
GeV the same occurs for $\sqrt{s} = 240$ GeV (for a broad range of
values of the soft supersymmetry breaking parameters). It is also worth
observing that a small increase in energy can radically change the range
of ($M_A,\tan\beta$) which can be explored. For instance, for $m_t =
140$ GeV, the increase from $\sqrt{s} = 190$ GeV to $\sqrt{s} = 200$ GeV
gives the result shown in Fig. 5a. This is easily understandable as
due to the discussed earlier approximate independence of the $M_h$ on
the $M_A$ and $\tan\beta$ for large values of those variables: changing
$\tan\beta$ from, say, 4 to 50 results in only a tiny increase in the
value of $M_h$ and a small change in energy is sufficient to reach the
discovery limit.

Finally, in Figs. 4 and 5 we show also the regions where the cross sections
$\sigma_{Zh(H)}$ are very close to the cross section for the SM Higgs
boson production. This actually occurs for most of the yet
unexplored parameter space. Similar conclusions hold for the decay
branching ratios \cite{KZ,MY4}. This is the general property of the MSSM: for a
sufficiently heavy pseudoscalar A, the effective low energy theory
(after decoupling of the heavier Higgs scalars) contains only one
SM-like Higgs boson. Detailed inspection of Fig. 5 shows that
for mass ranges not excluded yet by the LEP measurements and for the
values of $\sigma_0$ considered here we have two possibilities in general.
First, only one
MSSM neutral scalar is observed via in the Bj{\o}rken process $e^+e^-
\rightarrow Z^0 h^0(H^0)$ and is almost always
indistinguishable from the SM Higgs boson (at least on the basis of the
cross section measurements). Second, both the lighter and the heavier scalars
are observed, giving us strong direct evidence for supersymmetry.

The authors wish to thank A. Dabelstein for careful checking of the large
part of the formulae and W. Hollik for helpful discussions.

The formulae for self-energies and vertex formfactors presented in the
paper have been prepared as the library of FORTRAN procedures. They are
available on requests at the e-mail addresses:

\noindent rosiek@fuw.edu.pl (or rosiek@plearn.bitnet)

\noindent chank@padova.infn.it (or chank@ipdinfn.bitnet)

\renewcommand{\thesection}{Appendix~\Alph{section}.}
\renewcommand{\thesubsection}{\Alph{section}.\arabic{subsection}}
\renewcommand{\theequation}{\Alph{section}.\arabic{equation}}

\setcounter{equation}{0}
\setcounter{section}{0}
\section{Two point Green's functions}
\vskip 0.2cm

In this Appendix we display the formulae for the unrenormalized two
point Green functions. In order to preserve supersymmetry of the ~$S-$matrix
elements all Green functions have been regularized using dimensional
reduction \cite{DRED}.

First, we define our conventions for vertices which appear in the
formulae below. We write
vector boson($\mu$,a)-incoming fermion(l)-outgoing fermion(m)
vertex as
\begin{eqnarray}
\label{vff_vert}
  -i~\gamma^{\mu}~(c_V^{alm} - \gamma^5 ~c_A^{alm})
\end{eqnarray}
For scalar(i) replacing vector boson we write
\begin{eqnarray}
\label{sff_vert}
  -i~(c^{ilm}_L P_L - c^{ilm}_R P_R)
\end{eqnarray}
where ~$P_{L,R}$ ~are left/right projectors
\footnote{Notice the difference with the ref. \cite{MY4}, where this
vertex is denoted as $-i(c^{ilm}_L P_L + c^{ilm}_R P_R)$}.
Interactions of three
and four scalar fields in our convention take the form
(eg. scalar two pseudoscalars vertex and two scalars two down-type
squarks vertex):
\begin{eqnarray}
  -i~V^{ijk}_{SPP}, ~~~~ -i~V^{ijmn}_{SSDD}
\end{eqnarray}
\noindent and couplings of vector boson ~$V^{\mu}$ ~to
sfermions, e.g.:
\begin{eqnarray}
  -i~V^{mn}_{VDD}(k+p)^{\mu},~~~~+i~V^{mn}_{V_1V_2UU}
\end{eqnarray}
\noindent (where ~$k^{\mu}$ ~is the incoming momentum of ~$D_m$ squark).
The specific form of all vertices can be found in ref.
\cite{ROS}
\footnote{ Notice that expressions for neutralino couplings to
scalars and pseudoscalars given in ref. \cite{ROS} should be symmetrized
(without dividing by 2) in fermionic indices. Also the factor 2
multiplying ~$\delta^{ij}$ in the ~$Z^0$ coupling to charginos
should be removed and in the ~$Z^0-$ neutralinos vertex factors
multiplying left and right projectors in the bracket should read:
{}~$Z^{4i*}_NZ^{4j}_N-Z^{3i*}_NZ^{3j}_N$ ~and
{}~$Z^{3i}_NZ^{3j*}_N-Z^{4i}_NZ^{4j*}_N$, respectively. In
expressions for the sfermion mass matrices one should change the sign of
the terms proportional to $v^2_1~-~v^2_2$.}

Our notation for (s)particle masses is as follows. We denote lepton, d-
and u-type quark masses as ~$m_{l_I}$, ~$m_{d_I}$ ~and ~$m_{u_I}$
respectively where ~$I$ ~is the generation index ($I=1,2,3$). Similar index
{}~$I$ ~is used to count 3 generations of sneutrinos whose masses are denoted
as ~$m_{{\tilde\nu}_I}$. Slepton and squark masses are written as
{}~$m_{L_n}$, ~$m_{D_n}$ and ~~$m_{U_n}$ ~and
indices ~$n$ ~vary from 1 to 6. The two chargino masses are denoted
as ~$m_{C_i}$ ~($i=1,2$) and the neutralino masses as ~$m_{N_j}$
{}~($j=1,2,3,4$). Let us stress that all masses we use are
the true eigenvalues of the corresponding mass matrices. Rotations from
the initial fields to the mass eigenstates are described in
\cite{ROS}.

We define also some auxiliary functions which help us
to make the formulae for self energies more concise:
\begin{eqnarray}
  B(p^2,~m^2_1,~m^2_2)&\equiv &{2\over 3}~a_0(m^2_1)-{1\over 3}~a_0(m^2_2)
  -{m^2_1-m^2_2\over 3~p^2}~\left[a_0(m^2_1)-a_0(m^2_2)\right]\nonumber\\
  &+&{1\over 3}~m^2_2~b_0(p^2,~m^2_1,~m^2_2)
  +{1\over 6}\left(m^2_1+m^2_2-{1\over3}~p^2\right)\nonumber\\
  &-&{\left(p^2-m^2_1+m^2_2\right)^2\over 3~p^2}~b_0(p^2,~m^2_1,~m^2_2)
\end{eqnarray}
For scalar contribution to ~$\Pi^B$ ~(see \ref{eq_pitb}) we define
{}~$S^B(p^2,m^2_1,m^2_2)$:
\begin{eqnarray}
  S^B&=&-~{1\over p^2}\left[4~B(p^2,~m^2_1,~m^2_2)
  -2~a_0(m^2_1)+2~a_0(m^2_2)\right.\nonumber\\
  &+&\left.\left(p^2-2~m^2_1+2~m^2_2\right)
  ~b_0(p^2,~m^2_1,~m^2_2)\right]
\end{eqnarray}
and for fermionic contribution to ~$\Pi^T$ and ~$\Pi^B$ we
use ~$F^{ab}(p^2,m^2_i,m^2_j)$:
\begin{eqnarray}
  F^{ab}_A&=&-2\left(c^{aij}_Vc^{bji}_V+c^{aij}_Ac^{bji}_A\right)
  \left[4b_{22}(p^2,m^2_i,m^2_j)+a_0(m^2_i)+a_0(m^2_j)\right.\nonumber\\
  &+&\left.\left(p^2-m^2_i-m^2_j\right)~b_0(p^2,m^2_i,m^2_j)\right]\nonumber\\
  &-&4\left(c^{aij}_Vc^{bji}_V-c^{aij}_Ac^{bji}_A\right)~m_i
m_j~b_0(p^2,m^2_i,m^2_j)
\end{eqnarray}
\begin{eqnarray}
  F^{ab}_B&=&4\left(c^{aij}_Vc^{bji}_V+c^{aij}_Ac^{bji}_A\right)
  {1\over p^2}\left[2~B(p^2,m^2_i,m^2_j)-a_0(m^2_i)
  +a_0(m^2_j)\right.\nonumber\\
  &+&\left.\left(p^2-m^2_i+m^2_j\right)~b_0(p^2,m^2_i,m^2_j)\right]
\end{eqnarray}
Chargino and neutralino contributions to the $Z^0$-pseudoscalar mixing
can be expressed in terms of
\begin{eqnarray}
  F^{af}_{Ci}&=&{i\over
  p^2}~\left(\left[\left(c^{alk}_V+c^{alk}_A\right)c_L^{ikl}
  -\left(c_V^{alk}-c_A^{alk}\right))c_R^{ikl}\right]~m_l\right.\nonumber\\
  &\times&\left[a_0(m^2_l)-a_0(m^2_k)-\left(p^2-m^2_l
  +m^2_k\right)~b_0(p^2,m^2_k,m^2_l)
  \right]\nonumber\\
  &+&\left[\left(c^{alk}_V-c^{alk}_A\right)c_L^{ikl}
  -\left(c_V^{alk}+c_A^{alk}\right)c_R^{ikl}\right]~m_k\nonumber\\
  &\times&\left.\left[a_0(m^2_l)-a_0(m^2_k)+\left(p^2-m^2_k
  +m^2_l\right)~b_0(p^2,m^2_k,m^2_l)\right]\right)
\end{eqnarray}
Finally, for chargino and neutralino contributions to the scalar and
pseudoscalar self energies the auxiliary function $F^{ij}_D$ is useful:
\begin{eqnarray}
  F^{ff}_{Dij}&=&-{1\over2}\left(c^{ikl}_L
c^{jlk}_R+c^{ikl}_Rc^{jlk}_L\right)\left[
  \left(p^2-m^2_l-m^2_k\right)b_0(p^2,m^2_l,m^2_k)+a_0(m^2_k)\right.\nonumber\\
  &+&\left.a_0(m^2_l)\right] -\left(c^{ikl}_Lc^{jlk}_L
  +c^{ikl}_Rc^{jlk}_R\right) m_l~m_k~b_0(p^2,m^2_l,m^2_k)
\end{eqnarray}

\vskip 0.3cm

In all expressions here and in the other appendices
we use the standard Veltman functions
{}~$a_0$, ~$b_{ij}$ ~and ~$c_{ij}$ ~\cite{PTVAX} defined as follows
\footnote{We use the standard metric $(+---)$.}:
\begin{eqnarray}
{\mu}^{4-d}\int {d^dk\over (2\pi)^d} {1\over k^2-m^2}= {-i\over
(4\pi)^2}~a_0(m^2)
\end{eqnarray}
\begin{eqnarray}
{\mu}^{4-d}\int {d^dk\over (2\pi)^d}~{1,k^{\mu},k^{\mu}k^{\nu}\over
 \left[k^2-m^2_1\right]\left[\left(k+p\right)^2-m^2_2\right]}=
\phantom{aaaaaaaaaaaa} \\
=~{i\over (4\pi)^2}~\left\{
\begin{array}{ll}
  b_0(p^2,m^2_1,m^2_2) \\
  p^{\mu}~b_1(p^2,m^2_1,m^2_2) \\
  p^{\mu}p^{\nu}~b_{21}(p^2,m^2_1,m^2_2)+g^{\mu\nu}~b_{22}(p^2,m^2_1,m^2_2)
 \end{array}
\right.\nonumber
\end{eqnarray}
\begin{eqnarray}
{\mu}^{4-d}\int {d^dk\over (2\pi)^d}~{1,k^{\mu},k^{\mu}k^{\nu}\over
 \left[k^2-m^2_1\right]\left[\left(k+p\right)^2-m^2_2\right]
 \left[\left(k+p+q\right)^2-m^2_3\right]}=
\phantom{aaaaaaa} \\
=~{-i\over (4\pi)^2}~\left\{
\begin{array}{ll}
  c_0(p,q,m^2_1,m^2_2,m^2_3) \\
  p^{\mu}~c_{11}(p,q,m^2_1,m^2_2,m^2_3)+q^{\mu}~c_{12}(p,q,m^2_1,m^2_2,m^2_3)\\
p^{\mu}p^{\nu}~c_{21}(p,q,m^2_1,m^2_2,m^2_3)+q^{\mu}q^{\nu}~
c_{22}(p,q,m^2_1,m^2_2,m^2_3)\\
  ~~~~~~+~(p^{\mu}q^{\nu}+p^{\nu}q^{\mu})~c_{23}(p,q,m^2_1,m^2_2,m^2_3)\\
  ~~~~~~+~g^{\mu\nu}~c_{24}(p,q,m^2_1,m^2_2,m^2_3)
 \end{array}
\right.\nonumber
\end{eqnarray}
1) Vector bosons self-energies (and propagators which mix gauge
bosons). Using the decomposition
\begin{eqnarray}
\label{eq_pitb}
  i~\Pi^{\mu\nu}_{ab}(p^2) = i~\Pi^T_{ab}(p^2)~g^{\mu\nu}
  + i~\Pi^B_{ab}(p^2)~p^{\mu}p^{\nu}
\end{eqnarray}
(where ~$\Pi^B(p^2)\equiv \left[\Pi^L(p^2)-\Pi^T(p^2)\right]/p^2$
{}~in the notation of Section 3) we have:
\vskip 0.3cm

\noindent a) ~$Z^0$ boson self energy (we suppress the momentum argument in
all functions):
\begin{eqnarray}
&\phantom{=}&\left(4\pi\right)^2\Pi^T_{Z}
{}~=~-{3\over 4}{g^2_2\over
c^2_W}\left[4b_{22}(0,0)+p^2~b_0(0,0)\right]\nonumber\\
  &+&\sum_{I=1}^3\left[F^{ZZ}_A\left(m^2_{l_I},m^2_{l_I}\right)
  +3~F^{ZZ}_A\left(m^2_{d_I},m^2_{d_I}\right)
  +3~F^{ZZ}_A\left(m^2_{u_I},m^2_{u_I}\right)\right]\nonumber\\
  &+&4~\sum_{I=1}^3|V^{I}_{Z\tilde\nu\tilde\nu}|^2
  ~b_{22}(m^2_{{\tilde\nu}_I},m^2_{{\tilde\nu}_I})
+4\sum_{m,n=1}^6\left[|V^{mn}_{ZLL}|^2~
b_{22}(m^2_{L_m},m^2_{L_n})\right.\nonumber\\
  &+&\left.3~|V^{mn}_{ZDD}|^2~b_{22}(m^2_{D_m},m^2_{D_n})
  +3~|V^{mn}_{ZUU}|^2~b_{22}(m^2_{U_m},m^2_{U_n})\right]
  \nonumber\\
  &+&\sum_{I=1}^3 V^I_{ZZ\tilde\nu\tilde\nu}~a_0(m^2_{{\tilde \nu}_I})
  +\sum_{n=1}^6\left[V^{nn}_{ZZLL}~a_0(m^2_{L_n})
  \right.\nonumber\\
  &+&\left.3~V^{nn}_{ZZDD}~a_0(m^2_{D_n})
  + 3~V^{nn}_{ZZUU}~a_0(m^2_{U_n})\right]\\
  &+&\sum_{i,j=1}^2F^{ZZ}_A(m^2_{C_i},m^2_{C_j})
  +{1\over2}\sum_{i,j=1}^4F^{ZZ}_A(m^2_{N_i},m^2_{N_j})\nonumber\\
  &+&2e^2\cot^22\theta_W\sum_{i=1}^2\left[2~b_{22}(m^2_{H^+_i},m^2_{H^+_i})
  +a_0(m^2_{H^+_i})\right]\nonumber\\
  &+&{g^2_2\over c^2_W}\sum_{i,j=1}^2 {\left(A^{ij}_M\right)}^2
b_{22}(m^2_{H^0_i},m^2_{H^0_{j+2}})
  +{g^2_2\over 4c^2_W}\sum_{j=1}^4a_0(m^2_{H^0_j})\nonumber\\
  &-&2 g^2_2 s^2_W~M^2_Z~b_0(M^2_W,M^2_W)
  -{g^4\over4c_W^4}\sum_{i=1}^2{\left(C^i_R\right)}^2
  ~b_0(M^2_Z,m^2_{H^0_i})\nonumber\\
  &+&2 g^2_2 c_W^2\left[2a^2_0(M^2_W)+\left(2p^2+M^2_W\right)
  b_0(M^2_W,M^2_W)+4b_{22}(M^2_W,M^2_W)\right]\nonumber
\end{eqnarray}
\begin{eqnarray}
&\phantom{=}&\left(4\pi\right)^2\Pi^B_{Z}
{}~=~{3\over 2}{g^2_2\over c^2_W}\left[{2\over p^2}~B(0,0)
  +b_0(0,0)\right]\nonumber\\
  &+&\sum_{I=1}^3\left[F^{ZZ}_B\left(m^2_{l_I},m^2_{l_I}\right)
  +3~F^{ZZ}_B\left(m^2_{d_I},m^2_{d_I}\right)
  +3~F^{ZZ}_B\left(m^2_{u_I},m^2_{u_I}\right)\right]\nonumber\\
  &+&\sum_{I=1}^3|V^{I}_{Z\tilde\nu\tilde\nu}|^2
  ~S^B(m^2_{{\tilde \nu}_I},m^2_{{\tilde \nu}_I})
  +\sum_{m,n=1}^6\left[
  |V^{nm}_{ZLL}|^2~S^B(m^2_{L_n},m^2_{L_m})\right.
  \nonumber\\
  &+&\left.3~|V^{nm}_{ZDD}|^2~S^B(m^2_{D_n},m^2_{D_m})
  +3~|V^{nm}_{ZUU}|^2~S^B(m^2_{U_n},m^2_{U_m})\right]\\
  &+&\sum_{i,j=1}^2F^{ZZ}_B(m^2_{C_i},m^2_{C_j})
  +{1\over2}\sum_{i,j=1}^4F^{ZZ}_B(m^2_{N_i},m^2_{N_j})\nonumber\\
  &+&e^2\cot^22\theta_W\sum_{i=1}^2S^B(m^2_{H^+_i},m^2_{H^+_i})
  +{g^2_2\over 4c^2_W}\sum_{i,j=1}^2{\left(A^{ij}_M\right)}^2
  ~S^B(m^2_{H^0_i},m^2_{H^0_{j+2}})\nonumber\\
  &-&g^2_2 c_W^2{2\over3p^2}\left[4a_0(M^2_W)
  +\left(5p^2+4M_W^2\right)b_0(M^2_W,M^2_W)
  +4M^2_W-{2\over3}p^2\right]\nonumber
\end{eqnarray}
\vskip 0.3cm

\noindent b) Photon self energy. In this case we must have
{}~$p^2\Pi^B_{\gamma}=-\Pi^T_{\gamma}$ ~so we write only ~$\Pi^T_{\gamma}$:
\begin{eqnarray}
&\phantom{=}&\left(4\pi\right)^2\Pi^A_{\gamma}
  ~= \sum_{I=1}^3\left[F^{\gamma\gamma}_A\left(m^2_{l_I},m^2_{l_I}\right)
  +3F^{\gamma\gamma}_A\left(m^2_{d_I},m^2_{d_I}\right)
  +3F^{\gamma\gamma}_A\left(m^2_{u_I},m^2_{u_I}\right)\right]\nonumber\\
  &+&4e^2~\sum_{n=1}^6\left[b_{22}(m^2_{L_n},m^2_{L_n})
  +{1\over3}~b_{22}(m^2_{D_n},m^2_{D_n})
  +{4\over3}~b_{22}(m^2_{U_n},m^2_{U_n})\right]\nonumber\\
  &+&e^2~\sum_{n=1}^6\left[2~a_0(m^2_{L_n})
  +{2\over3}~a_0(m^2_{D_n})
  +{8\over3}~a_0(m^2_{U_n})\right]\\
  &+&\sum_{i=1}^2F^{\gamma \gamma}_A(m^2_{C_i},m^2_{C_i})
  +2 e^2\sum_{i=1}^2\left[2~b_{22}(m^2_{H^+_i},m^2_{H^+_i})
  +a_0(m^2_{H^+_i})\right]\nonumber\\
  &+&4~e^2~\left[a_0(M^2_W)+\left(p^2+2~M^2_W\right)
  ~b_0(M^2_W,M^2_W)+2~b_{22}(M^2_W,M^2_W)\right]\nonumber
\end{eqnarray}

\vskip 0.3cm

\noindent c) ~$Z^0$ mixing with the photon:
\begin{eqnarray}
&\phantom{=}&\left(4\pi\right)^2\Pi^T_{\gamma Z}~=
  \sum_{I=1}^3\left[F^{\gamma Z}_A\left(m^2_{l_I},m^2_{l_I}\right)
  +3F^{\gamma Z}_A\left(m^2_{d_I},m^2_{d_I}\right)
  +3F^{\gamma Z}_A\left(m^2_{u_I},m^2_{u_I}\right)\right]\nonumber\\
  &-&4e~\sum_{n=1}^6\left[V^{nn}_{ZLL}~b_{22}(m^2_{L_n},m^2_{L_n})
  +V^{nn}_{ZDD}~b_{22}(m^2_{D_n},m^2_{D_n})\right.\nonumber\\
  &-&\left.2~V^{nn}_{ZUU}~b_{22}(m^2_{U_n},m^2_{U_n})\right] \\
  &+&\sum_{n=1}^6\left[V^{nn}_{\gamma ZLL}~a_0(m^2_{L_n})
  +3~V^{nn}_{\gamma ZDD}~a_0(m^2_{D_n})
  +3~V^{nn}_{\gamma ZUU}~a_0(m^2_{U_n})\right]\nonumber\\
  &+&\sum_{i=1}^2F^{\gamma Z}_A(m^2_{C_i},m^2_{C_i})
  +2~e^2~c_W~s_W~M^2_Z~b_0(M^2_W,M^2_W)\nonumber\\
  &+&2 e g_2 c_W\left[2a_0(M^2_W)+\left(2p^2+M^2_W\right)
  b_0(M^2_W,M^2_W)+4b_{22}(M^2_W,M^2_W)\right]\nonumber\\
  &+&2 e^2~\cot2\theta_W\sum_{i=1}^2
  \left[2~b_{22}(m^2_{H^+_i},m^2_{H^+_i})+a_0(m^2_{H^+_i})\right]\nonumber
\end{eqnarray}
\begin{eqnarray}
&\phantom{=}&\left(4\pi\right)^2\Pi^B_{\gamma Z}
  ~=\sum_{I=1}^3\left[F^{\gamma Z}_B\left(m^2_{l_I},m^2_{l_I}\right)
  +3F^{\gamma Z}_B\left(m^2_{d_I},m^2_{d_I}\right)
  +3F^{\gamma Z}_B\left(m^2_{u_I},m^2_{u_I}\right)\right]\nonumber\\
  &-&e~\sum_{n=1}^6\left[
  V^{nn}_{ZLL}~S^B(m^2_{D_n},m^2_{D_n})
  +V^{nn}_{ZDD}~S^B(m^2_{D_n},m^2_{D_n})
  \right.\nonumber\\
  &-&\left.2~V^{nn}_{ZUU}~S^B(m^2_{U_n},m^2_{U_n})\right]
  +\sum_{i=1}^2F^{\gamma Z}_B(m^2_{C_i},m^2_{C_i})\\
  &+&e^2~\cot2\theta_W\sum_{i=1}^2S^B(m^2_{H^+_i},m^2_{H^+_i})
  -e~g_2~c_W{1\over3p^2}\left[8~a_0(M^2_W)\right.\nonumber\\
  &+&\left.\left(10~p^2+8~M_W^2\right)b_0(M^2_W,M^2_W)
  +8~M^2_W-{4\over3}~p^2\right]\nonumber
\end{eqnarray}

\noindent d) $W^{\pm}$ boson self energy :
\begin{eqnarray}
&\phantom{=}&\left(4\pi\right)^2\Pi^T_{W}~=~
  \sum_{I=1}^3\left[F^{WW}_A\left(m^2_{l_I},0\right)
  +3F^{WW}_A\left(m^2_{d_I},m^2_{u_I}\right)\right]\nonumber\\
  &+&4~\sum_{I=1}^3\sum_{n=1}^6|V^{In}_{W\tilde\nu L}|^2
  ~b_{22}(m^2_{{\tilde \nu}_I},m^2_{L_n})
  +12~\sum_{n,m=1}^6|V^{nm}_{WDU}|^2~b_{22}(m^2_{D_n},m^2_{U_m})\nonumber\\
  &+&\sum_{I=1}^3 V^I_{WW\tilde\nu\tilde\nu}~a_0(m^2_{{\tilde \nu}_I})
  +\sum_{n=1}^6\left[V^{nn}_{WWLL}~a_0(m^2_{L_n})
  +3~V^{nn}_{WWDD}~a_0(m^2_{D_n})\right.\nonumber\\
  &+&\left.3~V^{nn}_{WWUU}~a_0(m^2_{U_n})\right]
  +\sum_{i=1}^2\sum_{j=1}^4F^{WW}_A(m^2_{C_i},m^2_{N_j})\nonumber\\
&+&g^2_2\sum_{i,j=1}^2~{\left(A^{ij}_M\right)}^2~
b_{22}(m^2_{H^0_i},m^2_{H^+_j})
  +g^2_2\sum_{i=1}^2~b_{22}(m^2_{H^0_{i+2}},m^2_{H^+_i})\\
  &+&{1\over4}~g^2_2\sum_{i=1}^4a_0(m^2_{H^0_i})
  +{1\over2}~g^2_2\sum_{i=1}^2a_0(m^2_{H^+_i})
  -e^2~M^2_W~b_0(M^2_W,0)\nonumber\\
  &-&{1\over4}~g^2_2\sum_{i=1}^2{\left(C^i_R\right)}^2~b_0(M^2_W,m^2_{H^0_i})
  -e^2~s^2_W~M^2_Z~b_0(M^2_W,M^2_Z)\nonumber\\
  &+&g^2_2~c^2_W\left[2a_0(M^2_Z)-a_0(M^2_W)+\left(4~p^2+M^2_Z+M^2_W\right)
  ~b_0(M^2_W,M^2_Z)\right.\nonumber\\
  &+&\left.8~b_{22}(M^2_Z,M^2_W)\right]+3~g^2_2~a_0(M^2_W)\nonumber\\
  &+&e^2\left[\left(4~p^2+M^2_W\right)b_0(M^2_W,0)-a_0(M^2_W)
  +8~b_{22}(M^2_W,0)\right]\nonumber
\end{eqnarray}
\begin{eqnarray}
&\phantom{=}&\left(4\pi\right)^2  \Pi^B_{W}~=~
  \sum_{I=1}^3\left[F^{WW}_B\left(m^2_{l_I},0\right)
  +3~F^{WW}_B\left(m^2_{d_I},m^2_{u_I}\right)\right]\nonumber\\
  &+&\sum_{I=1}^3\sum_{n=1}^6|V^{In}_{W\tilde\nu L}|^2
  ~S^B(m^2_{{\tilde \nu}_I},m^2_{L_n})
  +3~\sum_{n,m=1}^6|V^{nm}_{WDU}|^2~S^B(m^2_{D_n},m^2_{U_m})
  \nonumber\\
&+&\sum_{i=1}^2\sum_{j=1}^4F^{WW}_B(m^2_{C_i},m^2_{N_j})
+{1\over4}g^2_2\sum_{i,j=1}^2
  ~{\left(A^{ij}_M\right)}^2~S^B(m^2_{H^0_i},m^2_{H^+_j})\nonumber\\
  &+&{1\over4}g^2_2\sum_{i=1}^2S^B(m^2_{H^0_{i+2}},m^2_{H^+_i})
  -g^2_2~c^2_W~{1\over3p^2}\left[4~a_0(M^2_W)+4~a_0(M^2_Z)\right.\nonumber\\
  &-&8~{M^2_W-M^2_Z\over p^2}\left(a_0(M^2_W)-a_0(M^2_Z)\right)
  -8~{\left(M^2_W-M^2_Z\right)^2\over p^2}b_0(M^2_W,M^2_Z)\nonumber\\
  &+&\left.\left(10p^2+2M^2_W+2M^2_Z\right)b_0(M^2_W,M^2_Z)
  +4M^2_W+4M^2_Z-{4\over3}~p^2\right]\nonumber\\
  &-&e^2~{1\over 3p^2}\left[4~a_0(M^2_W)-8~{M^2_W\over p^2}~a_0(M^2_W)
  -8~{M^4_W\over p^2}~b_0(M^2_W,0)\right.\nonumber\\
  &+&\left.\left(10p^2+2M^2_W\right)~b_0(M^2_W,0)
  +4~M^2_W-{4\over3}~p^2\right]
\end{eqnarray}
\vskip 0.5cm

2) Mixing of vector bosons with pseudoscalars. We denote the Green's
function of the incoming ~$Z^0$ ~with four-momentum ~$p^{\mu}$ ~and
the outgoing pseudoscalar as:
\begin{eqnarray}
  -~p^{\mu}~\Sigma_{ZP}^j(p^2)
\end{eqnarray}
\noindent Using this convention we have:
\begin{eqnarray}
&\phantom{=}&\left(4\pi\right)^2  \Sigma_{ZP}^j
  ~=~{g_2\over c_W}\sum_{I=1}^3
  \left[Z^{1j}_H~{m^2_{l_I}\over v_1}~b_0(m^2_{l_I},m^2_{l_I})
  +3~Z^{1j}_H~{m^2_{d_I}\over v_1}~b_0(m^2_{d_I},m^2_{d_I})
  \right.\nonumber\\
  &-&\left.3~Z^{2j}_H~{m^2_{u_I}\over
  v_2}~b_0(m^2_{u_I},m^2_{u_I})\right]\nonumber\\
  &+&{i\over p^2}\sum_{n,m=1}^6\left(
  V^{nm}_{ZLL}V^{jmn}_{PLL} \left[a_0(m^2_{L_n})
  -a_0(m^2_{L_m})+\left(m^2_{L_n}-m^2_{L_m}\right)
  b_0(m^2_{L_n},m^2_{L_m})\right]\right.\nonumber\\
  &+&3~V^{nm}_{ZUU}V^{jmn}_{PUU}\left[a_0(m^2_{U_n})-a_0(m^2_{U_m})
  +\left(m^2_{U_n}-m^2_{U_m}\right)b_0(m^2_{U_n},m^2_{U_m})\right]\nonumber\\
  &+&\left.3~V^{nm}_{ZDD}V^{jmn}_{PDD}\left[a_0(m^2_{D_n})-a_0(m^2_{D_m})
+\left(m^2_{D_n}-m^2_{D_m}\right)
b_0(m^2_{D_n},m^2_{D_m})\right]\right)\nonumber\\
  &-&{1\over8}{g^3_2\over c^3_W}\sum_{i=1}^2~C^i_R~A^{ij}_M
  ~{1\over p^2}\left[\left(3~p^2-M^2_Z+m^2_{H^0_i}\right)b_0(M^2_Z,m^2_{H^0_i})
  \right.\nonumber\\
  &-&\left.a_0(M^2_Z)+a_0(m^2_{H^0_i})\right]
  +{1\over2}{g_2\over c_W}\sum_{i,k=1}^2A^{ik}_M~V^{ikj}_{SPP}
  ~{1\over p^2}\left[a_0(m^2_{H^0_i})
  \right.\nonumber\\
  &-&\left.a_0(m^2_{H^0_{k+2}})
  +\left(m^2_{H^0_i}-m^2_{H^0_{j+2}}\right)
  b_0(m^2_{H^0_i},m^2_{H^0_{j+2}})\right]\nonumber\\
  &+&{1\over2}{g^2_2\over c_W}(3~s^2_W-c^2_W)~M_W~\delta^{2j}
  ~b_0(M^2_W,M^2_W)\nonumber\\
  &-&~\sum_{l,k=1}^2F^{ZC}_{Cj}\left(m^2_{C_k},m^2_{C_l}\right)
  -{1\over2}~\sum_{l,k=1}^4F^{ZN}_{Cj}\left(m^2_{N_k},m^2_{N_l}\right)
\end{eqnarray}

\noindent The matrix $Z_H^{ij}$ is defined in Section 2.

\noindent In the same convention, for the ~$\gamma - G^0$ ~mixing we get:
\begin{eqnarray}
\left(4\pi\right)^2  \Sigma_{\gamma P}^2=-~2eg_2M_W~b_0(M^2_W,M^2_W)
\end{eqnarray}
\noindent At 1-loop there are no graphs which mix ~$\gamma$ ~with ~$A^0$.
{}~ Thus, if the finite part of the counterterm to this mixing
vanishes (as is the case in our renormalization scheme because
{}~$\delta v_1/v_1=\delta v_2/v_2$) photon does not mix with ~$A^0$.
\vskip 0.5cm

3) Scalar and pseudoscalar Higgs self energies. We write them as:
\begin{eqnarray}
  -i~\Sigma_{PP}^{ij}(p^2),~~~-i~\Sigma_{SS}^{ij}(p^2)
\end{eqnarray}
\noindent We then have:
\begin{eqnarray}
&\phantom{=}&\left(4\pi\right)^2  \Sigma_{PP}^{ij}~=~
  -\sum_{I=1}^3\left[2~{m^2_{l_I}\over v^2_1}~Z^{1i}_HZ^{1j}_H
  \left(2~a_0(m^2_{l_I})+p^2~b_0(m^2_{l_I},m^2_{l_I})\right)
  \right.\nonumber\\
  &+&6~{m^2_{d_I}\over v^2_1}~Z^{1i}_HZ^{1j}_H
  \left(2~a_0(m^2_{d_I})+p^2~b_0(m^2_{d_I},m^2_{d_I})\right)
  \nonumber\\
  &+&\left.6~{m^2_{u_I}\over v^2_2}~Z^{2i}_HZ^{2j}_H
  \left(2~a_0(m^2_{u_I})+p^2~b_0(m^2_{u_I},m^2_{u_I})\right)
  \right]\nonumber\\
  &+&\sum_{I=1}^3V^{ijII}_{PP\tilde\nu\tilde\nu}~a_0(m^2_{{\tilde\nu}_I})
  +\sum_{n=1}^6\left[V^{ijnn}_{PPLL}~a_0(m^2_{L_n})
  +3~V^{ijnn}_{PPDD}~a_0(m^2_{D_n})\right.\nonumber\\
  &+&\left.3~V^{ijnn}_{PPUU}~a_0(m^2_{U_n})\right]
  -\sum_{m,n=1}^6\left[3~V^{imn}_{PUU}~V^{jnm}_{PUU}
  ~b_0(m^2_{U_n},m^2_{U_m})\right.\nonumber\\
  &+&\left.3~V^{imn}_{PDD}~V^{jnm}_{PDD} ~b_0(m^2_{D_n},m^2_{D_m})
  +V^{imn}_{PLL}~V^{jnm}_{PLL} ~b_0(m^2_{L_n},m^2_{L_m})\right]\nonumber\\
  &-&2\sum_{k,l=1}^2 F^{CC}_{Dij}\left(m^2_{C_k},m^2_{C_l}\right)
 -\sum_{k,l=1}^4 F^{NN}_{Dij}\left(m^2_{N_k},m^2_{N_l}\right)\nonumber\\
  &+&{1\over4}{g^2_2\over c^2_W}\sum_{k=1}^2A^{ki}_M A^{kj}_M
  \left[\left(2~p^2+2~m^2_{H^0_k}-M^2_Z\right)
  b_0(M^2_Z,m^2_{H^0_k})\right.\\
  &-&\left.2~a_0(M^2_Z)+a_0(m^2_{H^0_k})\right]
  +{1\over2}g^2_2~\delta^{ij}
  \left[a_0(m^2_{H^+_i})-2~a_0(M^2_W)\right.\nonumber\\
  &+&\left.\left(2~p^2+2~m^2_{H^+_i}-M^2_W\right)
  ~b_0(M^2_W,m^2_{H^+_i})\right]\nonumber\\
  &+&{g^2_2\over c^2_W}~\delta^{ij}\left[a_0(M^2_Z)+2~c^2_W~a_0(M^2_W)
  \right]-{1\over2}g^2_2~\delta^{ij}~M^2_W~b_0(M^2_W,m^2_{H^+_i})\nonumber\\
  &-&\sum_{k,l=1}^2V_{SPP}^{lki}~V_{SPP}^{lkj}
  ~b_0(m^2_{H^0_{k+2}},m^2_{H^0_l})\nonumber\\
  &+&\sum_{k=1}^2\left[{1\over2}V^{ijkk}_{PPPP}~a_0(m^2_{H^0_{k+2}})
  +{1\over2}V^{kkij}_{SSPP}~a_0(m^2_{H^0_k})
  +V^{ijkk}_{PP+-}~a_0(m^2_{H^+_k})\right]\nonumber
\end{eqnarray}
\begin{eqnarray}
&\phantom{=}&\left(4\pi\right)^2\Sigma_{SS}^{ij}~=~
  -\sum_{I=1}^3\left[2{m^2_{l_I}\over v^2_1}Z^{1i}_R Z^{1j}_R
  \left(2a_0(m^2_{l_I})+\left(p^2-4m^2_{l_I}\right)
  b_0(m^2_{l_I},m^2_{l_I})\right)\right.\nonumber\\
  &+&6~{m^2_{d_I}\over v^2_1}~Z^{1i}_RZ^{1j}_R
  \left(2~a_0(m^2_{d_I})+\left(p^2-4~m^2_{d_I}\right)
  b_0(m^2_{d_I},m^2_{d_I})\right)\nonumber\\
  &+&\left.6~{m^2_{u_I}\over v^2_2}~Z^{2i}_RZ^{2j}_R
  \left(2~a_0(m^2_{u_I})+\left(p^2-4~m^2_{u_I}\right)
  b_0(m^2_{u_I},m^2_{u_I})\right)\right]\nonumber\\
  &+&\sum_{I=1}^3\left[V^{ijII}_{SS\tilde\nu\tilde\nu}~a_0(m^2_{{\tilde\nu}_I})
  -V^{iII}_{S\tilde\nu\tilde\nu}V^{jII}_{S\tilde\nu\tilde\nu}
  ~b_0(m^2_{{\tilde\nu}_I},m^2_{{\tilde\nu}_I})\right]\nonumber\\
  &+&\sum_{n=1}^6\left[V^{ijnn}_{SSLL}~a_0(m^2_{L_n})
  +3~V^{ijnn}_{SSDD}~a_0(m^2_{D_n})
  +3~V^{ijnn}_{SSUU}~a_0(m^2_{U_n})\right]\nonumber\\
  &-&\sum_{m,n=1}^6\left[3~V^{imn}_{SUU}~V^{jnm}_{SUU}
  ~b_0(m^2_{U_n},m^2_{U_m})\right.\nonumber\\
  &+&\left.3~V^{imn}_{SDD}~V^{jnm}_{SDD}
  ~b_0(m^2_{D_n},m^2_{D_m})
  +V^{imn}_{SLL}~V^{jnm}_{SLL}
  ~b_0(m^2_{L_n},m^2_{L_m})\right]\nonumber\\
  &-&2\sum_{k,l=1}^2 F^{CC}_{Dij}\left(m^2_{C_k},m^2_{C_l}\right)
 -\sum_{k,l=1}^4 F^{NN}_{Dij}\left(m^2_{N_k},m^2_{N_l}\right)\nonumber\\
  &+&{1\over4}{g^2_2\over c^2_W}\sum_{k=1}^2A^{ik}_M A^{jk}_M
  \left[\left(2~p^2+2~m^2_{H^0_{k+2}}-M^2_Z\right)
  b_0(M^2_Z,m^2_{H^0_{k+2}})\right.\nonumber\\
  &-&\left.2~a_0(M^2_Z)+a_0(m^2_{H^0_{k+2}})\right]
  -{1\over2}g^2_2\sum_{k=1}^2A^{ik}_M A^{jk}_M
  \left[2~a_0(M^2_W)\right.\\
  &-&\left.a_0(m^2_{H^+_k})-\left(2~p^2+2~m^2_{H^+_k}-M^2_W\right)
  b_0(M^2_W,m^2_{H^+_k})\right]\nonumber\\
  &+&{g^2_2\over c^2_W}~\delta^{ij}\left[a_0(M^2_Z)+2~c^2_W~a_0(M^2_W)
  \right]-{7\over16}{g^4_2\over c^4_W}C^i_R C^j_R\left[
  b_0(M^2_Z,M^2_Z)\right.\nonumber\\
  &+&\left.2~c^4_W~b_0(M^2_W,M^2_W)\right]
  -\sum_{k,l=1}^2\left[{1\over2}~V^{ikl}_{SSS}~V^{lkj}_{SSS}
  ~b_0(m^2_{H^0_k},m^2_{H^0_l})\right.\nonumber\\
  &+&\left.V^{ikl}_{S+-}~V^{jlk}_{S+-}~b_0(m^2_{H^+_k},m^2_{H^+_l})
  +{1\over2}~V^{ikl}_{SPP}~V^{jlk}_{SPP}
  ~b_0(m^2_{H^0_{k+2}},m^2_{H^0_{l+2}})\right]\nonumber\\
  &+&\sum_{k=1}^2\left[{1\over2}V^{ijkk}_{SSSS}~a_0(m^2_{H^0_k})
  +V^{ijkk}_{SS+-}~a_0(m^2_{H^+_k})
  +{1\over2}V^{ijkk}_{SSPP}~a_0(m^2_{H^0_{k+2}})\right]\nonumber
\end{eqnarray}
\vskip 0.5cm

4) Finally we give formulae for the scalar Higgs boson tadpoles
which are needed to compute counterterms. We write them as:
\begin{eqnarray}
  -i~{\cal T}^i
\end{eqnarray}
and we have:
\begin{eqnarray}
&\phantom{=}&\left(4\pi\right)^2  {\cal T}^i~=~-\sum_{I=1}^3\left[
  4~{m^2_{l_I}\over v_1}~Z^{1i}_R~a_0(m^2_{l_I})
  +12~{m^2_{d_I}\over v_1}~Z^{1i}_R~a_0(m^2_{d_I})\right.\nonumber\\
  &+&\left.12~{m^2_{u_I}\over v_2}~Z^{2i}_R~a_0(m^2_{u_I})\right]
+\sum_{I=1}^3V^{iII}_{S\tilde\nu\tilde\nu}~a_0(m^2_{{\tilde\nu}_I})\nonumber\\
  &+&\sum_{n=1}^6\left[V^{inn}_{SLL}~a_0(m^2_{L_n})
  +3~V^{inn}_{SDD}~a_0(m^2_{D_n})
  +3~V^{inn}_{SUU}~a_0(m^2_{U_n})\right]\\
  &-&2\sum_{k=1}^2\left(c^{Likk}_{SCC}-c^{Rikk}_{SCC}\right)
m_{C_k}a_0(m^2_{C_k})
  -\sum_{k=1}^4\left(c^{Likk}_{SNN}-c^{Rikk}_{SNN}\right)
m_{N_k}a_0(m^2_{N_k})\nonumber\\
  &+&{3\over4}{g^2_2\over c^2_W}~C^i_R\left[a_0(M^2_Z)
  +2~c^2_W~a_0(M^2_W)\right]\nonumber\\
  &+&\sum_{k=1}^2\left[{1\over2}~V^{ikk}_{SSS}~a_0(m^2_{H^0_k})
  +V^{ikk}_{S+-}~a_0(m^2_{H^+_k})
  +{1\over2}~V^{ikk}_{SPP}~a_0(m^2_{H^0_{k+2}})\right]\nonumber
\end{eqnarray}

\setcounter{equation}{0}
\section{Vertex formfactors}
\vskip 0.2 cm

\indent In this Appendix we collect formulae for bare vertex formfactors
used in 1-loop cross section calculations described in Section 4.
Renormalized formfactors can be obtained be adding to the expressions
displayed below the counterterms defined in Sections 2 and 3.
Momentum and Lorentz index conventions are shown on Figure (...). We
neglect momenta in the $c$-functions arguments. The notation
$c_0(m_1^2,m_2^2,m_3^2)$ is equivalent to the
$c_0(p,q,m_1^2,m_2^2,m_3^2)$ and the same for the other $c_i, c_{ij}$.

\subsection{$Z^0$-$Z^0$-scalar vertex}

\noindent 1. Formfactor proportional to $g^{\mu\nu}$. As in Appendix A,
it is very useful to define auxiliary functions describing
contributions given by the fermion triangle loops. Lets define for matter
fermions:
\begin{eqnarray}
f_{1a}^{ff}(m^2)&=&
-3-2(p^2+m^2)c_0(m^2,m^2,m^2)-16c_{24}(m^2,m^2,m^2)\nonumber\\
&-&2p^2(4c_{11}(m^2,m^2,m^2)+3c_{21}(m^2,m^2,m^2))\nonumber\\
&-&2q^2(2c_{12}(m^2,m^2,m^2)+3c_{22}(m^2,m^2,m^2))\nonumber\\
&-&2pq(c_0(m^2,m^2,m^2)+2c_{11}(m^2,m^2,m^2)\nonumber\\
&+&4c_{12}(m^2,m^2,m^2)+6c_{23}(m^2,m^2,m^2))
\end{eqnarray}
\begin{eqnarray}
f_{1b}^{ff}(m^2)&=& -1 + 2(p^2+m^2)c_0(m^2,m^2,m^2)-2p^2
c_{21}(m^2,m^2,m^2)\nonumber\\
&-&2q^2 c_{22}(m^2,m^2,m^2)\nonumber\\
&+&2pq~(c_0(m^2,m^2,m^2)-2c_{23}(m^2,m^2,m^2))
\end{eqnarray}
Contribution of charginos and neutralinos are much more complicated due
to complex structure of their couplings. Using notation for the vector
boson-fermion-fermion and scalar/pseudoscalar-fermion-fermion vertices
defined in (\ref{vff_vert}, \ref{sff_vert}) we can define auxiliary function
$g_{1f}^{iklm}$ as:
\begin{eqnarray}
g_{1i}^f&=&m_k m_l m_m~{\cal R}e\left[(c_V^{Zkl}
c_V^{Zmk}-c_A^{Zkl} c_A^{Zmk}) (c_L^{lmi}-c_R^{lmi}) \right.\nonumber\\
&-&\left.(c_A^{Zkl} c_V^{Zmk} -c_V^{Zkl} c_A^{Zmk})
(c_L^{lmi}+c_R^{lmi})\right] c_0(m^2_k,m^2_m,m^2_l)\nonumber\\
&+&m_k~{\cal R}e\left[(c_V^{Zkl} c_V^{Zmk} -c_A^{Zkl}
c_A^{Zmk})(c_L^{lmi}-c_R^{lmi})\right.\nonumber\\
&+&\left.(c_A^{Zkl}c_V^{Zmk}-c_V^{Zkl}
c_A^{Zmk})(c_L^{lmi}+c_R^{lmi})\right]\left[{1\over 2} + (p^2+pq)
c_0(m^2_k,m^2_m,m^2_l)\right.\nonumber\\
&+&(2p^2+pq) c_{11}(m^2_k,m^2_m,m^2_l)+(q^2+2 pq)
c_{12}(m^2_k,m^2_m,m^2_l)\nonumber\\
&+&p^2 c_{21}(m^2_k,m^2_m,m^2_l)+q^2
c_{22}(m^2_k,m^2_m,m^2_l)+2pq~c_{23}(m^2_k,m^2_m,m^2_l)\nonumber\\
&+&\left.4 c_{24}(m^2_k,m^2_m,m^2_l)\right]\nonumber\\
&-&m_l~{\cal R}e\left[(c_V^{Zkl} c_V^{Zmk}+c_A^{Zkl}
c_A^{Zmk})(c_L^{lmi}-c_R^{lmi})\right.\nonumber\\
&-&\left.(c_A^{Zkl}c_V^{Zmk}+c_V^{Zkl} c_A^{Zmk})(c_L^{lmi}+c_R^{lmi})\right]
\left[{1\over 2} + p^2 c_{11}(m^2_k,m^2_m,m^2_l)\right.\nonumber\\
&+&pq~c_{12}(m^2_k,m^2_m,m^2_l)+p^2 c_{21}(m^2_k,m^2_m,m^2_l)+q^2
c_{22}(m^2_k,m^2_m,m^2_l)\nonumber\\
&+&\left.2 pq~c_{23}(m^2_k,m^2_m,m^2_l)+2
c_{24}(m^2_k,m^2_m,m^2_l)\right]\nonumber\\
&-&m_m~{\cal R}e\left[(c_V^{Zkl} c_V^{Zmk}+c_A^{Zkl}
c_A^{Zmk})(c_L^{lmi}-c_R^{lmi})\right.\nonumber\\
&+&\left.(c_A^{Zkl}c_V^{Zmk}+c_V^{Zkl}c_A^{Zmk})(c_L^{lmi}+c_R^{lmi})\right]
\left[{1\over 2} + (p^2+pq) c_{11}(m^2_k,m^2_m,m^2_l)\right.\nonumber\\
&+&(q^2+pq) c_{12}(m^2_k,m^2_m,m^2_l)+p^2 c_{21}(m^2_k,m^2_m,m^2_l)+q^2
c_{22}(m^2_k,m^2_m,m^2_l)\nonumber\\
&+&\left.2 pq~c_{23}(m^2_k,m^2_m,m^2_l)+2c_{24}(m^2_k,m^2_m,m^2_l)\right]
\end{eqnarray}
Then we have:
\begin{eqnarray}
&-&\left(4\pi\right)^2F^i_1=
-{e^2\over 4 s^2_W c_W^2}\sum_{K=1}^3 \left[ m_{l_K}^2 {Z_R^{1i}\over v_1}
{}~(f_{1a}^{ll}(m_{l_K}^2)+(1-4s_W^2)^2
f_{1b}^{ll}(m_{l_K}^2))\right.\nonumber\\
&+& 3m_{u_K}^2 {Z_R^{2i}\over v_2}
{}~(f_{1a}^{uu}(m_{u_K}^2)+(1-{8\over
3}s_W^2)^2f_{1b}^{uu}(m_{u_K}^2))\nonumber\\
&+&\left. 3m_{d_K}^2 {Z_R^{1i}\over v_1}
{}~(f_{1a}^{dd}(m_{d_K}^2)+(1-{4\over
3}s_W^2)^2f_{1b}^{dd}(m_{d_K}^2))\right]\nonumber\\
&+&4\sum_{k,l,m=1}^2 g_{1i}^C(m^2_{C_k},m^2_{C_m},m^2_{C_l})
+2\sum_{k,l,m=1}^4g_{1i}^N(m^2_{N_k},m^2_{N_m},m^2_{N_l})\nonumber\\
&+&8\sum_{k,l,m=1}^6\left[{\cal R}e
(V_{ZLL}^{lk}V_{ZLL}^{km}V_{SLL}^{iml})~
c_{24}(m^2_{L_k},m^2_{L_m},m^2_{L_l})\right.\nonumber\\
&+&3~{\cal R}e
(V_{ZUU}^{lk}V_{ZUU}^{km}V_{SUU}^{iml})~
c_{24}(m^2_{U_k},m^2_{U_m},m^2_{U_l})\nonumber\\
&+&\left.3~{\cal R}e
(V_{ZDD}^{lk}V_{ZDD}^{km}V_{SDD}^{iml})~
c_{24}(m^2_{D_k},m^2_{D_m},m^2_{D_l})\right]\nonumber\\
&+&{e^4\over 2 s_W^4 c_W^4} B_R^i \sum_{K=1}^3
{}~c_{24}(m^2_{{\tilde \nu}_K},m^2_{{\tilde \nu}_K},m^2_{{\tilde
\nu}_K})\nonumber\\
&+&{e^4\over 4 s_W^4 c_W^4} \sum_{k,l=1}^2 A_M^{kl} A_M^{il} C_R^k
{}~c_{24}(m^2_{H^0_k},M^2_Z,m^2_{H^0_{l+2}})\nonumber\\
&+&{e^2\over s_W^2 c_W^2} \sum_{j,k,l=1}^2  A_M^{kj} A_M^{lj} V_{SSS}^{ikl}
{}~c_{24}(m^2_{H^0_{j+2}},m^2_{H^0_l},m^2_{H^0_k})\nonumber\\
&+&{e^4\over 4 s_W^4 c_W^4} \sum_{j,k,l=1}^2 A_M^{jk} A_M^{jl} A_H^{kl} B_R^i
{}~c_{24}(m^2_{H^0_j},m^2_{H^0_{l+2}},m^2_{H^0_{k+2}})\nonumber\\
&+&{2e^2(c_W^2-s_W^2)^2\over s^2_W c_W^2} \sum_{j=1}^2 V_{S+-}^{ijj}
{}~c_{24}(m^2_{H^+_j},m^2_{H^+_j},m^2_{H^+_j})\nonumber\\
&-&{e^4 M_Z^2\over 2 c_W^2} A_H^{22} B_R^i~c_0(M^2_W,M^2_W,M^2_W)\nonumber\\
&+&{e^4 \over 2 c_W^2} C_R^i~(b_0(p^2,M^2_W,M^2_W)
+b_0((p+q)^2,M^2_W,M^2_W)\nonumber\\
&+&2c_{24}(M^2_W,M^2_W,M^2_W)-2 M_W^2 c_0(M^2_W,M^2_W,M^2_W))\nonumber\\
&-&{e^4 \over 2 s_W^2}
C_R^i~((2p^2+2pq-q^2)c_0(M^2_W,M^2_W,M^2_W)
-2p^2c_{21}(M^2_W,M^2_W,M^2_W)-1\nonumber\\
&-&2q^2c_{22}(M^2_W,M^2_W,M^2_W)
-4pq~c_{23}(M^2_W,M^2_W,M^2_W)-4c_{24}(M^2_W,M^2_W,M^2_W))\nonumber\\
&-&{e^6\over 8 s_W^6 c_W^6} C_R^i \sum_{k=1}^2 \left(C_R^k\right)^2
{}~c_0(m^2_{H^0_k},M^2_Z,M^2_Z)\nonumber\\
&-&{e^4\over 4 s_W^4 c_W^4} \sum_{j,k=1}^2 C_R^k C_R^j V_{SSS}^{ijk}
{}~c_0(M^2_Z,m^2_{H^0_k},m^2_{H^0_j})\nonumber\\
&+&{e^4\over 4 s_W^4 c_W^4} \sum_{j,k=1}^2 C_R^j A_M^{jk} A_M^{ik}
{}~c_{24}(m^2_{H^0_j},m^2_{H^0_{k+2}},M^2_Z)\nonumber\\
&+&{e^4 c_W^2 \over s_W^4} C_R^i~(5(p^2+pq)c_0(M^2_W,M^2_W,M^2_W)
+(2p^2+pq)c_{11}(M^2_W,M^2_W,M^2_W)\nonumber\\
&+&2p^2c_{21}(M^2_W,M^2_W,M^2_W)+(q^2+2pq)c_{12}(M^2_W,M^2_W,M^2_W)\nonumber\\
&+&2q^2~c_{22}(M^2_W,M^2_W,M^2_W)+4pq~c_{23}(M^2_W,M^2_W,M^2_W)\nonumber\\
&+&17c_{24}(M^2_W,M^2_W,M^2_W))+3b_0(q^2,M^2_W,M^2_W)-1)\nonumber\\
&+&\sum_{k,l=1}^6\left[V_{ZZLL}^{lk}V_{SLL}^{ikl}~b_0(q^2,m^2_{L_l},m^2_{L_k})
+3V_{ZZUU}^{lk}V_{SUU}^{ikl}~b_0(q^2,m^2_{U_l},m^2_{U_k})\right.\nonumber\\
&+&\left.3V_{ZZDD}^{lk}V_{SDD}^{ikl}~b_0(q^2,m^2_{D_l},m^2_{D_k})\right]
+{e^4\over 8 s_W^4 c_W^4} B_R^i \sum_{K=1}^3 b_0(q^2,m^2_{{\tilde
\nu}_K},m^2_{{\tilde \nu}_K})\nonumber\\
&+&{e^2(c_W^2-s_W^2)^2\over 2 s^2_W c_W^2} \sum_{j=1}^2
V_{S+-}^{ijj}~b_0(q^2,m^2_{H^+_j},m^2_{H^+_j})\nonumber\\
&+&{e^2\over 4 s_W^2 c_W^2} \sum_{j=1}^2 V_{SSS}^{ijj}
{}~b_0(q^2,m^2_{H^0_j},m^2_{H^0_j})\\
&+&{e^4\over 16 s_W^4 c_W^4} B_R^i\sum_{j=1}^2 A_H^{jj}
{}~b_0(q^2,m^2_{H^0_{j+2}},m^2_{H^0_{j+2}})\nonumber\\
&+&{e^4\over 4 s_W^4 c_W^4} C_R^i~(b_0(p^2,M^2_Z,m^2_{H^0_i})
+b_0((p+q)^2,M^2_Z,m^2_{H^0_i}))\nonumber
\end{eqnarray}

\noindent 2. Formfactor proportional to $q^{\mu}q^{\nu}$. Lets define:
\begin{eqnarray}
g_{2i}^f&=&m_l~{\cal R}e\left[(c_V^{Zkl} c_V^{Zmk}+c_A^{Zkl}
c_A^{Zmk})(c_L^{lmi}-c_R^{lmi})\right.\nonumber\\
&-&\left.(c_A^{Zkl}c_V^{Zmk}+c_V^{Zkl}
c_A^{Zmk})(c_L^{lmi}+c_R^{lmi})\right]c_{22}(m^2_k,m^2_m,m^2_l)\nonumber\\
&+&m_m~{\cal R}e\left[(c_V^{Zkl}
c_V^{Zmk}+c_A^{Zkl}c_A^{Zmk})(c_L^{lmi}-c_R^{lmi})\right.\nonumber\\
&+&\left.(c_A^{Zkl}c_V^{Zmk}+c_V^{Zkl}c_A^{Zmk})(c_L^{lmi}+c_R^{lmi})\right]
\left[c_{12}(m^2_k,m^2_m,m^2_l)\right.\nonumber\\
&+&\left.c_{22}(m^2_k,m^2_m,m^2_l)\right]
\end{eqnarray}
Then:
\begin{eqnarray}
&-&\left(4\pi\right)^2F^i_3=
-{e^2\over s^2_W c_W^2}\sum_{K=1}^3\left[m_{l_K}^2 {Z_R^{1i}\over v_1}
(1+(1-4s_W^2)^2)
{}~(c_{12}(m^2_{l_K},m^2_{l_K},m^2_{l_K})\right.\nonumber\\
&+&2c_{22}(m^2_{l_K},m^2_{l_K},m^2_{l_K}))\nonumber\\
&+&3m_{u_K}^2 {Z_R^{2i}\over v_2} (1+(1-{8\over 3}s_W^2)^2)
{}~(c_{12}(m^2_{u_K},m^2_{u_K},m^2_{u_K})
+2c_{22}(m^2_{u_K},m^2_{u_K},m^2_{u_K}))\nonumber\\
&+&\left.3m_{d_K}^2 {Z_R^{1i}\over v_1} (1+(1-{4\over 3}s_W^2)^2)
{}~(c_{12}(m^2_{d_K},m^2_{d_K},m^2_{d_K})
+2c_{22}(m^2_{d_K},m^2_{d_K},m^2_{d_K}))\right]\nonumber\\
&+&8\sum_{k,l,m=1}^2 g_{2i}^C(m^2_{C_k},m^2_{C_m},m^2_{C_l})
+4 \sum_{k,l,m=1}^4 g_{2i}^N(m^2_{N_k},m^2_{N_m},m^2_{N_l})\nonumber\\
&+&4\sum_{k,l,m=1}^6\left[{\cal R}e (V_{ZLL}^{lk}V_{ZLL}^{km}V_{SLL}^{iml})
{}~(c_{12}(m^2_{L_k},m^2_{L_m},m^2_{L_l})
+2c_{22}(m^2_{L_k},m^2_{L_m},m^2_{L_l}))\right.\nonumber\\
&+&~3{\cal R}e (V_{ZUU}^{lk}V_{ZUU}^{km}V_{SUU}^{iml})
{}~(c_{12}(m^2_{U_k},m^2_{U_m},m^2_{U_l})
+2c_{22}(m^2_{U_k},m^2_{U_m},m^2_{U_l}))\nonumber\\
&+&\left.3~{\cal R}e (V_{ZDD}^{lk}V_{ZDD}^{km}V_{SDD}^{iml})
{}~(c_{12}(m^2_{D_k},m^2_{D_m},m^2_{D_l})
+2c_{22}(m^2_{D_k},m^2_{D_m},m^2_{D_l}))\right]\nonumber\\
&+&{e^4\over 4 s_W^4 c_W^4}B_R^i\sum_{K=1}^3 (c_{12}(m^2_{{\tilde
\nu}_K},m^2_{{\tilde \nu}_K},m^2_{{\tilde \nu}_K})
+2c_{22}(m^2_{{\tilde \nu}_K},m^2_{{\tilde \nu}_K},m^2_{{\tilde
\nu}_K}))\nonumber\\
&+&{e^4\over 8 s_W^4 c_W^4}\sum_{k,l=1}^2 A_M^{kl} A_M^{il}
C_R^k~(2c_0(m^2_{H^0_k},M^2_Z,m^2_{H^0_{l+2}})\nonumber\\
&+&5c_{12}(m^2_{H^0_k},M^2_Z,m^2_{H^0_{l+2}})
+2c_{22}(m^2_{H^0_k},M^2_Z,m^2_{H^0_{l+2}}))\nonumber\\
&+&{e^2\over 2 s_W^2 c_W^2}\sum_{j,k,l=1}^2 A_M^{kj} A_M^{lj}
V_{SSS}^{ikl}~(c_{12}(m^2_{H^0_{j+2}},m^2_{H^0_l},m^2_{H^0_k})\nonumber\\
&+&2c_{22}(m^2_{H^0_{j+2}},m^2_{H^0_l},m^2_{H^0_k}))\nonumber\\
&+&{e^4\over 8 s_W^4 c_W^4} B_R^i\sum_{j,k,l=1}^2 A_M^{jk} A_M^{jl}
A_H^{kl}~(c_{12}(m^2_{H^0_j},m^2_{H^0_{l+2}},m^2_{H^0_{k+2}})\nonumber\\
&+&2c_{22}(m^2_{H^0_j},m^2_{H^0_{l+2}},m^2_{H^0_{k+2}}))\nonumber\\
&+&{e^2(c_W^2-s_W^2)^2\over s_W^2 c_W^2}\sum_{j=1}^2
V_{S+-}^{ijj}~(c_{12}(m^2_{H^+_j},m^2_{H^+_j},m^2_{H^+_j})\nonumber\\
&+&2c_{22}(m^2_{H^+_j},m^2_{H^+_j},m^2_{H^+_j}))\nonumber\\
&-&{e^4 \over 2
s_W^2}C_R^i~(c_0(M^2_W,M^2_W,M^2_W)+2c_{22}(M^2_W,M^2_W,M^2_W))\nonumber\\
&-&{e^4(c_W^2-s_W^2)\over 4 s_W^2 c_W^2} C_R^i (2c_0(M^2_W,M^2_W,M^2_W)
+4c_{12}(M^2_W,M^2_W,M^2_W)\nonumber\\
&+&3c_{22}(M^2_W,M^2_W,M^2_W))\\
&+&{e^4\over 4 s_W^4 c_W^4}\sum_{j,k=1}^2 C_R^j A_M^{jk} A_M^{ik}
{}~(c_{22}(m^2_{H^0_j},m^2_{H^0_{k+2}},M^2_Z)\nonumber\\
&-&c_{12}(m^2_{H^0_j},m^2_{H^0_{k+2}},M^2_Z))\nonumber\\
&+&{9 e^4 c_W^2\over 2 s_W^4}
C_R^i~(c_{12}(M^2_W,M^2_W,M^2_W)+2c_{22}(M^2_W,M^2_W,M^2_W))\nonumber
\end{eqnarray}

\noindent 3. Formfactor proportional to $q^{\nu}p^{\mu}$.
Lets define:
\begin{eqnarray}
f_{2a}^{ff}(m^2)&=&{1\over 2}c_0(m^2,m^2,m^2)+c_{11}(m^2,m^2,m^2)\nonumber\\
&+&c_{12}(m^2,m^2,m^2)+2c_{23}(m^2,m^2,m^2)
\end{eqnarray}
\begin{eqnarray}
f_{2b}^{ff}(m^2)&=&-{1\over 2}c_0(m^2,m^2,m^2)+c_{12}(m^2,m^2,m^2)\nonumber\\
&+&2c_{23}(m^2,m^2,m^2)
\end{eqnarray}
\begin{eqnarray}
g_{3i}^f&=&-m_k~{\cal R}e\left[(c_V^{Zkl} c_V^{Zmk} -c_A^{Zkl}
c_A^{Zmk})(c_L^{lmi}-c_R^{lmi})\right.\nonumber\\
&+&\left.(c_A^{Zkl}c_V^{Zmk}-c_V^{Zkl}
c_A^{Zmk})(c_L^{lmi}+c_R^{lmi})\right]
\left[c_0(m^2_k,m^2_m,m^2_l)\right.\nonumber\\
&+&\left.c_{11}(m^2_k,m^2_m,m^2_l)\right]\nonumber\\
&+&m_l~{\cal R}e\left[(c_V^{Zkl} c_V^{Zmk}+c_A^{Zkl}
c_A^{Zmk})(c_L^{lmi}-c_R^{lmi})\right.\nonumber\\
&-&\left.(c_A^{Zkl}c_V^{Zmk}+c_V^{Zkl} c_A^{Zmk})(c_L^{lmi}+c_R^{lmi})\right]
\left[c_{12}(m^2_k,m^2_m,m^2_l)\right.\nonumber\\
&+&\left.2c_{23}(m^2_k,m^2_m,m^2_l)\right]\nonumber\\
&+&m_m~{\cal R}e\left[(c_V^{Zkl} c_V^{Zmk}+c_A^{Zkl}
c_A^{Zmk})(c_L^{lmi}-c_R^{lmi})\right.\nonumber\\
&+&\left.(c_A^{Zkl}c_V^{Zmk}+c_V^{Zkl}c_A^{Zmk})(c_L^{lmi}+c_R^{lmi})\right]
\left[c_{11}(m^2_k,m^2_m,m^2_l)\right.\nonumber\\
&+&\left.c_{12}(m^2_k,m^2_m,m^2_l)+2 c_{23}(m^2_k,m^2_m,m^2_l)\right]
\end{eqnarray}
Then:
\begin{eqnarray}
&-&\left(4\pi\right)^2F^i_4=
-{e^2\over s^2_W c_W^2}\sum_{K=1}^3\left[m_{l_K}^2 {Z_R^{1i}\over v_1}
{}~(f_{2a}^{ll}(m^2_{l_K})
+(1-4s_W^2)^2f_{2b}^{ll}(m^2_{l_K}))\right.\nonumber\\
&+&3m_{u_K}^2 {Z_R^{2i}\over v_2}
{}~(f_{2a}^{uu}(m^2_{u_K})+(1-{8\over
3}s_W^2)^2f_{2b}^{uu}(m^2_{u_K}))\nonumber\\
&+&\left.3m_{d_K}^2 {Z_R^{1i}\over v_1}
{}~(f_{2a}^{dd}(m^2_{d_K})+(1-{4\over
3}s_W^2)^2f_{2b}^{dd}(m^2_{d_K}))\right]\nonumber\\
&+&4\sum_{k,l,m=1}^2 g_{3i}^C(m^2_{C_k},m^2_{C_m},m^2_{C_l})
 +2\sum_{k,l,m=1}^4 g_{3i}^N(m^2_{N_k},m^2_{N_m},m^2_{N_l})\nonumber\\
&+&4\sum_{k,l,m=1}^6\left[{\cal R}e (V_{ZLL}^{lk}V_{ZLL}^{km}V_{SLL}^{iml})
{}~(c_{12}(m^2_{L_k},m^2_{L_m},m^2_{L_l})
+2c_{23}(m^2_{L_k},m^2_{L_m},m^2_{L_l}))\right.\nonumber\\
&+&3~{\cal R}e (V_{ZUU}^{lk}V_{ZUU}^{km}V_{SUU}^{iml})
{}~(c_{12}(m^2_{U_k},m^2_{U_m},m^2_{U_l})
+2c_{23}(m^2_{U_k},m^2_{U_m},m^2_{U_l}))\nonumber\\
&+&\left.3~{\cal R}e (V_{ZDD}^{lk}V_{ZDD}^{km}V_{SDD}^{iml})
{}~(c_{12}(m^2_{D_k},m^2_{D_m},m^2_{D_l})
+2c_{23}(m^2_{D_k},m^2_{D_m},m^2_{D_l}))\right]\nonumber\\
&+&{e^4\over 4 s_W^4 c_W^4} B_R^i\sum_{K=1}^3
{}~(c_{12}(m^2_{{\tilde \nu}_K},m^2_{{\tilde \nu}_K},m^2_{{\tilde \nu}_K})
+2c_{23}(m^2_{{\tilde \nu}_K},m^2_{{\tilde \nu}_K},m^2_{{\tilde
\nu}_K}))\nonumber\\
&+&{e^4\over 8 s_W^4 c_W^4} \sum_{k,l=1}^2 A_M^{kl} A_M^{il} C_R^k
{}~(2c_0(m^2_{H^0_k},M^2_Z,m^2_{H^0_{l+2}})\nonumber\\
&+&c_{12}(m^2_{H^0_k},M^2_Z,m^2_{H^0_{l+2}})
+4c_{11}(m^2_{H^0_k},M^2_Z,m^2_{H^0_{l+2}})
+2c_{23}(m^2_{H^0_k},M^2_Z,m^2_{H^0_{l+2}}))\nonumber\\
&+&{e^2\over 2 s_W^2 c_W^2} \sum_{j,k,l=1}^2 A_M^{kj} A_M^{lj} V_{SSS}^{ikl}
{}~(c_{12}(m^2_{H^0_{j+2}},m^2_{H^0_l},m^2_{H^0_k})\nonumber\\
&+&2c_{23}(m^2_{H^0_{j+2}},m^2_{H^0_l},m^2_{H^0_k}))\nonumber\\
&+&{e^4\over 8 s_W^4 c_W^4}B_R^i \sum_{j,k,l=1}^2 A_M^{jk} A_M^{jl} A_H^{kl}
{}~(c_{12}(m^2_{H^0_j},m^2_{H^0_{l+2}},m^2_{H^0_{k+2}})\nonumber\\
&+&2c_{23}(m^2_{H^0_j},m^2_{H^0_{l+2}},m^2_{H^0_{k+2}}))\nonumber\\
&+&{e^2(c_W^2-s_W^2)^2\over s^2_W c_W^2} \sum_{j=1}^2 V_{S+-}^{ijj}
{}~(c_{12}(m^2_{H^+_j},m^2_{H^+_j},m^2_{H^+_j})
+2c_{23}(m^2_{H^+_j},m^2_{H^+_j},m^2_{H^+_j}))\nonumber\\
&-&{e^4 \over 2 s_W^2} C_R^i~(2c_{23}(M^2_W,M^2_W,M^2_W)
-7c_0(M^2_W,M^2_W,M^2_W)\nonumber\\
&-&6c_{11}(M^2_W,M^2_W,M^2_W))\nonumber\\
&-&{e^4(c_W^2-s_W^2)\over 4 s_W^2 c_W^2} C_R^i~(2c_0(M^2_W,M^2_W,M^2_W)
+4c_{11}(M^2_W,M^2_W,M^2_W)\nonumber\\
&+&3c_{12}(M^2_W,M^2_W,M^2_W)+4c_{23}(M^2_W,M^2_W,M^2_W))\\
&+&{e^4\over 4 s_W^4 c_W^4} \sum_{j,k=1}^2 C_R^j A_M^{jk} A_M^{ik}
{}~(c_{12}(m^2_{H^0_j},m^2_{H^0_{k+2}},M^2_Z)
+c_{23}(m^2_{H^0_j},m^2_{H^0_{k+2}},M^2_Z))\nonumber\\
&+&{e^4 c_W^2\over 2 s_W^4} C_R^i
{}~(2c_{11}(M^2_W,M^2_W,M^2_W)-8c_0(M^2_W,M^2_W,M^2_W)\nonumber\\
&+&9c_{12}(M^2_W,M^2_W,M^2_W)+18c_{23}(M^2_W,M^2_W,M^2_W))\nonumber
\end{eqnarray}

\subsection{$Z^0$-scalar-pseudoscalar vertex}

\noindent 1. Formfactor proportional to the momentum of the
pseudoscalar. In order to distinguish between scalar and pseudoscalar
couplings we have added additional subscripts $S$ or $P$ in the symbols
denoting scalar/pseudoscalar-2~chargino/neutralino vertices. Lets define
("${\cal I}m$" in $g_{4ij}^f$ is artificial - we are using uniform
convention (\ref{sff_vert}) for the $S$ and $P$ vertices, but actually
pseudoscalar couplings do not contain $i$):
\begin{eqnarray}
f_3^{ff}(m^2)&=&b_0(q^2,m^2,m^2)-2m^2c_0(m^2,m^2,m^2)\nonumber\\
&+&p^2(c_{11}(m^2,m^2,m^2)+2c_{21}(m^2,m^2,m^2))
-q^2c_{12}(m^2,m^2,m^2)\nonumber\\
&+&2pq~c_{23}(m^2,m^2,m^2)+2c_{24}(m^2,m^2,m^2))
\end{eqnarray}
\begin{eqnarray}
g_{4ij}^f&=&{\cal I}m\left[c_V^{Zkm}(c_{RS}^{lki}c_{LP}^{mlj}
+c_{LS}^{lki}c_{RP}^{mlj})+c_A^{Zkm}(c_{RS}^{lki}c_{LP}^{mlj}
-c_{LS}^{lki}c_{RP}^{mlj})\right]\times\nonumber\\
&\times&\left[(p^2+m_m^2)c_{11}(m^2_m,m^2_l,m^2_k)
-q^2 c_{12}(m^2_m,m^2_l,m^2_k)\right.\nonumber\\
&+&2p^2 c_{21}(m^2_m,m^2_l,m^2_k)+2pq~c_{23}(m^2_m,m^2_l,m^2_k)
+2c_{24}(m^2_m,m^2_l,m^2_k)\nonumber\\
&+&\left.b_0(q^2,m^2_l,m^2_k))\right]\nonumber\\
&-& m_k m_l {\cal I}m\left[c_V^{Zkm}(c_{RS}^{lki}c_{RP}^{mlj}
+c_{LS}^{lki}c_{LP}^{mlj}))-c_A^{Zkm}(c_{RS}^{lki}c_{RP}^{mlj}
-c_{LS}^{lki}c_{LP}^{mlj})\right]\times\nonumber\\
&\times&c_{11}(m^2_m,m^2_l,m^2_k)\nonumber\\
&+&m_k m_m {\cal I}m\left[c_V^{Zkm}(c_{RS}^{lki}c_{LP}^{mlj}
+c_{LS}^{lki}c_{RP}^{mlj})
-c_A^{Zkm}(c_{RS}^{lki}c_{LP}^{mlj}
-c_{LS}^{lki}c_{RP}^{mlj})\right]\times\nonumber\\
&\times&\left[c_0(m^2_m,m^2_l,m^2_k)
+c_{11}(m^2_m,m^2_l,m^2_k)\right]\nonumber\\
&-&m_m m_l {\cal I}m\left[c_V^{Zkm}(c_{RS}^{lki}c_{RP}^{mlj}
+c_{LS}^{lki}c_{LP}^{mlj})
+c_A^{Zkm}(c_{RS}^{lki}c_{RP}^{mlj}
-c_{LS}^{lki}c_{LP}^{mlj})\right]\times\nonumber\\
&\times&\left[c_0(m^2_m,m^2_l,m^2_k)+c_{11}(m^2_m,m^2_l,m^2_k))\right]
\end{eqnarray}
Then:
\begin{eqnarray}
&\phantom{-}&\left(4\pi\right)^2F^{ij}_P=
-{2e\over s_W c_W}\sum_{K=1}^3\left[{m_{l_K}^2\over v_1^2} Z_H^{1j}Z_R^{1i}
{}~f_3^{ll}(m^2_{l_K}) - {3m_{u_K}^2\over
v_2^2}Z_H^{2j}Z_R^{2i}~f_3^{uu}(m^2_{u_K})\right.\nonumber\\
&+&\left.{3m_{d_K}^2\over
v_1^2}Z_H^{1j}Z_R^{1i}~f_3^{dd}(m^2_{d_K})\right]\nonumber\\
&+&4\sum_{k,l,m=1}^2 g_{4ij}^C(m^2_{C_m},m^2_{C_l},m^2_{C_k})
+2\sum_{k,l,m=1}^4 g_{4ij}^N(m^2_{N_m},m^2_{N_l},m^2_{N_k})\nonumber\\
&+&2\sum_{k,l,m=1}^6\left[{\cal I}m  (V_{ZLL}^{lk}V_{PLL}^{jkm}V_{SLL}^{iml})
{}~(c_0(m^2_{L_k},m^2_{L_l},m^2_{L_m})
+2c_{11}(m^2_{L_k},m^2_{L_l},m^2_{L_m}))\right.\nonumber\\
&-&3~{\cal I}m
(V_{ZUU}^{lk}V_{PUU}^{jkm}V_{SUU}^{iml})~(c_0(m^2_{U_k},m^2_{U_l},m^2_{U_m})
+2c_{11}(m^2_{U_k},m^2_{U_l},m^2_{U_m}))\nonumber\\
&+&\left.3~{\cal I}m
(V_{ZDD}^{lk}V_{PDD}^{jkm}V_{SDD}^{iml})~(c_0(m^2_{D_k},m^2_{D_l},m^2_{D_m})
+2c_{11}(m^2_{D_k},m^2_{D_l},m^2_{D_m}))\right]\nonumber\\
&+&{e^2(c_W^2-s_W^2)\over 2s_W^2}M_Z\sum_{k,l=1}^2\varepsilon_{kl}V_{S+-}^{ikl}
{}~(c_0(m^2_{H^+_k},m^2_{H^+_l},m^2_{H^+_k})\nonumber\\
&+&2c_{11}(m^2_{H^+_k},m^2_{H^+_l},m^2_{H^+_k}))\nonumber\\
&+&{e^3\over 8 s_W^3 c_W^3}\sum_{k,l,m=1}^2A_M^{lk}A_H^{jk}
B_R^mV_{SSS}^{ilm}~(c_0(m^2_{H^0_{k+2}},m^2_{H^0_m},m^2_{H^0_l})\nonumber\\
&+&2c_{11}(m^2_{H^0_{k+2}},m^2_{H^0_m},m^2_{H^0_l}))\nonumber\\
&-&{e^5\over 32 s_W^5 c_W^5}B_R^i\sum_{k,l,m=1}^2A_M^{kl} A_H^{jm}
A_H^{lm} B_R^k ~(c_0(m^2_{H^0_k},m^2_{H^0_{m+2}},m^2_{H^0_{l+2}})\nonumber\\
&+&2c_{11}(m^2_{H^0_k},m^2_{H^0_{m+2}},m^2_{H^0_{l+2}}))\nonumber\\
&+&{e^3 c_W\over 2 s_W^3}A_M^{ij}~(1+6c_{24}(M^2_W,m^2_{H^+_j},M^2_W)
+q^2(2c_0(M^2_W,m^2_{H^+_j},M^2_W)\nonumber\\
&+&c_{11}(M^2_W,m^2_{H^+_j},M^2_W)
-c_{12}(M^2_W,m^2_{H^+_j},M^2_W)+2c_{22}(M^2_W,m^2_{H^+_j},M^2_W)\nonumber\\
&+&2c_{23}(M^2_W,m^2_{H^+_j},M^2_W))+pq(6c_0(M^2_W,m^2_{H^+_j},M^2_W)
+c_{12}(M^2_W,m^2_{H^+_j},M^2_W)\nonumber\\
&+&7c_{11}(M^2_W,m^2_{H^+_j},M^2_W)
+2c_{21}(M^2_W,m^2_{H^+_j},M^2_W)+2c_{23}(M^2_W,m^2_{H^+_j},M^2_W)))\nonumber\\
&+&{e^4 M_Z c_W^2 \over 8 s_W^4} C_R^i~\delta^{j2}
c_0(M^2_W,M^2_W,M^2_W)\nonumber\\
&+&{e^4 M_Z\over 4 s_W^2} C_R^i \delta^{j2}~(c_0(M^2_W,M^2_W,M^2_W)
-c_{11}(M^2_W,M^2_W,M^2_W))\nonumber\\
&+&e^2 M_Z s_W V_{S+-}^{i2j}~(c_{11}(M^2_W,m^2_{H^+_j},M^2_W)
+2c_0(M^2_W,m^2_{H^+_j},M^2_W))\nonumber\\
&+&{e^3 M^2_W\over 4 s_W c_W} A_M^{ij}
\delta^{j1}~(c_{11}(M^2_W,m^2_{H^+_j},M^2_W)
+c_0(M^2_W,m^2_{H^+_j},M^2_W))\nonumber\\
&-&{e^3\over 4 s_W^3 c_W^3}\sum_{k,l=1}^2 C_R^k A_M^{lj}
V_{SSS}^{kli}~(c_{11}(M^2_Z,m^2_{H^0_l},m^2_{H^0_k})
+2c_0(M^2_Z,m^2_{H^0_l},m^2_{H^0_k}))\nonumber\\
&-&{e^5\over 16 s_W^5 c_W^5}\sum_{k,l=1}^2 C_R^k B_R^k
A_H^{lj} A_M^{il}~(c_{11}(m^2_{H^0_k},m^2_{H^0_{l+2}},M^2_Z)
+c_0(m^2_{H^0_k},m^2_{H^0_{l+2}},M^2_Z))\nonumber\\
&+&{e^5\over 8 s_W^5 c_W^5} C_R^i \sum_{k=1}^2 C_R^k
A_M^{kj}~(c_{11}(m^2_{H^0_k},M^2_Z,M^2_Z)
-c_0(m^2_{H^0_k},M^2_Z,M^2_Z))\nonumber\\
&+&{e^3 (c_W^2-s_W^2)\over 4 s_W^3 c_W}A_M^{ij}~({1\over 2}
+2b_0(q^2,M^2_W,m^2_{H^+_j})+4c_{24}(m^2_{H^+_j},M^2_W,m^2_{H^+_j})\nonumber\\
&+&2m^2_{H^+_j}c_{11}(m^2_{H^+_j},M^2_W,m^2_{H^+_j})
+p^2(c_{21}(m^2_{H^+_j},M^2_W,m^2_{H^+_j})\nonumber\\
&-&c_0(m^2_{H^+_j},M^2_W,m^2_{H^+_j})
-2c_{11}(m^2_{H^+_j},M^2_W,m^2_{H^+_j}))\nonumber\\
&+&q^2(2c_{12}(m^2_{H^+_j},M^2_W,m^2_{H^+_j})
+c_{22}(m^2_{H^+_j},M^2_W,m^2_{H^+_j})\nonumber\\
&+&4c_{23}(m^2_{H^+_j},M^2_W,m^2_{H^+_j}))
+2pq(2c_{21}(m^2_{H^+_j},M^2_W,m^2_{H^+_j})\nonumber\\
&+&c_{23}(m^2_{H^+_j},M^2_W,m^2_{H^+_j})-c_0(m^2_{H^+_j},M^2_W,m^2_{H^+_j})
-c_{11}(m^2_{H^+_j},M^2_W,m^2_{H^+_j})))\nonumber\\
&-&{e^3\over 8 s_W^3 c_W^3} \sum_{k,l=1}^2 A_M^{il} A_M^{kj} A_M^{kl}
({1\over 2} + 2b_0(q^2,M^2_Z,m^2_{H^0_{l+2}})\nonumber\\
&+&4c_{24}(m^2_{H^0_k},M^2_Z,m^2_{H^0_{l+2}})
+2m^2_{H^0_k}c_{11}(m^2_{H^0_k},M^2_Z,m^2_{H^0_{l+2}})\nonumber\\
&+&p^2(c_{21}(m^2_{H^0_k},M^2_Z,m^2_{H^0_{l+2}})
-c_0(m^2_{H^0_k},M^2_Z,m^2_{H^0_{l+2}})\nonumber\\
&-&2c_{11}(m^2_{H^0_k},M^2_Z,m^2_{H^0_{l+2}}))
+q^2(2c_{12}(m^2_{H^0_k},M^2_Z,m^2_{H^0_{l+2}})\nonumber\\
&+&c_{22}(m^2_{H^0_k},M^2_Z,m^2_{H^0_{l+2}})
+4c_{23}(m^2_{H^0_k},M^2_Z,m^2_{H^0_{l+2}}))\nonumber\\
&-&2pq(c_0(m^2_{H^0_k},M^2_Z,m^2_{H^0_{l+2}})
+c_{11}(m^2_{H^0_k},M^2_Z,m^2_{H^0_{l+2}})\nonumber\\
&-&2c_{21}(m^2_{H^0_k},M^2_Z,m^2_{H^0_{l+2}})
-c_{23}(m^2_{H^0_k},M^2_Z,m^2_{H^0_{l+2}})))\nonumber\\
&+&{e^3\over 4 s_W^3 c_W^3}A_M^{ij}~(2b_0(p^2,M^2_Z,m^2_{H^0_i})
+b_1(p^2,M^2_Z,m^2_{H^0_i}))\\
&+&{e^3\over 2 s_W c_W}A_M^{ij}~(2b_0(p^2,M^2_W,m^2_{H^+_j})
+b_1(p^2,M^2_W,m^2_{H^+_j}))\nonumber\\
&-&{e^3 \over 8 s_W^3 c_W^3} \sum_{k,l=1}^2 A_M^{lk} A_R^{il} A_H^{kj}
{}~(b_0((p+q)^2,m^2_{H^0_{k+2}},m^2_{H^0_l})\nonumber\\
&+&2b_1((p+q)^2,m^2_{H^0_{k+2}},m^2_{H^0_l}))\nonumber
\end{eqnarray}

\noindent 2. Formfactor proportional to the momentum of the scalar.
Lets define:
\begin{eqnarray}
f_4^{ff}(m^2)&=&-b_0(q^2,m^2,m^2)-b_1(q^2,m^2,m^2)\nonumber\\
&+&p^2(c_{11}(m^2,m^2,m^2)+2c_{23}(m^2,m^2,m^2)+c_{12}(m^2,m^2,m^2))\nonumber\\
&+&2pq(c_{12}(m^2,m^2,m^2)+c_{22}(m^2,m^2,m^2))
\end{eqnarray}
\begin{eqnarray}
g_{5ij}^f&=&{\cal I}m\left[c_V^{Zkm}(c_{RS}^{lki}c_{LP}^{mlj}
+c_{LS}^{lki}c_{RP}^{mlj})+c_A^{Zkm}(c_{RS}^{lki}c_{LP}^{mlj}
-c_{LS}^{lki}c_{RP}^{mlj})\right]\times\nonumber\\
&\times&\left[{1\over 2} + p^2 c_{11}(m^2_m,m^2_l,m^2_k)
+p^2 c_{21}(m^2_m,m^2_l,m^2_k)\right.\nonumber\\
&+&(p^2+2pq+m_m^2)c_{12}(m^2_m,m^2_l,m^2_k)
+(q^2+2pq)c_{22}(m^2_m,m^2_l,m^2_k)\nonumber\\
&+&\left.2(p^2+pq)c_{23}(m^2_m,m^2_l,m^2_k)+4c_{24}(m^2_m,m^2_l,m^2_k)
-b_1(q^2,m^2_l,m^2_k)\right]\nonumber\\
&-& m_k m_l {\cal I}m\left[c_V^{Zkm}(c_{RS}^{lki}c_{RP}^{mlj}
+c_{LS}^{lki}c_{LP}^{mlj}))-c_A^{Zkm}(c_{RS}^{lki}c_{RP}^{mlj}
-c_{LS}^{lki}c_{LP}^{mlj})\right]\times\nonumber\\
&\times& c_{12}(m^2_m,m^2_l,m^2_k)\nonumber\\
&+&m_k m_m {\cal I}m\left[c_V^{Zkm}(c_{RS}^{lki}c_{LP}^{mlj}
+c_{LS}^{lki}c_{RP}^{mlj})
-c_A^{Zkm}(c_{RS}^{lki}c_{LP}^{mlj}
-c_{LS}^{lki}c_{RP}^{mlj})\right]\times\nonumber\\
&\times& c_{12}(m^2_m,m^2_l,m^2_k)\nonumber\\
&-&m_m m_l {\cal I}m\left[c_V^{Zkm}(c_{RS}^{lki}c_{RP}^{mlj}
+c_{LS}^{lki}c_{LP}^{mlj})
+c_A^{Zkm}(c_{RS}^{lki}c_{RP}^{mlj}
-c_{LS}^{lki}c_{LP}^{mlj})\right]\times\nonumber\\
&\times&\left[c_0(m^2_m,m^2_l,m^2_k)+c_{12}(m^2_m,m^2_l,m^2_k))\right]
\end{eqnarray}
Then:
\begin{eqnarray}
&\phantom{-}&\left(4\pi\right)^2F^{ij}_S=
-{2e\over s_W c_W}\sum_{K=1}^3\left[{m_{l_K}^2\over v_1^2} Z_H^{1j} Z_R^{1i}
{}~f_4^{ll}(m^2_{l_K}) - {3 m_{u_K}^2\over v_2^2} Z_H^{2j}
Z_R^{2i}~f_4^{uu}(m^2_{u_K})\right.\nonumber\\
&+&\left.{3 m_{d_K}^2\over v_1^2} Z_H^{1j}
Z_R^{1i}~f_4^{dd}(m^2_{d_K})\right]\nonumber\\
&+&4\sum_{k,l,m=1}^2 g_{5ij}^C(m^2_{C_m},m^2_{C_l},m^2_{C_k})+2\sum_{k,l,m=1}^4
g_{5ij}^N(m^2_{N_m},m^2_{N_l},m^2_{N_k})\nonumber\\
&+&2\sum_{k,l,m=1}^6\left[{\cal I}m
(V_{ZLL}^{lk}V_{PLL}^{jkm}V_{SLL}^{iml})~(c_0(m^2_{L_k},m^2_{L_l},m^2_{L_m})
+2c_{12}(m^2_{L_k},m^2_{L_l},m^2_{L_m}))\right.\nonumber\\
&-&3~{\cal I}m
(V_{ZUU}^{lk}V_{PUU}^{jkm}V_{SUU}^{iml})~(c_0(m^2_{U_k},m^2_{U_l},m^2_{U_m})
+2c_{12}(m^2_{U_k},m^2_{U_l},m^2_{U_m}))\nonumber\\
&+&\left.3~{\cal I}m
(V_{ZDD}^{lk}V_{PDD}^{jkm}V_{SDD}^{iml})~(c_0(m^2_{D_k},m^2_{D_l},m^2_{D_m})
+2c_{12}(m^2_{D_k},m^2_{D_l},m^2_{D_m}))\right]\nonumber\\
&+&{e^2(c_W^2-s_W^2)M_Z\over 2 s_W^2}\sum_{k,l=1}^2 \varepsilon_{kl}
V_{S+-}^{ikl}
{}~(c_0(m^2_{H^+_k},m^2_{H^+_l},m^2_{H^+_k})\nonumber\\
&+&2c_{12}(m^2_{H^+_k},m^2_{H^+_l},m^2_{H^+_k}))\nonumber\\
&+&{e^3 \over 8 s_W^3 c_W^3}\sum_{k,l,m=1}^2 A_M^{lk} A_H^{jk} B_R^m
V_{SSS}^{ilm}
{}~(c_0(m^2_{H^0_{k+2}},m^2_{H^0_m},m^2_{H^0_l})\nonumber\\
&+&2c_{12}(m^2_{H^0_{k+2}},m^2_{H^0_m},m^2_{H^0_l}))\nonumber\\
&-&{e^5 \over 32 s_W^5 c_W^5}B_R^i\sum_{k,l,m=1}^2 A_M^{kl} A_H^{jm} A_H^{lm}
B_R^k
{}~(c_0(m^2_{H^0_k},m^2_{H^0_{m+2}},m^2_{H^0_{l+2}})\nonumber\\
&+&2c_{12}(m^2_{H^0_k},m^2_{H^0_{m+2}},m^2_{H^0_{l+2}}))\nonumber\\
&-&{e^3 c_W\over 2
s_W^3}A_M^{ij}~(1+6c_{24}(M^2_W,m^2_{H^+_j},M^2_W)
+p^2(6c_0(M^2_W,m^2_{H^+_j},M^2_W)\nonumber\\
&+&7c_{11}(M^2_W,m^2_{H^+_j},M^2_W)
+c_{12}(M^2_W,m^2_{H^+_j},M^2_W)+2c_{21}(M^2_W,m^2_{H^+_j},M^2_W)\nonumber\\
&+&2c_{23}(M^2_W,m^2_{H^+_j},M^2_W))
+pq(2c_0(M^2_W,m^2_{H^+_j},M^2_W)+c_{11}(M^2_W,m^2_{H^+_j},M^2_W)\nonumber\\
&-&c_{12}(M^2_W,m^2_{H^+_j},M^2_W)
+2c_{22}(M^2_W,m^2_{H^+_j},M^2_W)+2c_{23}(M^2_W,m^2_{H^+_j},M^2_W)))\nonumber\\
&+&{e^4 M_Z c_W^2 \over 8 s_W^4} C_R^i
\delta^{j2}~c_0(M^2_W,M^2_W,M^2_W)\nonumber\\
&-&{e^4 M_Z\over 4 s_W^2} C_R^i \delta^{j2}~c_{12}(M^2_W,M^2_W,M^2_W)
+e^2 s_W M_Z V_{S+-}^{i2j}~c_{12}(M^2_W,m^2_{H^+_j},M^2_W)\nonumber\\
&+&{e^3\over 4 s_W c_W} M^2_W A_M^{ij}
\delta^{j1}~(c_{12}(M^2_W,m^2_{H^+_j},M^2_W)
-c_0(M^2_W,m^2_{H^+_j},M^2_W))\nonumber\\
&-&{e^3 \over 4 s_W^3 c_W^3}\sum_{k,l=1}^2 C_R^k A_M^{lj} V_{SSS}^{kli}
{}~c_{12}(M^2_Z,m^2_{H^0_l},m^2_{H^0_k})\nonumber\\
&-&{e^5 \over 16 s_W^5 c_W^5}\sum_{k,l=1}^2 C_R^k B_R^k A_H^{lj} A_M^{il}
{}~(c_{12}(m^2_{H^0_k},m^2_{H^0_{l+2}},M^2_Z)
-c_0(m^2_{H^0_k},m^2_{H^0_{l+2}},M^2_Z))\nonumber\\
&+&{e^5 \over 8 s_W^5 c_W^5}C_R^i\sum_{k,l=1}^2 C_R^k C_R^i
A_M^{kj}~c_{12}(m^2_{H^0_k},M^2_Z,M^2_Z)\nonumber\\
&+&{e^3 (c_W^2-s_W^2)\over 4 s_W^3  c_W} A_M^{ij}
{}~({1\over
2}+2m^2_{H^+_j}c_{12}(m^2_{H^+_j},M^2_W,m^2_{H^+_j})
-2b_1(q^2,M^2_W,m^2_{H^+_j})\nonumber\\
&+&8c_{24}(m^2_{H^+_j},M^2_W,m^2_{H^+_j})
+p^2(c_{21}(m^2_{H^+_j},M^2_W,m^2_{H^+_j})
-c_0(m^2_{H^+_j},M^2_W,m^2_{H^+_j})\nonumber\\
&-&2c_{12}(m^2_{H^+_j},M^2_W,m^2_{H^+_j}))
-2pq(c_0(m^2_{H^+_j},M^2_W,m^2_{H^+_j})\nonumber\\
&-&c_{11}(m^2_{H^+_j},M^2_W,m^2_{H^+_j})
+2c_{12}(m^2_{H^+_j},M^2_W,m^2_{H^+_j})-3c_{23}(m^2_{H^+_j},M^2_W,m^2_{H^+_j}))\nonumber\\
&+&q^2(2c_{12}(m^2_{H^+_j},M^2_W,m^2_{H^+_j})
+5c_{22}(m^2_{H^+_j},M^2_W,m^2_{H^+_j})))\nonumber\\
&-&{e^3 \over 8 s_W^3 c_W^3}\sum_{k,l=1}^2 A_M^{il} A_M^{kj} A_M^{kl}
{}~({1\over 2}-2b_1(q^2,M^2_Z,m^2_{H^0_{l+2}})\nonumber\\
&+&8c_{24}(m^2_{H^0_k},M^2_Z,m^2_{H^0_{l+2}})
+2m^2_{H^0_k}c_{12}(m^2_{H^0_k},M^2_Z,m^2_{H^0_{l+2}})\nonumber\\
&+&p^2(c_{21}(m^2_{H^0_k},M^2_Z,m^2_{H^0_{l+2}})
-c_0(m^2_{H^0_k},M^2_Z,m^2_{H^0_{l+2}})
-2c_{12}(m^2_{H^0_k},M^2_Z,m^2_{H^0_{l+2}}))\nonumber\\
&+&q^2(2c_{12}(m^2_{H^0_k},M^2_Z,m^2_{H^0_{l+2}})
+5c_{22}(m^2_{H^0_k},M^2_Z,m^2_{H^0_{l+2}}))\nonumber\\
&-&2pq(c_0(m^2_{H^0_k},M^2_Z,m^2_{H^0_{l+2}})
-c_{11}(m^2_{H^0_k},M^2_Z,m^2_{H^0_{l+2}})\nonumber\\
&+&2c_{12}(m^2_{H^0_k},M^2_Z,m^2_{H^0_{l+2}})
-3c_{23}(m^2_{H^0_k},M^2_Z,m^2_{H^0_{l+2}})))\nonumber\\
&-&{e^3 \over 4 s_W^3 c_W^3}
A_M^{ij}~(2b_0(q^2,M^2_Z,m^2_{H^0_{j+2}})+b_1(q^2,M^2_Z,m^2_{H^0_{j+2}}))\\
&-&{e^3 \over 2 s_W c_W}
A_M^{ij}~(2b_0(q^2,M^2_W,m^2_{H^+_j})+b_1(q^2,M^2_W,m^2_{H^+_j}))\nonumber\\
&-&{e^3 \over 8 s_W^3 c_W^3}\sum_{k,l=1}^2 A_M^{lk} A_R^{il} A_H^{kj}
{}~(b_0((p+q)^2,m^2_{H^0_{k+2}},m^2_{H^0_l})\nonumber\\
&+&2b_1((p+q)^2,m^2_{H^0_{k+2}},m^2_{H^0_l}))\nonumber
\end{eqnarray}

\subsection{Photon-$Z^0$-scalar vertex}
\noindent 1. Formfactor proportional to $g^{\mu\nu}$. Contribution from
the fermion triangle loops can be expressed in terms of $f_{1a}^{ff}(m^2)$,
$f_{1b}^{ff}(m^2)$ and (simpler then in the $Z^0$ case) auxiliary
function $g_{6i}^f$:
\begin{eqnarray}
g_{6i}^f&=&m_k~{\cal R}e\left[c_V^{Zmk} (c_L^{kmi}-c_R^{kmi})
-c_A^{Zmk}(c_L^{kmi}+c_R^{kmi})\right]\left[2
c_{24}(m^2_k,m^2_m,m^2_k)\right.\nonumber\\
&+&(p^2+pq) (c_0(m^2_k,m^2_m,m^2_k)+c_{11}(m^2_k,m^2_m,m^2_k))\nonumber\\
&+&\left.(q^2 + pq) c_{12}(m^2_k,m^2_m,m^2_k)\right]\nonumber\\
&-&m_m~{\cal R}e\left[c_V^{Zmk}(c_L^{kmi}-c_R^{kmi})
+c_A^{Zmk}(c_L^{kmi}+c_R^{kmi})\right]
\left[2c_{24}(m^2_k,m^2_m,m^2_k)\right.\nonumber\\
&+&{1\over 2} + (p^2+pq) c_{11}(m^2_k,m^2_m,m^2_k)
+(q^2+pq) c_{12}(m^2_k,m^2_m,m^2_k)\nonumber\\
&+&p^2 c_{21}(m^2_k,m^2_m,m^2_k)+q^2 c_{22}(m^2_k,m^2_m,m^2_k)
+2 pq~c_{23}(m^2_k,m^2_m,m^2_k)\nonumber\\
&-&\left.m_k^2 c_0(m^2_k,m^2_m,m^2_k)\right]
\end{eqnarray}
Using these functions one can write down expression for the $G^i_1$
formfactor as:
\begin{eqnarray}
&-&\left(4\pi\right)^2G^i_1=4e\sum_{k,m=1}^2
g_{6i}^C(m^2_{C_k},m^2_{C_m})\nonumber\\
&-&{e^2\over s_Wc_W}\sum^3_{K=1}\left[(1-4s_W^2)m^2_{l_K}{Z_R^{1i}\over v_1}
{}~f_{1b}^{ll}(m_{l_K}^2)+2(1-{8\over 3}s_W^2)m^2_{u_K}{Z_R^{2i}\over
v_2}~f_{1b}^{uu}(m_{u_K}^2)\right.\nonumber\\
&+&\left.(1-{4\over 3}s_W^2)m^2_{d_K}{Z_R^{1i}\over
v_1}~f_{1b}^{dd}(m_{d_K}^2)\right]\nonumber\\
&-&8e\sum_{k,l=1}^6\left[{\cal R}e
(V_{ZLL}^{lk}V_{SLL}^{ikl})~
c_{24}(m^2_{L_l},m^2_{L_k},m^2_{L_l})\right.\nonumber\\
&-&2~{\cal R}e
(V_{ZUU}^{kl}V_{SUU}^{ikl})~c_{24}(m^2_{U_l},m^2_{U_k},m^2_{U_l})\nonumber\\
&+&\left.{\cal R}e
(V_{ZDD}^{lk}V_{SDD}^{ikl})~
c_{24}(m^2_{D_l},m^2_{D_k},m^2_{D_l})\right]\nonumber\\
&+&{4e^2(c_W^2-s_W^2)\over s_W c_W}\sum_{j=1}^2
V_{S+-}^{ijj}~c_{24}(m^2_{H^+_j},m^2_{H^+_j},m^2_{H^+_j})\nonumber\\
&+&{e^4M^2_Z\over 2s_W c_W}~A_H^{22}~B_R^i~c_0(M^2_W,M^2_W,M^2_W)\nonumber\\
&-&{e^4\over 2 s_W
c_W}C_R^i~(p^2(c_0(M^2_W,M^2_W,M^2_W)-c_{21}(M^2_W,M^2_W,M^2_W))\nonumber\\
&+&q^2(2c_{12}(M^2_W,M^2_W,M^2_W)
-c_0(M^2_W,M^2_W,M^2_W)-c_{22}(M^2_W,M^2_W,M^2_W))\nonumber\\
&+&2pq~(c_{11}(M^2_W,M^2_W,M^2_W)
-c_{23}(M^2_W,M^2_W,M^2_W))-{1\over2}\nonumber\\
&+&b_0(p^2,M^2_W,M^2_W)+b_0((p+q)^2,M^2_W,M^2_W))\nonumber\\
&+&{e^4 c_W \over 2 s_W}C_R^i M_Z^2 c_0(M^2_W,M^2_W,M^2_W)\nonumber\\
&+&{e^4 c_W \over 2
s_W^3}C_R^i~(32c_{24}(M^2_W,M^2_W,M^2_W)+6b_0(q^2,M^2_W,M^2_W)-{5\over
2}\nonumber\\
&+&p^2(11c_0(M^2_W,M^2_W,M^2_W)
+4c_{11}(M^2_W,M^2_W,M^2_W)+3c_{21}(M^2_W,M^2_W,M^2_W))\nonumber\\
&+&3q^2c_{22}(M^2_W,M^2_W,M^2_W)+2pq(6c_0(M^2_W,M^2_W,M^2_W)\nonumber\\
&+&2c_{12}(M^2_W,M^2_W,M^2_W)+3c_{22}(M^2_W,M^2_W,M^2_W)))\nonumber\\
&-&2e\sum^6_{k,l=1}\left[V_{ZLL}^{lk}V_{SLL}^{ikl}~
b_0(q^2,m^2_{L_l},m^2_{L_k})\right.\nonumber\\
&-&\left.2V_{ZUU}^{lk}V_{SUU}^{ikl}~b_0(q^2,m^2_{U_l},m^2_{U_k})
+V_{ZDD}^{lk}V_{SDD}^{ikl}~b_0(q^2,m^2_{D_l},m^2_{D_k})\right]\nonumber\\
&+&{e^2(c_W^2-s_W^2)\over s_W c_W}\sum^2_{j=1}
V_{S+-}^{ijj}~b_0(q^2,m^2_{H^+_j},m^2_{H^+_j})
\end{eqnarray}

\noindent 2. Formfactor proportional to $q^{\mu}q^{\nu}$. Lets define:
\begin{eqnarray}
g_{7i}^C&=&m_k~{\cal R}e\left[c_V^{Zmk} (c_L^{kmi}-c_R^{kmi})
-c_A^{Zmk}(c_L^{kmi}+c_R^{kmi})\right]c_{22}(m^2_k,m^2_m,m^2_k)\nonumber\\
&+&m_m~{\cal R}e\left[c_V^{Zmk}(c_L^{kmi}-c_R^{kmi})
+c_A^{Zmk}(c_L^{kmi}+c_R^{kmi})\right]
\left[c_{12}(m^2_k,m^2_m,m^2_k)\right.\nonumber\\
&+&\left.c_{22}(m^2_k,m^2_m,m^2_k)\right]
\end{eqnarray}
Then:
\begin{eqnarray}
&-&\left(4\pi\right)^2G^i_3=8e\sum_{k,m=1}^2
g_{7i}^C(m^2_{C_k},m^2_{C_m})\nonumber\\
&-&{4e^2\over s_Wc_W}\sum^3_{K=1}\left[m^2_{l_K}{Z_R^{1i}\over v_1}(1-4s_W^2)
{}~(c_{12}(m^2_{l_K},m^2_{l_K},m^2_{l_K})
+2c_{22}(m^2_{l_K},m^2_{l_K},m^2_{l_K}))\right.\nonumber\\
&+&2m^2_{u_K}{Z_R^{2i}\over v_2}(1-{8\over 3}s_W^2)
{}~(c_{12}(m^2_{u_K},m^2_{u_K},m^2_{u_K})
+2c_{22}(m^2_{u_K},m^2_{u_K},m^2_{u_K}))\nonumber\\
&+&\left.m^2_{d_K}{Z_R^{1i}\over v_1}(1-{4\over 3}s_W^2)
{}~(c_{12}(m^2_{d_K},m^2_{d_K},m^2_{d_K})
+2c_{22}(m^2_{d_K},m^2_{d_K},m^2_{d_K}))\right]\nonumber\\
&-&4e\sum_{k,l=1}^6\left[{\cal R}e
(V_{ZLL}^{lk}V_{SLL}^{ikl})~(c_{12}(m^2_{L_l},m^2_{L_k},m^2_{L_l})
+2c_{22}(m^2_{L_l},m^2_{L_k},m^2_{L_l}))\right.\nonumber\\
&-&2~{\cal R}e
(V_{ZUU}^{kl}V_{SUU}^{ikl})~(c_{12}(m^2_{U_l},m^2_{U_k},m^2_{U_l})
+2c_{22}(m^2_{U_l},m^2_{U_k},m^2_{U_l}))\nonumber\\
&+&\left.{\cal R}e
(V_{ZDD}^{lk}V_{SDD}^{ikl})~(c_{12}(m^2_{D_l},m^2_{D_k},m^2_{D_l})
+2c_{22}(m^2_{D_l},m^2_{D_k},m^2_{D_l}))\right]\nonumber\\
&+&{2e^2(c_W^2-s_W^2)\over s_W c_W}\sum^2_{j=1}\left[
V_{S+-}^{ijj}~(c_{12}(m^2_{H^+_j},m^2_{H^+_j},m^2_{H^+_j})\right.\nonumber\\
&+&\left.2c_{22}(m^2_{H^+_j},m^2_{H^+_j},m^2_{H^+_j}))\right]\nonumber\\
&-&{e^4 \over 2 s_W c_W} C_R^i~(3c_0(M^2_W,M^2_W,M^2_W)
+2c_{12}(M^2_W,M^2_W,M^2_W)\nonumber\\
&+&4c_{22}(M^2_W,M^2_W,M^2_W))\nonumber\\
&+&{5 e^4c_W \over s_W^3}C_R^i~(c_{12}(M^2_W,M^2_W,M^2_W)
+2c_{22}(M^2_W,M^2_W,M^2_W))
\end{eqnarray}

\noindent 3. Formfactor proportional to $p^{\mu}q^{\nu}$. Lets define:
\begin{eqnarray}
g_{8i}^C&=&m_k~{\cal R}e\left[c_V^{Zmk} (c_L^{kmi}-c_R^{kmi})
-c_A^{Zmk}(c_L^{kmi}+c_R^{kmi})\right]
\left[c_{12}(m^2_k,m^2_m,m^2_k)\right.\nonumber\\
&+&\left.2 c_{23}(m^2_k,m^2_m,m^2_k)-c_0(m^2_k,m^2_m,m^2_k)
- c_{11}(m^2_k,m^2_m,m^2_k)\right]\nonumber\\
&+&m_m~{\cal R}e\left[c_V^{Zmk}(c_L^{kmi}-c_R^{kmi})
+c_A^{Zmk}(c_L^{kmi}+c_R^{kmi})\right]
\left[c_{11}(m^2_k,m^2_m,m^2_k)\right.\nonumber\\
&+&\left. c_{12}(m^2_k,m^2_m,m^2_k) + 2 c_{23}(m^2_k,m^2_m,m^2_k)\right]
\end{eqnarray}
Then
\begin{eqnarray}
&-&\left(4\pi\right)^2 G^i_4=4e\sum_{k,m=1}^2
g_{8i}^C(m^2_{C_k},m^2_{C_m})\nonumber\\
&-&{4e^2\over s_W c_W}\sum_{K=1}^3\left[(1-4s_W^2) m_{l_K}^2 {Z_R^{1i}\over
v_1}
{}~f_{2b}^{ll}(m^2_{l_K}) + 2 (1-{8\over 3}s_W^2) m_{u_K}^2 {Z_R^{2i}\over
v_2}~f_{2b}^{uu}(m^2_{u_K})\right.\nonumber\\
&+&\left. (1-{4\over 3}s_W^2) m_{d_K}^2 {Z_R^{1i}\over
v_1}~f_{2b}^{dd}(m^2_{d_K})\right]\nonumber\\
&-& 4e \sum_{k,l=1}^6\left[{\cal R}e (V_{ZLL}^{lk}V_{SLL}^{ikl})
{}~(c_{12}(m^2_{L_l},m^2_{L_k},m^2_{L_l})
+2c_{23}(m^2_{L_l},m^2_{L_k},m^2_{L_l}))\right.\nonumber\\
&-&2~{\cal R}e (V_{ZUU}^{kl}V_{SUU}^{ikl})
(c_{12}(m^2_{U_l},m^2_{U_k},m^2_{U_l})
+2c_{23}(m^2_{U_l},m^2_{U_k},m^2_{U_l}))\nonumber\\
&+&{\cal R}e (V_{ZDD}^{lk}V_{SDD}^{ikl})
{}~(c_{12}(m^2_{D_l},m^2_{D_k},m^2_{D_l})
+2c_{23}(m^2_{D_l},m^2_{D_k},m^2_{D_l}))\nonumber\\
&+&{2e^2(c_W^2-s_W^2)\over s_W c_W}\sum_{j=1}^2 V_{S+-}^{ijj}
{}~(c_{12}(m^2_{H^+_j},m^2_{H^+_j},m^2_{H^+_j})
+2c_{23}(m^2_{H^+_j},m^2_{H^+_j},m^2_{H^+_j}))\nonumber\\
&-&{e^4\over 2 s_W c_W} C_R^i~(2c_{12}(M^2_W,M^2_W,M^2_W)
+4c_{23}(M^2_W,M^2_W,M^2_W)\nonumber\\
&-&c_0(M^2_W,M^2_W,M^2_W))\nonumber\\
&+&{e^4 c_W\over s_W^3} C_R^i~(-6c_0(M^2_W,M^2_W,M^2_W)
+5c_{12}(M^2_W,M^2_W,M^2_W)\nonumber\\
&+&10c_{23}(M^2_W,M^2_W,M^2_W))
\end{eqnarray}

\subsection{Photon-scalar-pseudoscalar vertex}

\noindent 1. Formfactor proportional to the momentum of the
pseudoscalar. Lets define:
\begin{eqnarray}
g_{9ij}^f&=&{\cal I}m\left[c_{RS}^{lki}c_{LP}^{klj}
+c_{LS}^{lki}c_{RP}^{klj}\right]\left[m_k^2
c_0(m^2_k,m^2_l,m^2_k)\right.\nonumber\\
&+&(p^2+2 m_k^2)c_{11}(m^2_k,m^2_l,m^2_k)-q^2
c_{12}(m^2_k,m^2_l,m^2_k)+2p^2 c_{21}(m^2_k,m^2_l,m^2_k)\nonumber\\
&+&\left.2pq~c_{23}(m^2_k,m^2_l,m^2_k)
+2c_{24}(m^2_k,m^2_l,m^2_k)+b_0(q^2,m^2_l,m^2_k))\right]\\
&-&m_l m_k {\cal I}m\left[c_{RS}^{lki}c_{RP}^{klj}
+c_{LS}^{lki}c_{LP}^{klj}\right]
\left[c_0(m^2_k,m^2_l,m^2_k)+2c_{11}(m^2_k,m^2_l,m^2_k))\right]\nonumber
\end{eqnarray}
Then:
\begin{eqnarray}
&\phantom{-}&\left(4\pi\right)^2G^{ij}_P= 4e\sum_{k,l=1}^2
g_{9ij}^C(m^2_{C_k},m^2_{C_l})\nonumber\\
&-&2e\sum_{k,m=1}^6\left[{\cal I}m
(V_{PLL}^{jkm}V_{SLL}^{imk})~(c_0(m^2_{L_k},m^2_{L_m},m^2_{L_k})
+2c_{11}(m^2_{L_k},m^2_{L_m},m^2_{L_k}))\right.\nonumber\\
&+&2~{\cal I}m
(V_{PUU}^{jkm}V_{SUU}^{imk})~(c_0(m^2_{U_k},m^2_{U_m},m^2_{U_k})
+2c_{11}(m^2_{U_k},m^2_{U_m},m^2_{U_k}))\nonumber\\
&+&\left.{\cal I}m
(V_{PDD}^{jkm}V_{SDD}^{imk})~(c_0(m^2_{D_k},m^2_{D_m},m^2_{D_k})
+2c_{11}(m^2_{D_k},m^2_{D_m},m^2_{D_k}))\right]\nonumber\\
&-&e^2M_W V_{S+-}^{i2j}~(c_{11}(M^2_W,m^2_{H^+_j},M^2_W)
+2c_0(M^2_W,m^2_{H^+_j},M^2_W))\nonumber\\
&+&{e^2M_W\over s_W}\sum_{k,l=1}^2
\varepsilon_{kl}V_{S+-}^{ikl}(c_0(m^2_{H^+_k},m^2_{H^+_l},m^2_{H^+_k})
+2c_{11}(m^2_{H^+_k},m^2_{H^+_l},m^2_{H^+_k}))\nonumber\\
&+&{e^3\over 2s^2_W}A_M^{ij}~(1+6c_{24}(M^2_W,m^2_{H^+_j},M^2_W)
+q^2(2c_0(M^2_W,m^2_{H^+_j},M^2_W)\nonumber\\
&+&c_{11}(M^2_W,m^2_{H^+_j},M^2_W)
-c_{12}(M^2_W,m^2_{H^+_j},M^2_W)+2c_{22}(M^2_W,m^2_{H^+_j},M^2_W)\nonumber\\
&+&2c_{23}(M^2_W,m^2_{H^+_j},M^2_W))
+pq(6c_0(M^2_W,m^2_{H^+_j},M^2_W)\nonumber\\
&+&7c_{11}(M^2_W,m^2_{H^+_j},M^2_W)+c_{12}(M^2_W,m^2_{H^+_j},M^2_W)\nonumber\\
&+&2c_{21}(M^2_W,m^2_{H^+_j},M^2_W)
+2c_{23}(M^2_W,m^2_{H^+_j},M^2_W)))\nonumber\\
&-&{e^4M_W\over 8s^3_W}C_R^i\delta^{j2}~(c_0(M^2_W,M^2_W,M^2_W)
-2c_{11}(M^2_W,M^2_W,M^2_W))\\
&-&{e^3M_W^2\over 4s_W^2}A_M^{ij}\delta^{j1}~(c_{11}(M^2_W,m^2_{H^+_j},M^2_W)
+c_0(M^2_W,m^2_{H^+_j},M^2_W))\nonumber\\
&+&{e^3\over 2s_W^2}A_M^{ij}~({1\over 2}
+2b_0(q^2,M^2_W,m^2_{H^+_j})+4c_{24}(m^2_{H^+_j},M^2_W,m^2_{H^+_j})\nonumber\\
&+&2m^2_{H^+_j}~c_{11}(m^2_{H^+_j},M^2_W,m^2_{H^+_j})
+p^2(c_{21}(m^2_{H^+_j},M^2_W,m^2_{H^+_j})\nonumber\\
&-&c_0(m^2_{H^+_j},M^2_W,m^2_{H^+_j})
-2c_{11}(m^2_{H^+_j},M^2_W,m^2_{H^+_j}))\nonumber\\
&+&q^2(2c_{12}(m^2_{H^+_j},M^2_W,m^2_{H^+_j})
+c_{22}(m^2_{H^+_j},M^2_W,m^2_{H^+_j})\nonumber\\
&+&4c_{23}(m^2_{H^+_j},M^2_W,m^2_{H^+_j}))
+2pq(2c_{21}(m^2_{H^+_j},M^2_W,m^2_{H^+_j})\nonumber\\
&+&c_{23}(m^2_{H^+_j},M^2_W,m^2_{H^+_j})
-c_0(m^2_{H^+_j},M^2_W,m^2_{H^+_j})
-c_{11}(m^2_{H^+_j},M^2_W,m^2_{H^+_j}))\nonumber\\
&-&2b_0(p^2,M^2_W,m^2_{H^+_j})-b_1(p^2,M^2_W,m^2_{H^+_j}))\nonumber
\end{eqnarray}

\noindent 2. Formfactor proportional to the momentum of the scalar. Lets
define:
\begin{eqnarray}
g_{10ij}^f&=& {\cal I}m\left[c_{RS}^{lki}c_{LP}^{klj}
+c_{LS}^{lki}c_{RP}^{klj}\right]\left[{1\over 2}
+ p^2 c_{11}(m^2_k,m^2_l,m^2_k)\right.\nonumber\\
&+&p^2
c_{21}(m^2_k,m^2_l,m^2_k)+(p^2+2pq+2m_k^2)c_{12}(m^2_k,m^2_l,m^2_k)\nonumber\\
&+&(q^2+2pq)c_{22}(m^2_k,m^2_l,m^2_k)
+2(p^2+pq)c_{23}(m^2_k,m^2_l,m^2_k)\nonumber\\
&+&\left.4c_{24}(m^2_k,m^2_l,m^2_k)-b_1(q^2,m^2_l,m^2_k)\right]\\
&-&m_l m_k {\cal I}m\left[c_{RS}^{lki}c_{RP}^{klj}
+c_{LS}^{lki}c_{LP}^{klj}\right]
\left[c_0(m^2_k,m^2_l,m^2_k)+2c_{12}(m^2_k,m^2_l,m^2_k))\right]\nonumber
\end{eqnarray}
Then:
\begin{eqnarray}
&\phantom{-}&\left(4\pi\right)^2G^{ij}_S=
 4e\sum_{k,l=1}^2 g_{10ij}^C(m^2_{C_k},m^2_{C_l})\nonumber\\
&-&2e\sum^6_{k,m=1}\left[
{\cal I}m  (V_{PLL}^{jkm}V_{SLL}^{imk})~(c_0(m^2_{L_k},m^2_{L_m},m^2_{L_k})
+2c_{12}(m^2_{L_k},m^2_{L_m},m^2_{L_k}))\right.\nonumber\\
&+&2~{\cal I}m  (V_{PUU}^{jkm}V_{SUU}^{imk})
{}~(c_0(m^2_{U_k},m^2_{U_m},m^2_{U_k})
+2c_{12}(m^2_{U_k},m^2_{U_m},m^2_{U_k}))\nonumber\\
&+&\left.{\cal I}m  (V_{PDD}^{jkm}V_{SDD}^{imk})
{}~(c_0(m^2_{D_k},m^2_{D_m},m^2_{D_k})
+2c_{12}(m^2_{D_k},m^2_{D_m},m^2_{D_k}))\right]\nonumber\\
&-&e^2M_WV_{S+-}^{i2j}~c_{12}(M^2_W,m^2_{H^+_j},M^2_W)\nonumber\\
&+&{e^2M_W\over s_W}\sum^2_{k,l=1}
\varepsilon_{kl}V_{S+-}^{ikl}~(c_0(m^2_{H^+_k},m^2_{H^+_l},m^2_{H^+_k})
+2c_{12}(m^2_{H^+_k},m^2_{H^+_l},m^2_{H^+_k}))\nonumber\\
&-&{e^3\over 2s_W^2}A_M^{ij}~(1+6c_{24}(M^2_W,m^2_{H^+_j},M^2_W)
+p^2(6c_0(M^2_W,m^2_{H^+_j},M^2_W)\nonumber\\
&+&7c_{11}(M^2_W,m^2_{H^+_j},M^2_W)
+c_{12}(M^2_W,m^2_{H^+_j},M^2_W)+2c_{21}(M^2_W,m^2_{H^+_j},M^2_W)\nonumber\\
&+&2c_{23}(M^2_W,m^2_{H^+_j},M^2_W))+pq(2c_0(M^2_W,m^2_{H^+_j},M^2_W)
+c_{11}(M^2_W,m^2_{H^+_j},M^2_W)\nonumber\\
&-&c_{12}(M^2_W,m^2_{H^+_j},M^2_W)
+2c_{22}(M^2_W,m^2_{H^+_j},M^2_W)+2c_{23}(M^2_W,m^2_{H^+_j},M^2_W)))\nonumber\\
&+&{e^4M_W\over
8s_W^3}C_R^i\delta^{j2}~(c_0(M^2_W,M^2_W,M^2_W)+2c_{12}(M^2_W,M^2_W,M^2_W)\\
&-&{e^3 M^2_W\over 4s_W^2}A_M^{ij}\delta^{j1}~(c_{12}(M^2_W,m^2_{H^+_j},M^2_W)
-c_0(M^2_W,m^2_{H^+_j},M^2_W))\nonumber\\
&+&{e^3\over 2s_W^2}A_M^{ij}~({1\over 2}
-b_1(q^2,M^2_W,m^2_{H^+_j})+8c_{24}(m^2_{H^+_j},M^2_W,m^2_{H^+_j})\nonumber\\
&+&2m^2_{H^+_j}~c_{12}(m^2_{H^+_j},M^2_W,m^2_{H^+_j})
+p^2(c_{21}(m^2_{H^+_j},M^2_W,m^2_{H^+_j})\nonumber\\
&-&c_0(m^2_{H^+_j},M^2_W,m^2_{H^+_j})
-2c_{12}(m^2_{H^+_j},M^2_W,m^2_{H^+_j}))\nonumber\\
&+&q^2(2c_{12}(m^2_{H^+_j},M^2_W,m^2_{H^+_j})
+5c_{22}(m^2_{H^+_j},M^2_W,m^2_{H^+_j}))\nonumber\\
&-&2pq(c_0(m^2_{H^+_j},M^2_W,m^2_{H^+_j})
-c_{11}(m^2_{H^+_j},M^2_W,m^2_{H^+_j})\nonumber\\
&+&2c_{12}(m^2_{H^+_j},M^2_W,m^2_{H^+_j})
-3c_{23}(m^2_{H^+_j},M^2_W,m^2_{H^+_j}))
+2b_0(q^2,M^2_W,m^2_{H^+_j}))\nonumber
\end{eqnarray}

\newpage
\pagestyle{empty}
{\Large Figure Captions}
\begin{description}
\item[Figure 1.] Diagrams contributing at the 1-loop level to the processes
$e^+e^-\rightarrow Z^0h^0(H^0)$ and $e^+e^-\rightarrow A^0h^0(H^0)$,
included in the calculations of Section~4.
\item [Figure 2a.] Allowed mass regions for $h^0$ and $H^0$ for the different
values of $m_t$ and supersymmetric parameters. Supersymmetric parameters
$(M_{sq}, M_{sl}, M_{gau}, \mu, A)$ read (in GeV): for $m_t=120$ GeV:
solid (1000, 300, 1000, 500, 0), long-dashed  (200, 100, 60, 50, 0),
medium-dashed (1000, 300, 1000, 100, 1000),
short-dashed (1000, 300,1000, 500, 1000);
for $m_t=180$ GeV:
solid (1000, 300, 500, 250, 500), long-dashed (200, 100, 100, 100, 0),
medium-dashed (1000, 300, 1000, 100, 1000).
\item [Figure 2b.] Allowed mass regions for $h^0$ and $H^0$ and contour lines
of
fixed $\tan\beta$.
\item [Figure 3.] Cross sections for the lighter scalar production as a
function
of $M_h$ mass for fixed values of $M_A$.
\item [Figure 4a.] Regions in the ($M_A,M_h$) plane
in which at least one of the cross sections
$e^+e^- \rightarrow Z^0 h^0$ or $e^+e^- \rightarrow A^0 h^0$ is larger
than assumed $\sigma_0$, for the top quark mass $m_t = 140$ GeV and for
the different values of the $\sigma_0$ and $CMS$ energy.
\item [Figure 4b.] Regions in the ($M_A,M_h$) plane
in which at least one of the cross sections
$e^+e^- \rightarrow Z^0 h^0$ or $e^+e^- \rightarrow A^0 h^0$ is larger
than assumed $\sigma_0$, for the top quark mass $m_t = 180$ GeV and for
the different values of the $\sigma_0$ and $CMS$ energy.
\item [Figure 4c.] Regions in the ($M_A,M_H$) plane
in which at least one of the cross sections
$e^+e^- \rightarrow Z^0 H^0$ or $e^+e^- \rightarrow A^0 H^0$ is larger
than assumed $\sigma_0$, for the top quark masses $m_t = $ 140 and 180
GeV and for different values of the $\sigma_0$ and $CMS$ energy.
\item [Figure 5a.] Regions in the ($M_A,\tan\beta$) plane
in which at least one of the cross sections
$e^+e^- \rightarrow Z^0 h^0$, $e^+e^- \rightarrow A^0 h^0$ or
$e^+e^- \rightarrow Z^0 H^0$, $e^+e^- \rightarrow A^0 H^0$
is larger than assumed $\sigma_0$, for the top quark mass $m_t = 140$ GeV
and for different values of the $\sigma_0$ and $CMS$ energy.
\item [Figure 5b.] Regions in the ($M_A,\tan\beta$) plane
in which at least one of the cross sections
$e^+e^- \rightarrow Z^0 h^0$, $e^+e^- \rightarrow A^0 h^0$ or
$e^+e^- \rightarrow Z^0 H^0$, $e^+e^- \rightarrow A^0 H^0$
is larger than assumed $\sigma_0$, for the top quark mass $m_t = 180$ GeV
and for different values of the $\sigma_0$ and $CMS$ energy.
\item [Figure 5c.] Regions in the ($M_A,\tan\beta$) plane
in which at least one of the cross sections
$e^+e^- \rightarrow Z^0 h^0$, $e^+e^- \rightarrow A^0 h^0$ or
$e^+e^- \rightarrow Z^0 H^0$, $e^+e^- \rightarrow A^0 H^0$
is larger than assumed $\sigma_0$, for the top quark mass $m_t = 140$
GeV and $m_t = 180$ GeV, $CMS$ energy 500 GeV
and for different values of the $\sigma_0$.
\end{description}

\newpage
\begin{center}
{\LARGE FIGURE 1}
\newpage
{\LARGE FIGURE 2a}
\newpage
{\LARGE FIGURE 2b}
\newpage
{\LARGE FIGURE 3}
\newpage
{\LARGE FIGURE 4a}
\newpage
{\LARGE FIGURE 4b}
\newpage
{\LARGE FIGURE 4c}
\newpage
{\LARGE FIGURE 5a}
\newpage
{\LARGE FIGURE 5b}
\newpage
{\LARGE FIGURE 5c}
\end{center}

\end{document}